\documentclass[aps,prx,twocolumn,showpacs,floatfix,superscriptaddress,graphics,longbibliography]{revtex4-1}

\usepackage{amsmath,amssymb,graphicx,color,braket,hyphenat,makeidx,xcolor}
\usepackage[ocgcolorlinks,colorlinks=true,linkcolor=blue,citecolor=red,linktocpage=true]{hyperref}

\usepackage{bm}
\usepackage{subfigure}
\usepackage{epstopdf}
\usepackage{braket}
\usepackage{enumerate}

\newcommand{\tr}{\mathrm{Tr}}
\newcommand{\E}{{\mathcal E}}
\newcommand{\D}{{\mathcal D}}
\newcommand{\HAT}{}

\begin{document}
 
\title{Quantum fluctuation theorems for arbitrary environments: \\ adiabatic and non-adiabatic entropy production}
\author{Gonzalo Manzano}
\email{gmanzano@ucm.es}
\affiliation{Departamento de F\'isica At\'omica, Molecular y Nuclear and  GISC, Universidad Complutense Madrid, 28040 Madrid, Spain}
\affiliation{IFISC (UIB-CSIC), Instituto de F\'isica Interdisciplinar y Sistemas Complejos, UIB Campus, E-07122 Palma de Mallorca, Spain}
\author{Jordan M. Horowitz}
\affiliation{Physics of Living Systems Group, Department of Physics,
Massachusetts Institute of Technology, 400 Technology Square, Cambridge, MA 02139}
\author{Juan M.R. Parrondo}
\affiliation{Departamento de F\'isica At\'omica, Molecular y Nuclear and  GISC, Universidad Complutense Madrid, 28040 Madrid, Spain}


\date{\today}

\begin{abstract}
We analyze the production of entropy along non-equilibrium processes in quantum systems coupled to generic environments. 
First, we show that the entropy production due to final measurements and the loss of correlations obeys a fluctuation theorem in detailed and integral forms. 
Second, we discuss the decomposition of the entropy production into two positive contributions, adiabatic and non-adiabatic, based on the existence of invariant states of the local dynamics. 
Fluctuation theorems for both contributions hold only for evolutions verifying a specific condition of quantum origin.
We illustrate our results with three relevant examples of quantum thermodynamic processes far from equilibrium.
\end{abstract}

\maketitle
 
\section{Introduction}

Classical thermodynamics and statistical mechanics provide a systematic approach to the phenomenology of a system immersed in a large environment.
Within these frameworks, two complementary strategies are employed.
The first is to explicitly model the environment ---often an equilibrium thermal reservoir--- to obtain an effective reduced dynamics for the system alone, which then can be analyzed.
The second is to derive fundamental constraints in the form of inequalities using the second law of thermodynamics and magnitudes like entropy, entropy production and free energy.
The recent introduction of an entropy for stochastic trajectories \cite{SeifertPRL} allows one to refine these inequalities with exact equalities for arbitrary nonequilibrium processes, 
results generically known as fluctuation theorems (FT's) \cite{SeifertREV, JarzynskiREV}.

These two strategies  have also been successfully applied to quantum systems.
Open quantum system dynamics  ---the determination and analysis of the system's reduced dynamics--- is a well-developed and active field~\cite{BreuerBook, Rivas}.
Complementing this approach, a variety of quantum FT's have been derived~\cite{CampisiREV,EspositoREV, LutzEP, Gaspard2013, Watanabe:2014fh, Imparato} to asses the statistics of the relevant quantities. 
Different proposals to obtain these statistics in the laboratory have been reported, using techniques related to quantum tomography \cite{Dorner:2013, Mazzola:2013, Campisi:2013, Goold:2013, Roncaglia:2014, Chiara:2015}, 
and some of them have already been used to carry out experimental verifications of FT's \cite{Batalhao:2014ta, An:2015}.
However, most of the research on quantum FT's is only valid for equilibrium reservoirs with a focus on the energy exchange between the system and the environment in the form of heat and work. 
By contrast, classical FT's have been formulated more generally for generic Markov systems~\cite{HatanoSasa,EspositoFaces,EspositoFacesI,EspositoFacesII, Bisker} using the entropy production instead 
of heat and work, which are only meaningful in physical situations where a system exchanges energy with equilibrium reservoirs.

In light of the success of classical FT's, it is desirable to obtain complementary FT's for generic quantum dynamics \cite{Vedral, Kafri2012, Chetrite2012, Albash:2013fq, Rastegin:2014hc, DeffnerPS, JordanParrondo, Funo2013, MHP, Alhambra, Park}.
They could be of particular relevance, since quantum mechanics allows for a richer phenomenology in finite baths \cite{PekolaCalorimeter, Gaspinaretti, Suomela}, as well as novel and interesting non-thermal environments such as 
coherent \cite{Scully, Superradiant}, correlated \cite{LutzCorr}, or squeezed \cite{JaposSqueez,LutzSqueez,CorreaSqz, SqzRes} reservoirs. Such environments induce an interesting phenomenology that goes beyond the 
thermodynamics of thermal equilibrium reservoirs, such as heat engines that outperform Carnot efficiency \cite{Klaers} and may exhibit new regimes of operation \cite{SqzRes, Niedenzu} or tighter bounds on Landauer's principle \cite{ReebWolf,GooldLand}.

The task of deriving FT's for generic quantum dynamics also implies a more detailed characterization of entropy production in nonequilibrium quantum contexts, a problem that has attracted a growing interest in recent years \cite{EspositoNJP, LutzEP, SagawaEntropies, JordanParrondo, JordanSagawa, Leggio, DeffnerPS, MHP, Auffeves, Auffeves2, Santos, Lewestein}.
In Ref.~\cite{MHP} we derived a FT for a class of completely-positive trace-preserving (CPTP) quantum maps, which model a variety of quantum processes. 
Through this analysis, we identified a quantity that coincides with the entropy production for thermalization processes and resembles the nonadiabatic entropy production introduced in the classical context \cite{EspositoFaces, EspositoFacesI,EspositoFacesII}. 
The purpose of this paper is to clarify and extend those results considering together the system and its surroundings.
By tracing over the environment, we can then recover the quantum map for the reduced system dynamics.
This setup allows us to unveil the origin of entropy production in arbitrary processes from coarse-graining, and derive corresponding FT's.
We also split the entropy production into an adiabatic and a nonadiabatic contribution, exactly as in classical stochastic thermodynamics. 
However, contrary to what happens in classical systems, the split is not always possible. A condition, derived in Ref.~\cite{MHP}, is necessary. We will explore the similarities and differences between classical and quantum FT's in concrete examples.
 
The paper is organized as follows. In Section \ref{S-Model}, we introduce a thermodynamic process for a generic bipartite system that models a system and its environment. We will define in this section the entropy production along the process and the concomitant reduced system dynamics. 
We develop a FT for this entropy production in Section \ref{S-TotalEP} using a time-reversed or backward thermodynamic process.
In Section \ref{S-DecompositionFTs}, FT's for the adiabatic and nonadiabatic entropy production are derived. 
Our results are also extended both to the case of concatenations of CPTP maps and multipartite environments. 
This is applied to the specific case of quantum trajectories unraveled from Lindblad Master Equations in \ref{S-MarkovianME}. 
Finally, relevant examples to illustrate our results are given in Section \ref{S-Examples} while concluding in Section \ref{S-Conclusions} with some final remarks.

\section{Quantum operations and entropy production} \label{S-Model}

\begin{figure*}[t]\label{Figprocess}
\begin{center}
\includegraphics[width=0.8 \textwidth]{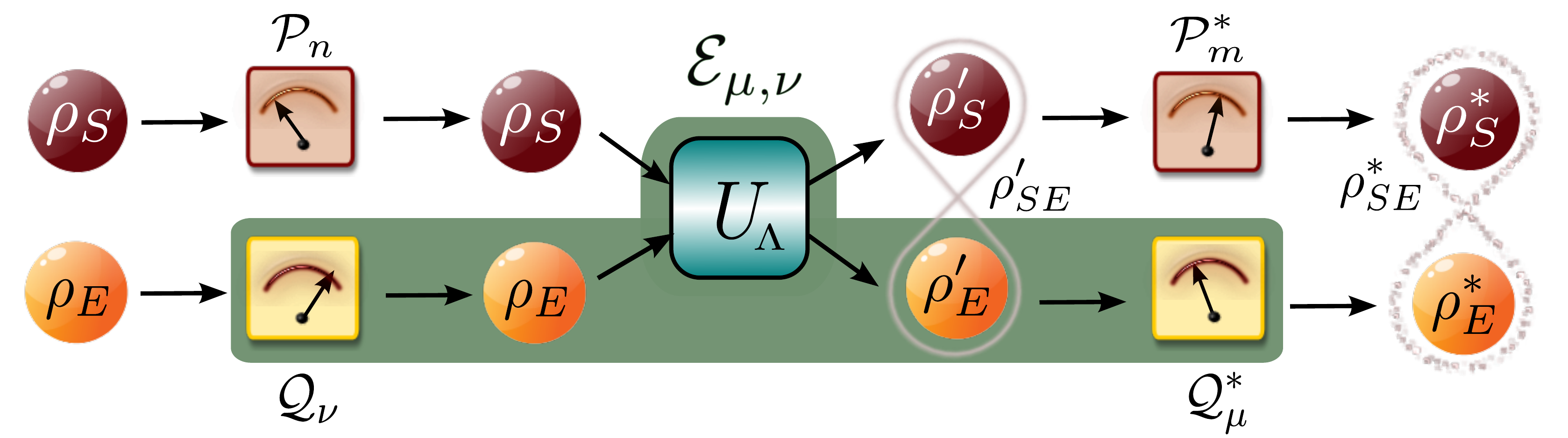}
\caption{Schematic picture of the forward process presented in the main text. 
System and environment start from an uncorrelated state $\rho_S \otimes \rho_E$. A local measurement of observables with projectors $\{\HAT{\mathcal{P}}_n, \HAT{\mathcal{Q}}_\nu \}$ is carried out, which does not alter the density matrix in the average evolution but selects a pure state $\ket{\psi_n}\otimes \ket{\phi_\nu}$ at the trajectory level. 
System and environment then  interact with each other according to the unitary evolution $\HAT{U}_\Lambda$, ending in an entangled state $\rho'_{SE}$. Finally, we measure again now using projectors $\{\HAT{\mathcal{P}}_m^\ast, \HAT{\mathcal{Q}}_\mu^\ast. \}$. 
In the last measurement quantum correlations in state $\rho_{SE}'$ are erased, while the final state $\rho_{SE}^\ast$ may still have in general non-zero classical correlations. 
The reduced evolution of the system conditioned to the measurement in the environment is described through the quantum operation $\E_{\mu \nu}$ (shaded area). } 
\end{center}
\end{figure*}

Along the paper we consider an isolated quantum system composed of two parts, system and environment (or ancilla), with Hilbert space $\mathcal{H} \equiv \mathcal{H}_{S} \otimes \mathcal{H}_E$, 
where $\mathcal{H}_{S}$ and $\mathcal{H}_{E}$ are the local Hilbert spaces of the system and the environment respectively. We focus our attention on the entropy production along the generic process depicted 
in Fig.~\ref{Figprocess}, consisting of initial and final local projective measurements that bracket a unitary evolution. The outcomes of the measurements constitute a quantum trajectory, which plays 
a crucial role in the formulation of FT's.

\subsection{The (forward) process}

The process begins with the global system in  an uncorrelated product state $\rho_{SE} = \rho_S \otimes \rho_E$. The spectral decomposition of the local states reads
\begin{equation}\label{i_state}
\rho_S = \sum_n p_n \HAT{\mathcal{P}}_n \qquad \rho_E = \sum_\nu q_\nu \HAT{\mathcal{Q}}_\nu,
\end{equation}
where $p_n$ and $q_\nu$ are the eigenvalues, and $\{\HAT{\mathcal{P}}_n\}$ and $\{\HAT{\mathcal{Q}}_\nu\}$ are orthogonal projectors onto their respective eigen-subpaces (for simplicity we assume they are rank-1 projectors).

At $t=0$ an initial projective measurement on the system and environment is performed using the eigenprojectors in Eq.~(\ref{i_state}) and outcomes $n$ and $\nu$ are obtained. 
This measurement projects the system and environment onto pure states $\ket{\psi_n}\bra{\psi_n}_S \equiv \HAT{\mathcal{P}}_n$ and $\ket{\phi_\nu}\bra{\phi_\nu}_E \equiv \HAT{\mathcal{Q}}_\nu$, without modifying the average or unconditional  state of the global system ($[{\mathcal P}_n{\mathcal Q}_\nu,\rho_{SE}]=0$).

Subsequently, we evolve the compound system during the time interval $[0, \tau]$. The corresponding unitary operator $U_{\Lambda}$ is generated by the Hamiltonian $H(t)=H(\lambda_t)$, which depends on time 
through an external parameter $\lambda_t$ that we vary according to a prescribed protocol $\Lambda = \{ \lambda_t : 0 \leq t \leq \tau \}$:
\begin{equation}\label{unitary}
 \HAT{U}_{\Lambda} \equiv \mathcal{T}_{+} \exp\left(- \frac{i}{\hbar}\int_{0}^{\tau} dt\, \HAT{H}(\lambda_t) \right),
\end{equation}
where $\mathcal{T}_{+}$ denotes the time-ordering operator. 
As a result, the compound system at time $t=\tau$ is described by the new density matrix
\begin{equation}\label{f_state}
\rho_{SE}^\prime = \HAT{U}_{\Lambda} (\rho_S \otimes \rho_E) \HAT{U}_{\Lambda}^\dagger,
\end{equation}
which in general contains classical and quantum correlations. 
The reduced (or local) states of the system and the environment can be obtained by partial tracing:
$\rho_S' = {\rm Tr}_E [\rho_{SE}']$ and $\rho_E^\prime = {\rm Tr}_S [\rho_{SE}']$. 

To complete the process, a second local projective measurement is performed at time $t=\tau$ on both the system and environment. 
The measurement operators are arbitrary (rank-1) orthogonal projectors, denoted as $\{\HAT{\mathcal{P}}_m^\ast\}$ and $\{\HAT{\mathcal{Q}}_\mu^\ast \}$. 
Unlike in the first measurement, in this case the average global state is disturbed, transforming into
\begin{align}\label{global_final}
\rho_{SE}^\ast & = \sum_{m, \mu} (\HAT{\mathcal{P}}_m^\ast \otimes \HAT{\mathcal{Q}}_\mu^\ast) \rho_{S E}^\prime 
(\HAT{\mathcal{P}}_m^\ast \otimes \HAT{\mathcal{Q}}_\mu^\ast) \nonumber \\ 
& = \sum_{m, \mu} \varrho_{m  \mu}^\ast (\HAT{\mathcal{P}}_m^\ast \otimes \HAT{\mathcal{Q}}_\mu^\ast).
\end{align}
Notice also that this  is not a product state: the final local measurement does not eliminate the classical correlations contained in $\rho_{SE}^\ast$ \cite{Groisman}.
However, the measurement collapses the local states of the system and environment into pure states $\ket{\psi_m^\ast}\bra{\psi_m^\ast}_S \equiv \HAT{\mathcal{P}}_m^\ast$ and $\ket{\phi_\mu^\ast}\bra{\phi_\mu^\ast}_E \equiv \HAT{\mathcal{Q}}_\mu^\ast$. 
Thus, the spectral decomposition of the reduced states after the final measurement is 
\begin{align}
\rho_S^\ast \equiv \tr_E(\rho^*_{SE}) & = \sum_m p_m^\ast \HAT{\mathcal{P}}_m^{\ast},  \\
\rho_E^\ast \equiv\tr_S(\rho^*_{SE}) & = \sum_\mu q_\mu^\ast \HAT{\mathcal{Q}}_\mu^\ast. 
\end{align}
where $p_m^\ast=\sum_\mu \varrho_{m  \mu}^\ast$ and  $q_\mu^\ast=\sum_m \varrho_{m  \mu}^\ast$ are the corresponding classical marginal distributions.

\subsection{Reduced dynamics: maps and operations}

The global manipulation described above induces a reduced dynamics on the system alone. The shaded area in Fig.~\ref{Figprocess} can be considered as an effective transformation of the state of the system, $\rho_S \rightarrow \rho_S^\prime$, described by the action of a 
quantum CPTP map $ \E$ that admits a Kraus representation \cite{Kraus}
\begin{equation}\label{f_map}
\rho_S^\prime = \E(\rho_S)=\sum_{\mu, \nu} \HAT{M}_{\mu \nu} \rho_S \HAT{M}_{\mu \nu}^\dagger,
\end{equation}
with a set of Kraus operators $\HAT{M}_{\mu \nu}$ satisfying $\sum_{\mu, \nu} \HAT{M}_{\mu \nu}^\dagger \HAT{M}_{\mu \nu} = \mathbb{I}$.

There exist many Kraus representations $\{\HAT{M}_{\mu \nu}\}$ that reproduce the reduced dynamics on the system. We choose
\begin{equation}\label{k_ops}
\HAT{M}_{\mu \nu} = \sqrt{q_\nu} \bra{\phi_\mu^\ast}_{E} \HAT{U}_{\Lambda} \ket{\phi_\nu^{~}}_{E}.
\end{equation} 
This specific representation retains all the details of the  evolution of the environment, relating unequivocally each Kraus operator $\HAT{M}_{\mu \nu}$ with a transition $\ket{\phi_\nu}_E \rightarrow \ket{\phi_\mu^\ast}_E$ in the environment. 
This is a key point in order to characterize the thermodynamics of the process at the trajectory level, as we will  see shortly. Let us finally define the {\it quantum operation}:
\begin{equation}\label{q_operation}
\E_{\mu \nu}(\rho_S) = M_{\mu \nu} \rho_S M_{\mu \nu}^\dagger, 
\end{equation}
which describes the conditioned evolution of the system when the environment starts in the pure state $\ket{\phi_\nu}_E$ and ends in the state $\ket{\phi_\mu^\ast}_E$ after measurement \cite{Wiseman}.

\subsection{Average entropy production}
\label{sec:entropy}

We now discuss the entropy change along the process.  We analyze here the von Neumann entropy, $S(\rho) =- {\rm Tr}[\rho \ln \rho]$ of the global system. Recall that the von Neumann entropy coincides with the thermodynamic entropy for equilibrium states (setting the Boltzmann constant $k=1$).  
For non-equilibrium states there are some situations where the von Neumann entropy still can be interpreted as a thermodynamic entropy \cite{ThermoInfo}. However, in this paper we will refrain from identifying $S(\rho)$ with a thermodynamic entropy and refer to it simply as the entropy or the quantum entropy of state $\rho$. 

Along the process described above, the quantum entropy of the global system changes as
\begin{equation} \label{sprod1}
\Delta_{\rm i} S_{\rm inc} \equiv S(\rho_{SE}^\ast) - S(\rho_{SE}).
\end{equation}
This quantity is the quantum entropy production along the process. We will refer to  $\Delta_{\rm i} S_{\rm inc}$ as the inclusive entropy production to distinguish it from the entropy production 
when the system and the environment are separated at the end of the process and the final classical correlations are lost (see below). 
The inclusive entropy production is always non-negative, since von Neumann entropy cannot decrease in a projective measurement and stays constant along any unitary evolution, i.e., $S(\rho_{SE}) = S(\rho'_{SE}) \leq S(\rho^\ast_{SE})$.
Notice also that $S(\rho)$ equals the classical Shannon entropy of the probability distribution of pure states in the eigenbasis of $\rho$. In particular we have
\begin{align}
S(\rho_{SE}) & = -\sum_{n,\nu} p_n q_\nu \ln (p_n q_\nu),  \\  
S(\rho_{SE}^\ast) & = -\sum_{m,\mu} \varrho^\ast_{m\mu}\ln \varrho^\ast_{m\mu}. 
\end{align}

To express the entropy of the global state in terms of local entropies and correlations, one can use the mutual information. 
For an arbitrary state  $\sigma_{SE}$ with reduced states $\sigma_S$ and $\sigma_E$, the mutual information is defined as
\begin{equation}\label{m_info}
 \mathcal{I}({\sigma_{SE}}) \equiv  S(\sigma_S) + S(\sigma_E) - S(\sigma_{SE})  = S(\sigma_{SE} || \sigma_S \otimes \sigma_E). 
\end{equation} 
Here we have introduced the quantum relative entropy,  $S(\rho || \sigma) \equiv {\rm Tr}[\rho (\ln \rho - \ln \sigma)]$, a non-symmetric and non-negative measure 
of the distinguishability  between  states $\rho$ and $\sigma$, which vanishes if and only if $\rho = \sigma$ \cite{NielsenChuang}. This property implies that  
mutual information becomes zero only for product (uncorrelated) states $\sigma_{SE} = \sigma_S \otimes \sigma_E$.
Using  mutual information, the inclusive entropy production can be rewritten as:
\begin{align}\label{sq2}
  \Delta_{\rm i} S_{\rm inc} & = S(\rho_S^\ast) - S( \rho_S) +  S(\rho_E^\ast) - S(\rho_E) - \mathcal{I}(\rho_{SE}^\ast) \nonumber \\
   & =  S(\rho_S^\ast) - S( \rho_S^\prime) +  S(\rho_E^\ast) - S(\rho_E^\prime)  \nonumber \\
   & ~~~ +  \mathcal{I}(\rho_{SE}^\prime) - \mathcal{I}(\rho_{SE}^\ast) \geq 0, 
\end{align}
where we have taken into account that the initial state is uncorrelated and, therefore, $\mathcal{I}(\rho_{SE})=0$. The second equality shows that there are two sources of entropy production. 
The first one is the measurement disturbance of the final local states $\rho_S^\prime \rightarrow \rho_S^\ast$ and $\rho_E^\prime \rightarrow \rho_E^\ast$, 
which can be  avoided  only by measuring in the eigenbasis of the reduced states $\rho_S^\prime$ and $\rho_E^\prime$. The second source, captured by the term 
$\mathcal{I}(\rho_{SE}^\prime) - \mathcal{I}(\rho_{SE}^\ast) \geq 0 $, is the erasure of quantum correlations in the state $\rho_{SE}^\prime$. This is due to the local character of 
the measurements, being zero only if the global interaction $U_{\Lambda}$ does not generate quantum correlations \cite{Luo08, ModiRev}.

In most situations the classical correlations remaining after the final measurement are irreversibly lost, with an entropic cost equal to the mutual information $\mathcal{I}(\rho_{SE}^\ast)$. 
This is the case if we separate system and environment after the process and all subsequent manipulations are local and do not incorporate any feedback based on the outcomes of the final measurements. 
The entropy production in those situations is
\begin{equation}\label{sprod2}
\Delta_{\rm i} S \equiv S(\rho_{S}^\ast) - S(\rho_{S}) +  S(\rho_{E}^\ast) - S(\rho_{E}).
\end{equation}
We will refer to $\Delta_{\rm i} S$ as the non-inclusive entropy production or simply entropy production. The positivity of the non-inclusive entropy production in Eq.~(\ref{sprod2}) has been 
already identified with the second law \cite{ReebWolf} and the existence of a thermodynamic arrow of time \cite{Par08, Jen10}. Notice that $\Delta_{\rm i} S \geq \Delta_{\rm i} S_{\rm inc} \geq 0$, 
since the mutual information $\mathcal{I}(\rho_{SE}^\ast) $ is always non-negative. 

The differences between inclusive and non-inclusive entropy production will be illustrated in a specific example in Sec.~\ref{sec:cnot}.
  
\section{Backward process and fluctuation theorem for the entropy production}\label{S-TotalEP}

\subsection{Forward and backward trajectories}

We now extend the previous analysis to stochastic entropy changes at the level of individual quantum trajectories. A trajectory $\gamma$ of the process introduced in the previous section (hereafter, we will call it the forward process) is simply given by the outcome of the four measurements, i.e.,  $\gamma = \{n,\nu,  \mu, m \}$. This trajectory corresponds to the following transition between pure states
\begin{equation}\label{qjumps}
 \ket{\psi_n}_S \otimes \ket{\phi_\nu}_E  \rightarrow \ket{\psi_m^{\ast}}_S  \otimes \ket{\phi_\mu^\ast}_E. 
\end{equation}
Notice that, in virtue of our choice of the Kraus representation for the reduced dynamics [Eq.~(\ref{k_ops})], a trajectory $\gamma$ is also a trajectory of the reduced dynamics, where the pair $(\nu,\mu)$ now indicates the Kraus operation affecting the system instead of the initial and final 
states of the environment (which is otherwise hidden in the reduced dynamics). 
The probability to observe that trajectory $\gamma$ is given by
\begin{equation}\label{m_pf}
P ({\gamma}) = p_n q_\nu \tr[(\HAT{\mathcal{P}}_m^{\ast} \otimes \HAT{\mathcal{Q}}_\mu^{\ast}) \HAT{U}_{\Lambda} (\HAT{\mathcal{P}}_n \otimes \HAT{\mathcal{Q}}_\nu) \HAT{U}_{\Lambda}^\dagger]. 
\end{equation}

To introduce the backward process, we make use of the anti-unitary time-reversal operator in quantum mechanics, $\HAT\Theta$, satisfying $\HAT\Theta \HAT\Theta^\dagger = \HAT\Theta^\dagger \HAT\Theta = \mathbb{I}$ and $\HAT\Theta i = -i \HAT\Theta$. 
This operator changes the sign of odd variables under time reversal, like linear and angular momenta or magnetic fields \cite{CampisiREV,haake}. 
We will consider the separate time reversal operators for system, $\HAT\Theta_S$, and environment, $\HAT\Theta_E$, as well as the one for the total bipartite system $\HAT\Theta = \HAT\Theta_S\otimes \HAT\Theta_E$.

The backward process is defined as follows. 
We start with a generic initial state of the form
\begin{equation}\label{rhoiniback}
 \tilde{\rho}_{SE} =  \sum_{m, \mu} \tilde{\varrho}_{m \mu}\, \HAT\Theta_{S}\HAT{\mathcal{P}}_m^{\ast} \HAT\Theta^\dagger_S\otimes  \HAT\Theta_{E}\HAT{\mathcal{Q}}_\mu^{\ast} \HAT\Theta_{E}^\dagger.
\end{equation}
As in the forward process, the first step at time $t=0$ is a local measurement of the family of projectors $\{\HAT\Theta_S \HAT{\mathcal{P}}_m^\ast \HAT\Theta_S^\dagger , \HAT\Theta_E \HAT{\mathcal{Q}}_\mu^\ast \HAT\Theta_E^\dagger\}$. 
According to Eq.~(\ref{rhoiniback}), the outcomes  $m$ and $\mu$ are obtained with probability $\tilde\varrho_{m\mu}$.
We then let the global system evolve under the Hamiltonian $\Theta H(\lambda_t)\Theta^\dagger$ inverting the time-dependent protocol as $\tilde{\Lambda} \equiv \{\tilde{\lambda}_{t} | ~0 \leq t \leq \tau \}$ with $\tilde{\lambda}_{t} = \lambda_{\tau - t}$. 
This evolution is given by the unitary transformation 
\begin{equation}
\HAT{U}_{\tilde{\Lambda}} \equiv \mathcal{T}_{+} \exp\left( -\frac{i}{\hbar}\int_{0}^{\tau} dt\, \HAT\Theta \HAT{H}(\tilde{\lambda}_{t}) \HAT\Theta^\dagger \right).  
\end{equation}
Finally, at time $t=\tau$ we perform new local measurements on the system and environment using projectors $\{\HAT\Theta_S \HAT{\mathcal{P}}_n \HAT\Theta_S^\dagger , \HAT\Theta_E \HAT{\mathcal{Q}}_\nu \HAT\Theta_E^\dagger \}$. 
The outcome induces a quantum jump
\begin{equation}\label{qjumps_b}
\HAT\Theta \left( \ket{\psi_m^{\ast}}_S \otimes  \ket{\phi_\mu^\ast}_E\right) \rightarrow  \HAT\Theta {\Big (}\ket{\psi_n}_S  \otimes \ket{\phi_\nu}_E {\Big )}, 
\end{equation}
and the corresponding backward trajectory $\tilde{\gamma} = \{m,\mu, \nu, n \}$ occurs with probability
\begin{equation}\label{m_pb}
\tilde{P}({\tilde{\gamma}}) = \tilde{\varrho}_{m \mu} {\rm Tr}[\HAT\Theta \left(\HAT{\mathcal{P}}_n \otimes \HAT{\mathcal{Q}}_\nu\right) \HAT\Theta^\dagger \HAT{U}_{\tilde{\Lambda}} 
\HAT\Theta \left(\HAT{\mathcal{P}}_m^{\ast} \otimes \HAT{\mathcal{Q}}_\mu^\ast\right) \HAT\Theta^\dagger \HAT{U}_{\tilde{\Lambda}}^\dagger]. 
\end{equation} 

\subsection{Fluctuation theorem}

The unitary transformations corresponding to the forward and the backward process satisfy the so-called micro-freversibility principle for non-autonomous systems
\cite{Andrieux-Gaspard, CampisiREV}:
\begin{equation}\label{microrev}
\HAT\Theta^\dagger \HAT{U}_{\tilde{\Lambda}} \HAT\Theta= \HAT{U}_{\Lambda}^{-1}= \HAT{U}_{\Lambda}^\dagger.
\end{equation} 
This is the key property that relates the probabilities of trajectories $\gamma$ and $\tilde{\gamma}$ in a quantum FT. By comparing the probabilities (\ref{m_pf}) and (\ref{m_pb}), using micro-reversibility (\ref{microrev}) and the 
cyclic property of the trace, we immediately get
\begin{equation}\label{m_logratio}
\Delta_{\rm i} s_\gamma \equiv \ln \frac{P({\gamma})}{\tilde{P}({\tilde{\gamma})}} =\ln \frac{{p}_n {q}_\nu}{\tilde\varrho_{m, \mu}}=\sigma^S_{n m} + \sigma^E_{\nu \mu} - \tilde{I}_{m \mu},
\end{equation}
where we have defined the quantities
\begin{align} \label{mutualfin1} 
 & \sigma^S_{n m} = \ln p_n - \ln \tilde{p}_m, ~~~~ \sigma_{\mu \nu}^E = \ln q_\nu - \ln \tilde{q}_\mu, \\ \label{mutualfin2}
 & \tilde{I}_{m, \mu} = \ln{\tilde\varrho_{m, \mu}} - \ln{\tilde{p}_m \tilde{q}_\mu}.  
 \end{align}
The terms in Eq.~(\ref{mutualfin1}) are related to entropy changes per trajectory in the system and the environment, whereas $\tilde{I}_{m, \mu}$ in Eq.~\eqref{mutualfin2} corresponds to the stochastic version of the mutual information \cite{StochasticMutual, Vedral} in the initial state of the backward process (\ref{rhoiniback}).
From the detailed FT in Eq.~(\ref{m_logratio}), we immediately have the integral version
\begin{equation}\label{inttotalFT}
 \langle e^{- \Delta_{\rm i} s_\gamma} \rangle = \sum_{\gamma} P(\gamma) e^{- \Delta_{\rm i} s_\gamma} = \sum_{\gamma}  \tilde{P}({\tilde{\gamma}}) = 1,
\end{equation}
and, using Jensen's inequality $\langle e^x \rangle \geq e^{\langle x \rangle}$, one obtains a second-law-like expression
$\langle  \Delta_{\rm i} s_\gamma \rangle = \langle \sigma^S \rangle + \langle \sigma^E \rangle - \langle \tilde{I} \rangle \geq 0$.

The interpretation of $\Delta_{\rm i} s_\gamma$ depends on the choice of $\tilde\rho_{SE}$, the initial global state of the backward process. 
If we set $\tilde{\rho}_{SE} = \HAT\Theta \rho_{SE}^\ast \HAT\Theta^\dagger$, then $\tilde \varrho_{m\mu}=\varrho^*_{m\mu}$ and $\Delta_{\rm i} s_\gamma$  becomes the inclusive entropy production per trajectory. 
Its average
\begin{eqnarray}\nonumber
\langle \Delta_{\rm i} s_\gamma \rangle &=& -\sum_{m,\mu} \varrho^*_{m\mu}\ln \varrho^*_{m\mu} +\sum_n p_n\ln p_n+\sum_\nu q_\nu\ln q_\nu \nonumber \\
&=& S(\rho^*_{SE})-S(\rho_S)-S(\rho_E)=\Delta_{\rm i} S_{\rm inc},
\end{eqnarray}
equals the inclusive entropy production defined in Eq.~(\ref{sprod1}). 
If the initial condition for the backward process is the uncorrelated state $\tilde{\rho}_{SE} = \HAT\Theta (\rho_{S}^\ast\otimes  \rho_{E}^\ast ) \HAT\Theta^\dagger$, 
then $\tilde \varrho_{m\mu}=p^*_{m}q^*_{\mu}$ and $\Delta_{\rm i} s_\gamma$ is the non-inclusive entropy production per trajectory, whose average yields the entropy 
production defined in Eq.~(\ref{sprod2})
\begin{equation}
\langle \Delta_{\rm i} s_\gamma\rangle=S(\rho^*_S)-S(\rho_S)+S(\rho^*_E)-S(\rho_E)=\Delta_{\rm i} S. 
\end{equation}
A third choice sets the environment in the (inverted) initial state of the forward process, $\tilde{\rho}_{SE} = \HAT\Theta (\rho_{S}^\ast \otimes  \rho_{E}) \HAT\Theta^\dagger$, which yields $\tilde \varrho_{m\mu}=p^*_{m}q_{\mu}$. 
In this case both initial and final local measurements in the environment are performed in the same basis $\mathcal{Q}_\mu^\ast = \mathcal{Q}_\mu$, and we obtain
\begin{equation} \label{eptrajectory-reset}
 \langle \Delta_{\rm i} s_\gamma\rangle = \Delta_{\rm i} S + S(\rho_E^\ast || \rho_E),
\end{equation}
which includes an extra contribution measuring the disturbance on the environment during the process. The term $S(\rho_E^\ast || \rho_E)$, unlike $S(\rho_E^\ast) - S(\rho_E)$, is negligible when the environmental state is modified only infinitesimally (see appendix \ref{appA}), 
as is the case e.g. of a large reservoir. Moreover, when $\rho_E$ is a Gibbs state, Eq.~(\ref{eptrajectory-reset}) is the entropy production proposed in Ref. \cite{EspositoNJP}, and $S(\rho_E^\ast || \rho_E)$ 
corresponds to the thermodynamic entropy production due to irreversibly reseting the ancilla back to $\rho_E$ in contact with an equilibrium reservoir at the same temperature. Finally, we stress that for equilibrium canonical initial conditions both in 
the forward and in the backward processes, the trajectory entropy production (\ref{m_logratio}) equals the stochastic dissipative work and one recovers the celebrated Crooks work theorem and the original Jarzynski equality \cite{CampisiREV, EspositoREV}.

\section{Dual processes: adiabatic and non-adiabatic entropy production}
\label{S-DecompositionFTs}

We now focus on the reduced dynamics. Our aim is to obtain FT's involving only the quantum trajectory defined in Sec. \ref{S-TotalEP} and the initial and final states of the system. To do that, we follow our previous work \cite{MHP}, where we derived a FT for CPTP maps based on the dual dynamics introduced by Crooks in Ref.~\cite{Crooks}. 
Remarkably, the resulting FT goes beyond the one that we have obtained considering the global dynamics, Eq.~(\ref{m_logratio}), as it will reveal an interesting split of the total entropy production into two terms: the adiabatic entropy production, which accounts for the irreversibility of the stationary regime, and the non-adiabatic entropy 
production, which measures how far the system is from that stationary state.

We apply the formalism in Ref.~\cite{MHP} to $\E$, the map governing the reduced dynamics of the process, as well as to the map corresponding to the backward dynamics.  Therefore, we first need to introduce the reduced dynamics in the backward process, which will be described by a new CPTP map $\tilde\E$. To do that, it is necessary that the system and the environment start the backward process in an uncorrelated state $\tilde\rho_{SE}=\tilde \rho_{S}\otimes \tilde\rho_E$, i.e., we have to impose $\tilde I_{m\mu}=0$ [see Eq.~(\ref{mutualfin2})]. Otherwise the CPTP map of the backward reduced dynamics would depend on the initial state of the system.
In that case, similarly to our choice (\ref{k_ops}) for the forward process, a useful representation of $\tilde \E$ is
\begin{equation}\label{b_ops}
\tilde{\E}_{\nu \mu}(\tilde{\rho}_S) = \HAT{\tilde{M}}_{\nu \mu} \tilde{\rho}_S \HAT{\tilde{M}}_{\nu \mu}^\dagger  
\end{equation}
where the  backward Kraus operators are given by
\begin{equation}\label{b_kraus}
 \HAT{\tilde{M}}_{\nu \mu} = \sqrt{\tilde{q}_\mu} \bra{\phi_\nu}_{E} \HAT\Theta_E^\dagger ~\HAT{U}_{\tilde{\Lambda}}~ \HAT\Theta_E \ket{\phi_\mu^\ast}_{E}. 
\end{equation}
Notice that here we have swapped subscripts with respect to the definition of the forward operators given by Eq.~(\ref{k_ops}). 
This can be done since the pair $(\mu,\nu)$ is just a label of the Kraus operator. The choice in Eq.~(\ref{b_kraus}) means that the operation $\tilde{\E}_{\nu \mu}$ is equivalent to obtaining $\mu$ in the initial measurement of the backward process and $\nu$ in the final one.
Now, micro-reversibility (\ref{microrev})  implies an intimate relationship between the forward and backward Kraus operators:
\begin{equation}\label{b_kraus2}
\HAT\Theta_S^\dagger \HAT{\tilde{M}}_{\nu \mu} \HAT\Theta_S= \sqrt{{\tilde{q}_\mu}} \bra{\phi_\nu}_{E} \HAT{U}^\dagger_{{\Lambda}}\ket{\phi_\mu^\ast}_{E}=e^{-\sigma^E_{\mu\nu}/2}\HAT M^\dagger_{\mu\nu}. 
\end{equation}
It is important to notice that the FT for the total entropy production (\ref{m_logratio}) can be derived directly from the above equation. 
In other words, Eq.~(\ref{b_kraus2}) expresses the fundamental symmetry under time reversal yielding  the FT.

\subsection{The dual-reverse process and non-adiabatic entropy production FT}

In order to go beyond the FT for the total entropy production, we proceed as in Refs.~\cite{Crooks,MHP}. 
These works, inspired by classical stochastic thermodynamics, introduce a quantum dual dynamics that reveals the irreversibility associated to a map at the steady state.
In the following we denote $\pi$ an invariant state of the forward map, $\E(\pi)=\pi$, which we indeed assume to be positive definite. 
The dual dynamics, which we will call here dual-reverse following the criterion used for classical systems \cite{EspositoFaces,EspositoFacesI,EspositoFacesII}, is defined as a map $\tilde{\mathcal D}(\rho)$ such that $\tilde\pi\equiv \Theta_S \pi\Theta^\dagger_S$ is an  invariant state, i.e., $\tilde\D(\tilde\pi)=\tilde\pi$. 
Furthermore, when the map is applied several times starting from the stationary state $\tilde\pi$, it generates trajectories $\tilde\gamma$ distributed as $\tilde P_{D}(\tilde\gamma|\tilde\pi)=P(\gamma|\pi)$. 
Here the trajectories are  $\gamma=\{n,(\nu_1,\mu_1),\dots,(\nu_N,\mu_N),m\}$ and $\tilde\gamma=\{m,(\mu_N,\nu_N),\dots,(\mu_1,\nu_1),n\}$, corresponding to $N$ applications of the map. 

Summarizing, in the stationary regime the dual-reverse generates the same ensemble of trajectories as the forward process, but reversed in time.
For instance, if the map describes the dynamics of a system in contact with a single thermal bath (thermalization), then the forward process generates reversible trajectories (indistinguishable from their reversal) and the dual-reverse coincides with the forward map. 
In nonequilibrium situations, the dual generically inverts flows. For instance, for a system in contact with two thermal baths at different temperatures, the dual-reverse is usually obtained by swapping the temperatures of the baths, hence inverting the flow of heat.

In any case, one can prove that a Kraus representation of the dual-reverse map is given by the operators \cite{Crooks,MHP}:
 \begin{equation}
\label{dualr_k}
 \HAT{\tilde{D}}_{\nu \mu}= \HAT\Theta_S \pi^{ \frac{1}{2}} \HAT{M}_{\mu \nu}^\dagger \pi^{-\frac{1}{2}}\HAT\Theta_S^\dagger.
\end{equation}
Finally, the dual-reverse process is the dual-reverse map complemented by a specific choice of the initial condition for the system (the environment does not appear explicitly in the dual map, which acts only on the system). 
The appropriate initial  condition for the dual-reverse process is $\tilde\rho_S$, i.e., the same as the backward process.

We now have three processes: the forward $\E$, the backward $\tilde\E$, and the dual-reverse $\tilde\D$, each one inducing an evolution in the system characterized by trajectories $\gamma=\{n,\nu,\mu,m\}$. 
We can compute the probability of observing a trajectory $\gamma$ or its reverse $\tilde\gamma=\{m,\mu,\nu,n\}$ in each of those evolutions. 
With a self-explanatory notation, these probabilities read
\begin{eqnarray}
P ({\gamma}) &=& p_n \tr[\HAT{\mathcal{P}}_m^{\ast} \HAT{{M}}_{\mu \nu} \HAT{\mathcal{P}}_n\HAT{{M}}_{\mu \nu}^\dagger] \label{forward_p} \\ 
\tilde{P} ({\tilde{\gamma}}) & =& \tilde{p}_m {\tr}[\HAT\Theta_S \HAT{\mathcal{P}}_n 
\HAT\Theta_S^\dagger \HAT{\tilde{M}}_{\nu \mu} \HAT\Theta_S \HAT{\mathcal{P}}_m^{\ast} \HAT\Theta_S^\dagger \HAT{\tilde{M}}_{\nu \mu}^\dagger ] \label{backward_p}\\
\label{dual_p}
\tilde{P}_D(\tilde{\gamma}) &=& \tilde{p}_m \tr[\HAT\Theta_S \HAT{\mathcal{P}}_n \HAT\Theta_S^\dagger \HAT{\tilde{D}}_{\nu \mu}  \HAT\Theta_S \HAT{\mathcal{P}}_m^\ast \HAT\Theta_S^\dagger\HAT{\tilde{D}}_{\nu \mu}^\dagger].
\end{eqnarray}
To obtain FT's from these expressions we need a condition of proportionality between operators $M^\dagger_{\mu\nu}$, and $\tilde D_{\nu\mu}$, similar to the relationship (\ref{b_kraus2}) between $M^\dagger_{\mu\nu}$ and $\tilde M_{\nu\mu}$.

In \cite{MHP} inspired by \cite{Fagnola:2007hj}, we found that a necessary and sufficient condition for that proportionality is the following. 
We first define the nonequilibrium potential $\HAT{\Phi} = - \ln \pi$, from the invariant state $\pi$. Its spectral decomposition reads:
\begin{equation}\label{noneqpotdecomposition}
\HAT{\Phi} =\sum_i \phi_i \ket{\pi_i}\bra{\pi_i} 
\end{equation}
where $\phi_i=-\ln \pi_i$, and $\pi_i$ and $\{\ket{\pi_i}\}$ are, respectively, the eigenvalues and eigenstates of the invariant density matrix $\pi$.
Now we require that each Kraus operator $\HAT{M}_{\mu \nu}$ is unambiguously related to a nonequilibrium potential change $\Delta \phi_{\mu \nu}$ (note 
however that the converse statement is not necessarily true, i.e. we may have for different values of $\mu$ and $\nu$ the same value of $\Delta \phi_{\mu \nu}$). In the invariant state eigenbasis this condition reads:
\begin{equation}\label{condition}
 \HAT{M}_{\mu \nu} = \sum_{i, j} m_{i j}^{\mu \nu} \ket{\pi_j} \bra{\pi_i},
\end{equation}
with $m_{i j}^{\mu \nu} = 0$ whenever $\phi_j - \phi_i \neq \Delta \phi_{\mu \nu}$.
As pointed in \cite{MHP}, this condition does not imply single jumps between pairs of $\pi$ eigenstates, but it could account for any set of correlated transitions between different pairs with 
same associated $\Delta \phi_{\mu \nu}$. An extreme example are unital maps, where $\pi$ is proportional to the identity matrix. 
In that case, $\Delta\phi_{\mu\nu}=0$ and any complex coefficients $m_{ij}^{\mu\nu}$ satisfy Eq.~(\ref{condition}).
It is not hard to show that condition (\ref{condition}) is equivalent to \cite{MHP}
\begin{eqnarray} \label{condition2}
[\HAT{\Phi}, \HAT{M}_{\mu \nu} ] = ~\Delta \phi_{\mu \nu} \HAT{M}_{\mu \nu}, ~~~ [\HAT{\Phi}, \HAT{M}_{\mu \nu}^\dagger ] = - \Delta \phi_{\mu \nu} \HAT{M}_{\mu \nu}^\dagger.~~
\end{eqnarray} 
This alternative formulation of  (\ref{condition}) indicates that, when $\Delta\phi_{\mu \nu}\neq 0$, $\HAT{M}_{\mu \nu}$ can be interpreted as ladder operators in the eigenbasis of the invariant state $\pi$.

For thermalization or Gibbs preserving  maps, with $\pi =e^{-\beta (H-F)}$, $\beta=(kT)^{-1}$ being the inverse temperature and $F$ the equilibrium free energy, the potential is $\HAT{\Phi}= \beta (H -F)$ and $kT\Delta\Phi$ 
is the energy transfer between the system and the environment, i.e., the heat. Condition (\ref{condition}) in this case implies that the Kraus operators produce jumps between levels with the same 
energy spacing or, equivalently, jumps  with a well defined value of the heat.
 
Introducing condition (\ref{condition}) in Eq.~(\ref{dualr_k}), one easily derives  the following relationship between the forward and the dual-reverse Kraus operators \cite{MHP}:
\begin{equation}\label{d_b_dr}
\HAT \Theta_S^\dagger \HAT{\tilde{D}}_{\nu \mu} \HAT\Theta_S = e^{~\Delta \phi_{\mu \nu} /2}  \HAT{M}_{\mu \nu}^\dagger 
\end{equation}
and, using Eq.~(\ref{b_kraus2}), one gets:
 \begin{equation}
\label{dualr_k2}
 \HAT{\tilde{D}}_{\nu \mu}= e^{(\sigma^E_{\mu\nu}+\Delta \phi_{\mu \nu})/2}  \HAT{\tilde M}_{\nu \mu}.
\end{equation}

Finally, inserting  (\ref{d_b_dr}) and (\ref{dualr_k2}) into the expressions for the probability of trajectories (\ref{forward_p}-\ref{dual_p}) we obtain the following FT's:
\begin{eqnarray}\label{dualr_dft}
 \Delta_{\rm i} s^{\rm na}_\gamma &\equiv  & \ln \frac{P(\gamma)}{\tilde{P}_D (\tilde{\gamma})} = \sigma^S_{n m} - \Delta \phi_{\mu \nu}  \\ \label{dual_dft}
 \Delta_{\rm i} s^{\rm a}_{\mu \nu} &\equiv  & \ln \frac{\tilde{P}_D (\tilde{\gamma})}{\tilde P (\tilde\gamma)} = \sigma^E_{\mu \nu} + \Delta \phi_{\mu \nu}.
 \end{eqnarray}
We will call  $\Delta_{\rm i} s^{\rm a}_{\mu \nu}$ the adiabatic entropy production and $\Delta_{\rm i} s^{\rm na}_{\mu \nu}$  the non-adiabatic entropy production, 
following the terminology used in classical stochastic thermodynamics \cite{EspositoFaces,EspositoFacesI,EspositoFacesII}. 
They contribute to the total entropy production per trajectory,
$ \Delta_{\rm i} s_{\gamma}=\Delta_{\rm i} s^{\rm a}_{\mu \nu} + \Delta_{\rm i} s^{\rm na}_\gamma$, as defined in Eq.~(\ref{m_logratio}). 
Below we  discuss the averages of the adiabatic and non-adiabatic entropy production in some cases, clarifying the origin of the terms.

\subsection{The dual process and adiabatic entropy production FT}\label{sec:4b}

Notice that (\ref{dual_dft}) is not a proper FT for the forward process. In particular, we cannot derive a Jarzynski-like equality for $\exp(\Delta_{\rm i} s^{\rm a}_{\mu \nu})$ averaged over forward trajectories, $P(\gamma)$. 
To achieve this goal we need a further assumption that will allow us to apply the results of Ref.~\cite{MHP} to the backward process. 
In this way, we will obtain the dual-reverse of the backward process, which we simply call the dual map $\D$. If condition (\ref{condition}) is satisfied, then, by virtue of (\ref{b_kraus2}), the backward Kraus operators can be written as:
\begin{align}
 \HAT{\tilde{M}}_{\nu \mu} & = e^{-\sigma^E_{\mu\nu}/2}  \sum_{i, j} (m_{i j}^{\mu \nu})^\ast \Theta_S\ket{\pi_i} \bra{\pi_j}\Theta_S^\dagger \nonumber \\
& =  \sum_{i, j} \tilde m_{i j}^{\nu \mu} \ket{\tilde\pi_j} \bra{\tilde\pi_i}, 
\end{align}
with $\tilde m_{ij}^{\nu\mu}\equiv e^{-\sigma^E_{\mu\nu}/2}(m_{ji}^{\mu \nu})^\ast$. 
We observe that, setting $\Delta\tilde \phi_{\nu\mu}= -\Delta\phi_{\mu\nu}$, condition (\ref{condition}) is recovered for the backward process. 
However, a requirement to apply the theoretical framework developed in Ref.~\cite{MHP} is that $\ket{\tilde\pi}\equiv\Theta_S\ket{\pi}$ is an invariant state of the backward map $\tilde\E$. 
This is not warranted by the definition of $\tilde\E$, not even when the Kraus operators are of the form  (\ref{condition}). 
Therefore, we have to add this extra assumption.  
In particular, it is satisfied when the driving protocol associated to the map is time-symmetric, the Hamiltonian of the environment is invariant under time reversal, and we perform the same 
measurements at the beginning and the end of the process on the environment. This is the case of the infinitesimal maps that govern the dynamics of a quantum Markov process since, even in the case of arbitrary driving, each map is generated by a constant Hamiltonian.

We now obtain the dual operators $D_{\mu\nu}$, applying transformation (\ref{dualr_k}) to the backward Kraus operators $\tilde M_{\nu\mu}$ (with the role of $\Theta_S$ and $\Theta_S^\dagger$ swapped \cite{MHP}). 
Similarly to (\ref{d_b_dr}), condition (\ref{condition}) on the backward operators imply
 \begin{equation}\label{d_b_dr2}
\HAT\Theta_S \HAT{{D}}_{\mu \nu} \HAT\Theta^*_S = e^{\Delta \tilde\phi_{\nu \mu} /2}  \HAT{\tilde M}_{\nu \mu}^\dagger 
 = e^{-\Delta \phi_{\mu \nu} /2}  \HAT{\tilde M}_{\nu \mu}^\dagger , 
\end{equation}
and, using Eq.~(\ref{b_kraus2}),
\begin{equation}
\label{dual_k2}
D_{\mu \nu}= e^{-(\sigma^E_{\mu\nu}+\Delta \phi_{\mu \nu})/2}  \HAT{ M}_{\mu \nu}.
\end{equation}
 
The dual process is given by the dual map with initial condition $\rho_S$. The trajectories generated by this process are distributed as
\begin{equation}\label{dualp}
P_D(\gamma)=p_n{\rm Tr}_S\left[ {\cal P}_m^* D_{\mu\nu} {\cal P}_nD_{\mu\nu}^\dagger\right].
\end{equation}

Combining Eqs.~(\ref{forward_p}) and (\ref{dualp}) and using condition (\ref{b_kraus2}), we get a new FT for the adiabatic entropy production:
\begin{equation}\label{sa2}
\Delta_{\rm i} s^{\rm a}_{\mu \nu}  = \ln \frac{P (\gamma)}{P_D (\gamma)} = \sigma^E_{\mu \nu} + \Delta \phi_{\mu \nu}.
\end{equation}

\subsection{Integral fluctuation theorems}

We can now derive integral FT's for the adiabatic and non-adiabatic entropy productions:
\begin{equation}\label{i-FT}
\langle e^{- \Delta_{\rm i} s^{\rm na}} \rangle = 1, \qquad \langle e^{- \Delta_{\rm i} s^{\rm a}} \rangle = 1,
\end{equation}
which follow from the detailed versions by averaging over trajectories $\gamma$. 
Finally, convexity of the exponential function provides the following two second-law-like inequalities as a corollary $\langle \Delta_{\rm i} s^{\rm na}_\gamma \rangle \ge 0$ and $\langle \Delta_{\rm i} s^{\rm a}_\gamma \rangle \ge 0$. 
As for the FT for the total entropy production (\ref{m_logratio}), the meaning of these average entropies becomes clearer if the initial condition of the backward process is specified. 
Setting $ \tilde{\rho}= \HAT\Theta (\rho_S^\ast \otimes \rho_E^*) \HAT\Theta^\dagger$, the average of the adiabatic and non-adiabatic entropy production defined by (\ref{d_b_dr}) and  (\ref{dualr_k2}) reads
\begin{eqnarray} \label{second-law-like1}
\Delta_{\rm i} S_{\rm na}\equiv \langle \Delta_{\rm i} s^{\rm na}_\gamma \rangle &=& S(\rho_S^\ast) - S(\rho_S) - \langle\Delta \phi\rangle \ge 0,  \\ \label{second-law-like2}
\Delta_{\rm i} S_{\rm a} \equiv\langle \Delta_{\rm i} s^{\rm a}_\gamma \rangle &=& S(\rho_E^\ast) - S(\rho_E) + \langle \Delta \phi \rangle \ge 0.
\end{eqnarray}
and the sum equals the total non-inclusive average entropy production $\Delta_{\rm i} S$ [see Eq.~(\ref{sprod2})].
It is interesting to notice that the average change of the potential,
\begin{equation}\label{deltaphi3}
\langle\Delta\phi\rangle = \sum_{\mu,\nu}P(\gamma)\Delta \phi_{\mu\nu} = \sum_{\mu,\nu}\tr[M_{\mu\nu}\rho_S M_{\mu\nu}^\dagger]\Delta \phi_{\mu\nu}, 
\end{equation}
can be alternatively written in terms of averages over the states of the system, $\rho'_S$ and $\rho_S$ if condition (\ref{condition}) is fulfilled. 
That condition implies $  [\HAT{\Phi}, \HAT{M}_{\mu \nu}]=\HAT{M}_{\mu \nu} \Delta \phi_{\mu \nu}$ (see also Ref.~\cite{MHP}). 
Introducing the commutator in (\ref{deltaphi3}),
\begin{align}\label{deltaphi4}
\langle\Delta\phi\rangle & = \sum_{\mu,\nu}\tr[(\HAT{\Phi} M_{\mu\nu}-M_{\mu\nu}\HAT{\Phi})\rho_S M_{\mu\nu}^\dagger] \nonumber \\
& = \tr [ \HAT{\Phi}\,(\rho_S'-\rho_S)],
\end{align}
where we have used the cyclic property of the trace and Eqs.~(\ref{f_map}) and  (\ref{k_ops}). 
Therefore, the average potential change $\langle\Delta\phi\rangle$ can be expressed as the change in the expected value of the operator $\HAT{\Phi}$ due to the map. 
Recall that the operator $\HAT{\Phi}$ acts on the Hilbert space of the system $\cal H_S$, i.e., is a local observable on the system. 

If the final measurement does not alter the state of the system, i.e., if $\rho_S^*=\rho'_S$, or if the final measurement is skipped, as it is the case when we concatenate maps and the system is measured only after the whole concatenation (see Sec.~\ref{S-Concatenations} below), we can write the average non-adiabatic entropy production in an appealing form:
\begin{align}\label{ex_meaning}
 \Delta_{\rm i} S_{\rm na} &= S(\rho_S')-S(\rho_S)-\langle \Delta \phi\rangle  \nonumber \\
 & = \tr [\rho_S(\ln\rho_S + \Phi)]-\tr [\rho_S'(\ln\rho_S'+ \Phi)] \nonumber \\ & = S(\rho_S ||\pi) - S(\rho_S'|| \pi) \geq 0.
\end{align}
where we have used the definition $\HAT{\Phi} =-\ln \pi$ of the potential operator in terms of the invariant state $\pi$.
Here we see  that the non-adiabatic entropy production is related to the distance between the state of the system and the invariant state $\pi$. 
During the evolution, the state of the system can only approximate the invariant state and the non-adiabatic entropy production is a measure of the irreversibility associated to such convergence. 
In fact, inequality in Eq.~(\ref{ex_meaning}) follows from direct application of Ulhman's inequality (monotonicity of quantum relative entropy) holding for general CPTP evolutions \cite{NielsenChuang,SagawaEntropies}.

\subsection{Multipartite environments}

The results obtained so far can be also applied to multipartite environment. 
The corresponding Hilbert space is decomposed as $\mathcal{H}_E = \bigotimes_{r= 1}^R \mathcal{H}_{r}$, corresponding to $R$ independent ancillas or reservoirs  interacting  
with the open system. We assume an uncorrelated  initial state of the environment, $\rho_E = \rho_1 \otimes ... \otimes \rho_R$, and that the measurements are performed locally in 
each environmental ancilla. 

In this case the adiabatic entropy production per trajectory and its average read (see details in appendix \ref{appB}):
\begin{eqnarray}
\Delta_\mathrm{i} s^{\mathrm{a}}_{\mu \nu} &=& \sum_{r=1}^R \sigma^r_{\mu^{(r)} \nu^{(r)}} + ~\Delta \phi_{\mu \nu},  \\
\Delta_\mathrm{i} S_\mathrm{ a} &=&  \sum_{r=1}^R  S(\rho_r^\ast) - S(\rho_r)  + \langle \Delta \phi \rangle \geq 0.
\end{eqnarray}

\subsection{Concatenation of CPTP maps} \label{S-Concatenations}

\begin{figure*}[t]
\includegraphics[width = 1.0 \linewidth]{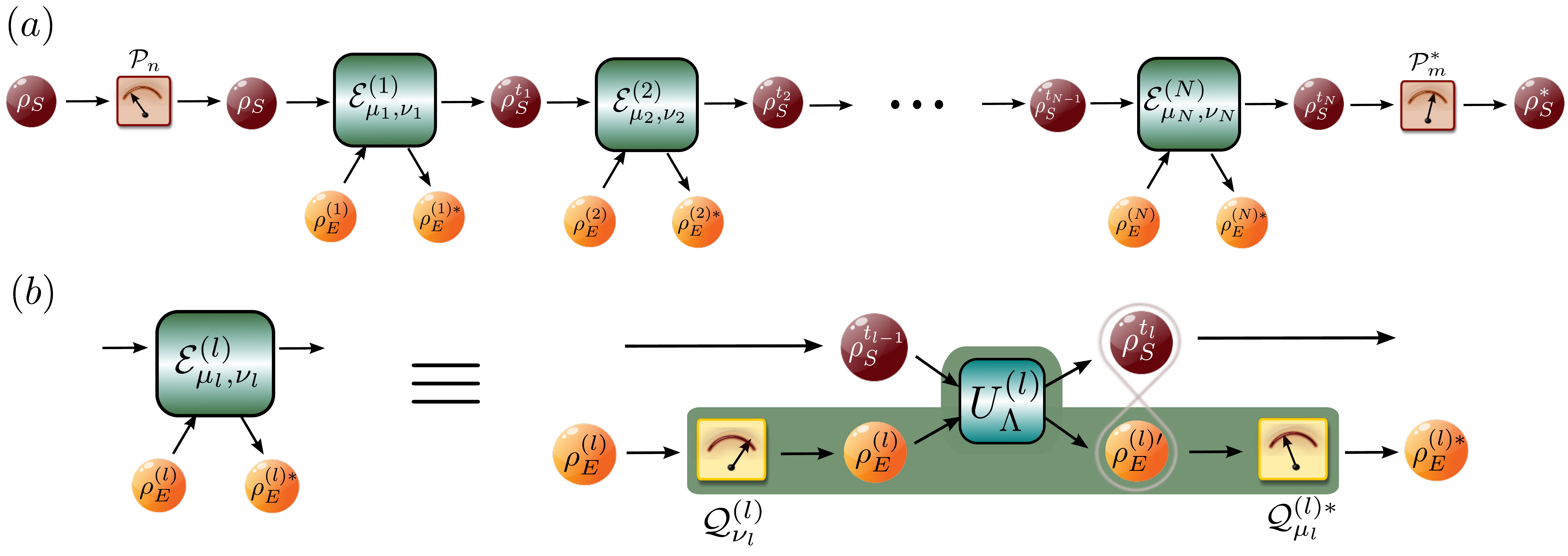}
\caption{(a) Schematic diagram of a trajectory generated by the maps concatenation. Projective measurements on the system are only performed at the begging and at the end of the 
concatenation. (b) Any operation $\E_{\mu_l, \nu_l}^{(l)}$ in the concatenation consists in the interaction of the system with an environmental ancilla in the state $\rho_E^{(l)}$ via the 
unitary $\hat{U}_\Lambda^{(l)}$ depending on the protocol $\Lambda_l$. The ancilla is measured before and after interaction generating outcomes $\nu_l$ and $\mu_l$ respectively.} 
\label{F-Concatenation}
\end{figure*}

Up to now, we have considered a single interaction of duration $\tau$ between the system and the environment [see Eq.~(\ref{unitary})]. 
The CPTP map $\E$ describes the evolution of the system when the environment is measured before and after interaction. 
This framework can be extended to  concatenations of CPTP maps, where the system interacts sequentially 
with the environment. Each single interaction in an time interval $[t,t + \tau]$ is described by a single CPTP map like $\E$. The map describing the reduced dynamical evolution for $N$ interactions, from $t=0$ to $t= N \tau$, is:
\begin{eqnarray}\label{Omega}
\HAT{\Omega} = \E^{(N)} \circ ... \circ~\E^{(l)} \circ ...\circ ~\E^{(1)},
\end{eqnarray}
where, in particular, each map $\E^{(l)}$ may have a different (positive-definite) invariant state $\pi^{(l)}$. We assume that the system interacts 
from time $t_{l-1} \equiv (l-1) \tau$ to time $t_l \equiv l \tau$ with a ``fresh" (uncorrelated) environment in a generic state 
$\rho_E^{(l)} \equiv \sum_{\alpha} q_{\alpha}^{(l)} \HAT{\mathcal{Q}}_{\alpha}^{(l)}$, and, as in the single map case, the environment is measured before and after interaction with 
the system by projective measurements. On the other hand, the system is only measured at the beginning and end of the the whole concatenation (\ref{Omega}), as depicted in Fig. \ref{F-Concatenation}.

In this case, trajectories are specified by the set of outcomes $\gamma = \{n , (\nu_1, \mu_1), ..., (\nu_N, \mu_N), m \}$, which can be compared to the backward trajectories 
$\tilde{\gamma} = \{m, (\nu_N, \mu_N), ... , (\nu_1, \mu_1), n \}$ generated by the reverse sequence of maps $\tilde{\Omega} = \tilde{\E}^{(1)} \circ ... \circ~\tilde{\E}^{(l)} \circ ...\circ ~\tilde{\E}^{(N)}$.
We find that all our above results apply as well to the concatenations setup (see appendix \ref{appC}) yielding the following three detailed fluctuation theorems:
\begin{eqnarray} \label{dualr_dft_c}
\Delta_{\rm i} s^{\rm na}_{\gamma} &=& \ln \frac{P(\gamma)}{\tilde{P}_D (\tilde{\gamma})} = ~\sigma_{n m}^S - \sum_{l=1}^N \Delta \phi_{\mu_l \nu_l}^{(l)},  \\ 
\Delta_{\rm i} s^{\rm a}_{\gamma} &=& \ln \frac{P (\gamma)}{P_D (\gamma)} = \sum_{l=1}^N  \left( \sigma^E_{\mu_l \nu_l}  + \Delta \phi_{\mu_l \nu_l}^{(l)} \right),  \\ \label{total_dft_c}
\Delta_{\rm i} s_\gamma &=& \ln \frac{P (\gamma)}{\tilde{P} (\tilde{\gamma})} ~=~ \Delta_{\rm i} s^{\rm na}_{\gamma} ~+~ \Delta_{\rm i} s^{\rm a}_{\gamma},
\end{eqnarray}
where $\sigma_{n m}^S$ is given by Eq.~(\ref{mutualfin1}), $\sigma_{\mu_l\nu_l}^E = \ln q_{\nu_l}^{(l)} - \ln \tilde{q}_{\mu_l}^{(l)}$ is the entropy change in the environment due to the 
$l$-th map, and $\Delta \phi^{(l)}_{\mu \nu} = - \ln \pi_{\mu}^{(l)} + \ln \pi_{\nu}^{(l)}$ is the change in non-equilibrium potential for the $l$-th map.

\section{Lindblad Master Equations}\label{S-MarkovianME}

The results of the last section can be applied to Lindblad master equations \cite{JordanParrondo,JordanSagawa}.
Consider the following master equation in Lindblad form \cite{BreuerBook, Wiseman,Lindblad}, depending on an external parameter $\lambda_t$:
\begin{equation}\label{ME}
\dot{\rho}_t = -\frac{i}{\hbar}[H, \rho] + \sum_{k=1}^K \left( \HAT{L}_k \rho_t \HAT{L}_k^\dagger - \frac{1}{2}\{\HAT{L}_k^\dagger \HAT{L}_k, \rho_t \} \right) \equiv \mathcal{L}_{\lambda_t} \rho_t,
\end{equation} 
where $H(\lambda_t)$ is the system Hamiltonian in the selected picture 
 and 
$\HAT{L}_k(\lambda_t)$ are positive Lindblad operators, which generally  depend on the control parameter $\lambda_t$ and describe jumps in some 
(possibly time-dependent) basis.  We assume that there exists an instantaneous invariant state  $\pi_\lambda$, which is the steady state of Eq.~(\ref{ME}) when the external control parameter is frozen: $\mathcal{L}_\lambda\pi_\lambda = 0$ \cite{Rivas}. 

The Lindblad  equation (\ref{ME}) can be written as a concatenation of CPTP maps
\begin{equation}\label{inf-time-ev}
 \rho_{t + dt} = \left( \mathbb{I}_S + dt \mathcal{L}_{\lambda_t} \right) \rho_t \equiv \E (\rho_t),
\end{equation}
with the  Kraus representation 
\begin{eqnarray}\label{Inf_k_op_0}
\HAT{M}_0(\lambda_t) &\equiv& ~\mathbb{I}_S - dt\left(\frac{i}{\hbar}H +  \frac{1}{2}\sum_{k=1}^K \HAT{L}_k^\dagger(\lambda_t) \HAT{L}_k(\lambda_t) \right) \\ \label{Inf_k_op}
\HAT{M}_k(\lambda_t) &\equiv& \sqrt{dt} \HAT{L}_k(\lambda_t),  ~~~~ k=1, ..., K.
\end{eqnarray}
Recall that this Kraus representation is not unique \cite{Wiseman}. As before, the representation (\ref{Inf_k_op_0}-\ref{Inf_k_op})  is related to a specific detection scheme for the jumps, that is, it implies a specific choice of the initial state and the local 
observables to be monitored  in the environment (the set of projectors $\{ \HAT{\mathcal{Q}}_\nu \}$ and $\{ \HAT{\mathcal{Q}}^\ast_\mu \}$).

The Kraus representation (\ref{Inf_k_op_0}-\ref{Inf_k_op}) is based on a family of operations $M_k$ with $k=1,\dots,K$ that induce jumps in the state of the system and occur with probabilities of 
order $dt$, and a single  operation $M_0$ that induces a smooth nonunitary evolution and occurs with probability of order 1. This implies that a trajectory $\gamma$ 
consists of a large number of 0's punctuated by a few jumps $M_k$ with $k=1,\dots,K$. An alternative way of describing the trajectory is to specify the jumps $k_j$ and the times $t_j$ 
where they occur, i.e.,  $\gamma = \{ n, (k_1,t_1), ..., (k_j,t_j), ..., (k_N, t_N), m\}$, where, as before, $n$ and $m$ denote the outcomes of the initial and final measurements 
in the system at times $t=0$ and $t = t_\mathrm{f}$. Jump $k$ is given by the operation $\E_k(\rho)\equiv\HAT{M}_k \rho \HAT{M}_k^\dagger$, whereas between two consecutive jumps at $t_j$ and $t_{j+1}$ the evolution is given by the repeated 
application of the operation corresponding to the Kraus operator $M_0(\lambda_t)$ in (\ref{Inf_k_op_0}). This results in a smooth evolution given by the operator:
\begin{equation} \label{ueff}
{U}_{\rm eff} (t_{j+1} ,t_j) = \mathcal{T}_{+} \exp \left( -\frac{i}{\hbar} \int_{t_j}^{t_{j+1}} ds ~ H_{\rm eff} ({\lambda}_{s})  \right), 
\end{equation}
with an effective non-hermitian Hamiltonian that reads
\begin{equation}
 H_{\rm eff}(\lambda_t) = H(\lambda_t)- \frac{i \hbar}{2} \sum_{k=1}^K L_k^\dagger(\lambda_t) L_k(\lambda_t). 
\end{equation} 
In this representation, the probability of a trajectory $\gamma = \{ n, (k_1,t_1), ..., (k_j,t_j), ..., (k_N, t_N), m\}$ is
\begin{align}\label{Pfor}
 P(\gamma) = & \tr[\mathcal{P}_m^\ast \mathcal{U}_{t_\mathrm{f}, t_N} \E_{k_N} \mathcal{U}_{t_N, t_{N-1}} ... ~\E_{k_l}~ \nonumber \\
& ~~~~~ ... ~\mathcal{U}_{t_2, t_1} \E_{k_1} \mathcal{U}_{t_1, 0} (\mathcal{P}_n \rho_0 \mathcal{P}_n)], ~~~~
\end{align}
with $ \mathcal{U}_{t_{j+1}, t_{j}}(\rho)={U}_{\rm eff} (t_{j+1} ,t_j)\rho\,{U}^\dagger_{\rm eff} (t_{j+1} ,t_j)$.

\subsection{Backward, dual and dual-reverse dynamics}

Consider now the backward dynamics. The time-inversion of the evolution of the global system  corresponds to a time-reversed version of the Lindblad master equation (\ref{ME}). 
As in the previous section, the backward process is generated by inverting the sequence of operations together with time-inversion of each operation in the sequence. 
The map corresponding to an infinitesimal time-step in the time-reversed dynamics, $\tilde{\rho}_{t + dt} = \tilde{\E}(\tilde{\rho}_t)$, admits a Kraus representation with 
Kraus operators $\tilde{M}_{k}(\lambda_t)$. To obtain the backward map, we would need to know details about the environment that induces the Markovian dynamics given by the Lindblad equation (\ref{ME}). However, in the previous sections we have derived a relationship between the forward and backward CPTP maps, namely  Eq.~(\ref{b_kraus2}):
\begin{align} \label{b_lind}
 \HAT{\tilde{M}}_{0} & = e^{-\sigma^E_{0}/2} \HAT{\Theta}_S \HAT{M}_{0}^{\dagger} \HAT{\Theta}_S^\dagger, \\ 
 \HAT{\tilde{M}}_{k} & = e^{-\sigma^E_{k}/2} \HAT{\Theta}_S \HAT{M}_{k}^{\dagger}\HAT{\Theta}_S^\dagger.
\end{align}
Imposing the backward maps to be trace-preserving, that is, $\tilde M_0^\dagger \tilde M_0+\sum_k \tilde M_k^\dagger \tilde M_k=\mathbb{I}$, we obtain $\sigma_0^E = 0$, and the consistency condition
\begin{eqnarray}\label{trace-pre}
 \sum_{k=1}^K \left(\HAT{L}_k^\dagger \HAT{L}_k -\HAT{L}_k \HAT{L_k}^\dagger e^{-\sigma_{k}^E} \right) = 0.
\end{eqnarray}
Any set of numbers $\{ \sigma_{k}^E\}$ satisfying Eq.~(\ref{trace-pre}) defines, through (\ref{b_lind}), an admissible backward process. The existence of such set is warranted, since any Lindblad equation can be derived from the interaction between the system and an ancilla.

For any trajectory $\gamma =\{ n, (k_1,t_1),..., (k_N, t_N), m\}$ generated in the forward process with probability $P(\gamma)$, there exist a backward 
trajectory $\tilde{\gamma} = \{m, (k_N,t_N), ..., (k_1,t_1), n\}$ occurring in the backward process with probability $\tilde{P}(\tilde{\gamma})$. The backward trajectory can also be identified by the times of successive jumps.
In this representation, the probability of trajectory $\tilde{\gamma}$ can be written as:
\begin{align}\label{Pbac}
 \tilde{P}(\tilde{\gamma}) = & \tr[ \Theta_S \mathcal{P}_n \Theta_S^\dagger ~\tilde{\mathcal{U}}_{t_1, 0} \tilde{\E}_{k_1} \tilde{\mathcal{U}}_{t_2, t_1}~ ... ~\tilde{\E}_{k_l}~ \nonumber \\
& ~~~~ ...  ~ \tilde{\mathcal{U}}_{t_N, t_{N-1}} \tilde{\E}_{k_N} \tilde{\mathcal{U}}_{t_{\rm f}, t_N} (\Theta_S \mathcal{P}_m^\ast \rho_{t_{\rm f}} \mathcal{P}_m^\ast \Theta_S^\dagger)],
\end{align}
where $\tilde\E_k(\tilde\rho)=\tilde M_k\tilde\rho\tilde M^\dagger_k$.
The smooth evolution between jumps $\tilde{\mathcal{U}}_{t' ,t} (\tilde{\rho}_{t}) = \tilde{U}_{\rm eff}(t',t) \tilde{\rho}_{t} \tilde{U}_{\rm eff}^\dagger(t',t)$ is given by the operator
\begin{equation} \label{drift-sym}
\tilde{U}_{\rm eff} (t' ,t) = \mathcal{T}_{+} \exp \left( \frac{i}{\hbar} \int_t^{t'} ds ~\Theta_S H_{\rm eff}^\dagger (\tilde{\lambda}_{s}) \Theta_S^\dagger \right), 
\end{equation}
where $\{ \tilde{\lambda}_t \}$ again corresponds to the inverse sequence of values for the control parameter. It can be shown that this smooth evolution obeys the micro-reversibility relationship
$\Theta_S^\dagger \tilde{U}_{\rm eff} (t' ,t)  \Theta_S = U_{\rm eff}(t', t)^\dagger$.

Let us discuss now the dual and dual-reverse dynamics.
The condition (\ref{condition2}), necessary to define the dual-reverse process, reads~\cite{JordanParrondo,JordanSagawa}:
\begin{eqnarray} \label{conditionrelations}
[\HAT{\Phi}, \HAT{L}_k ] = ~\Delta \phi_k \HAT{L}_k, ~~~ [\HAT{\Phi}, \HAT{L}_k^\dagger ] = - \Delta \phi_k \HAT{L}_k^\dagger.
\end{eqnarray} 
These commutation relationships indicate that the Lindblad operators $\HAT{L}_k(\lambda_t)$ promote jumps between the eigenstates of $\pi_{\lambda_t}$. Furthermore,
as the condition must be fulfilled for the operator $M_0$ in Eq.~(\ref{Inf_k_op_0}) as well, we need $[H, \sum_k L_k^\dagger L_k] = [H, \HAT{\Phi}] = 0$, which 
in turn implies $\Delta \phi_0 = 0$.  This means that the instantaneous  steady state of the dynamics must be diagonal in the basis of the Hamiltonian term appearing in Eq.~(\ref{ME}). 
This condition is fulfilled by equilibrium Lindblad equations and in situations in which the operator $H$ becomes the identity operator in an appropriate interaction picture 
(see e.g. Refs. \cite{SqzRes, Szczygielski:2013wc}). However, the condition can be broken in nonequilibrium situations, a genuine quantum effect. In section \ref{S-cavity} we present an example of a periodically driven cavity mode where the adiabatic entropy production can be negative.
Finally, as discussed in section \ref{sec:4b}, we recall that the fluctuation theorem for the adiabatic entropy production can be 
stated when the backward maps $\tilde{\E}$ admits $\tilde{\pi}_\lambda \equiv \Theta_S \pi_\lambda \Theta_S^\dagger$ as an invariant state.

If these conditions are fulfilled, the dual process  is defined by the dual operations $\D_{k}(\cdot) = D_k (\cdot) D_k^\dagger$ with Kraus 
operators $\{ D_{k} \}$ as defined in Eq.~(\ref{dual_k2}), whereas the dual-reverse process is given by operations $\tilde{\D}_k = \tilde{D}_k (\cdot) \tilde{D}_k^\dagger$ 
with Kraus operators $\{ \tilde{D}_k \}$ defined in Eq.~(\ref{d_b_dr}) (see also Eqs. (\ref{d_b_b_c}-\ref{d_b_b_c2}) in App. \ref{appC}). The probability of a trajectory $\gamma$ in 
the dual process, $P_D(\gamma)$, can be calculated from Eq.~(\ref{Pfor}) by using the same map $\mathcal{U}_{t',t}$ for 
the no-jump time evolution intervals, and replacing the operations $\E_{k}$ by the dual operations $\D_{k}$.
Analogously, for the dual-reverse process the probability of trajectory $\tilde{\gamma}$, $\tilde{P}_D (\tilde{\gamma})$, can be constructed from Eq.~(\ref{Pbac}) with 
$\tilde{\mathcal{U}}_{t', t}$ for the no-jump evolution, and dual-reverse operations $\tilde{\D}_k$. 
We further notice that in general $D_k \neq M_k$, and $\tilde{D}_k \neq \tilde{M}_k$, that is, $\sigma_k^E \neq - \Delta \phi_k$. 

In many applications, the Lindblad operators come in pairs and the corresponding pair of terms in the sum~(\ref{trace-pre}) cancel. That occurs if, for a specific pair of operators $\{\HAT{L}_i,\HAT{L}_j\}$, 
we have $\HAT{L}_i=\sqrt{\Gamma_i} \HAT{L}$ and $\HAT{L}_j =\sqrt{\Gamma_j} \HAT{L}^\dagger$,  $\Gamma_i(\lambda_t)$ and $\Gamma_j(\lambda_t)$ being positive transition rates, and 
$\HAT{L}(\lambda_t)$ some arbitrary (possibly time-dependent) system operator. Then, condition (\ref{trace-pre}) implies (cf.~\cite{JordanPRA})
\begin{align}\label{entpairs}
\sigma_{i}^E(\lambda_t) & = \ln (\Gamma_i/\Gamma_j), \\ 
\sigma_{j}^E(\lambda_t) & = \ln (\Gamma_j/\Gamma_i)= - \sigma_{i}^E(\lambda_t), \nonumber
\end{align}
and the (inverted) Kraus operators of the backward map are also operators of the forward map:
\begin{align}
\Theta_S^\dagger \tilde{M}_i \Theta_S & = e^{-\sigma_i^E/2} M_i^\dagger = \sqrt{dt} e^{-\sigma_i^E/2} L_i^\dagger \nonumber \\ 
& = \sqrt{dt} L_j = M_j, \label{mpairs}
\end{align}
where we have used the detailed-balance relation (\ref{b_kraus2}). Moreover, $\tilde{\pi}_\lambda \equiv \Theta \pi_\lambda \Theta^\dagger$  is invariant under the backward map:
\begin{align}
\tilde\E (\tilde\pi_\lambda) & = \sum_k \tilde{M}_k \Theta_S \pi_\lambda \Theta_S^\dagger \tilde{M}_k^\dagger \nonumber \\ 
& = \sum_k \Theta_S M_k \pi_\lambda M_k^\dagger \Theta_S^\dagger = \tilde{\pi}_\lambda.
\end{align}

\subsection{Entropy production rates}

The above considerations lead us to reproduce the three different detailed FT's in Eqs.~(\ref{total_dft_c}) for quantum trajectories 
generated by Lindblad master equations.
From the integral fluctuation theorems we can derive second-law-like inequalities analogous to Eqs. (\ref{second-law-like1}) and (\ref{second-law-like2})
for the entropy production rates~\cite{JordanParrondo}:
\begin{align} \label{dynamical-sl1}
& \dot{S}_{\rm i}  = \dot{S}_{\rm na} + \dot{S}_{\rm a} = \dot{S} + \langle \dot{\sigma}^E \rangle \geq 0, \\ \label{dynamical-sl2}
& \dot{S}_{\rm na}  = \dot{S} - \dot{\phi} \geq 0, ~~~~
\dot{S}_{\rm a} = \langle \dot{\sigma}^E \rangle + \dot{\phi} \geq 0, 
\end{align}
where $\dot{S} = - \tr[\dot{\rho}_t \ln \rho_t]$ is the derivative of the von Neumann entropy of the system, $\dot{\phi} \equiv \tr[\dot{\rho}_t \HAT{\Phi}(\lambda_t)]=-\tr [\dot\rho_t \ln\pi_{\lambda_t}]$ is the nonequilibrium 
potential change rate, and $\langle \dot{\sigma}^E(\lambda_t) \rangle dt = \sum_{k_l} \tr[\E_{k_l}(\rho_t)]\sigma_{k_l}^E(\lambda_t)$ the entropy change in the monitored environment during $dt$~\cite{JordanSagawa}.
The three above equations guarantee the monotonicity of the average entropy production, 
$\Delta_{\rm i} S$, and the adiabatic and non-adiabatic contributions, $\Delta_{\rm i} S_{\rm na}$ and $\Delta_{\rm i} S_{\rm a}$, during the whole evolution.

The physical interpretation of the adiabatic and non-adiabatic entropy production now becomes clear. 
The non-adiabatic part can be written as
\begin{equation}\label{non-adiab_lind}
\dot{S}_{\rm na}= \tr \left[ \dot\rho_t (\ln \pi_{\lambda_t}-\ln\rho_t) \right]
\end{equation}
which is the continuous time version of Eq.~(\ref{ex_meaning}).
If the control parameter changes quasi-statically, we have $\rho_t \simeq \pi_{\lambda_t}$ and, therefore, the non-adiabatic entropy production vanishes. This is analogous to the classical non-adiabatic entropy production introduced in Refs.~\cite{HatanoSasa,EspositoFaces, EspositoFacesI,EspositoFacesII}. 
On the other hand, the adiabatic contribution $\dot{S}_{\rm a}$ is in general different from zero even if the driving is extremely slow. In a physical system, this term accounts for the entropy production required to keep the system out of equilibrium when $\lambda$ is fixed, and the associated dissipated energy is usually referred to as housekeeping heat \cite{HatanoSasa}. 

At this point, it is worth it to remark an important difference between classical and quantum systems.
In classical systems, the split of the entropy production in two terms, adiabatic and non-adiabatic, can always be done at the level of trajectories, and both terms obey fluctuations theorem that ensure the positivity of their respective averages. This is possible for quantum systems only if  (\ref{condition2}) [or (\ref{conditionrelations}) for Lindblad operators] is met. One can still use (\ref{non-adiab_lind}) as a definition for the average non-adiabatic entropy production $\dot S_{\rm na}$ and $\dot S_{\rm a}=\dot S_{\rm i}-\dot S_{\rm na}$ for the average adiabatic entropy production rate. However, these definitions cannot be extended to single trajectories and, furthermore, they do not obey a fluctuation theorem. In the next section, we discuss a specific example where the condition is not fulfilled and, as a consequence, the average adiabatic entropy production rate can be negative.

Finally, it is also important to notice that $\dot S$ and $\langle \dot{\sigma}^E \rangle $ in Eqs.~(\ref{dynamical-sl1}) are exact differentials, i.e., can be written as the time derivative of the system and the environment entropy, respectively. On the other hand, the term $\dot{\phi} \equiv \tr[\dot{\rho}_t \HAT{\Phi}(\lambda_t)]$, as well as the adiabatic and non-adiabatic entropy production rates in Eq. \eqref{dynamical-sl2}, cannot be expressed, in general, as a time derivative. One important exception is the case of a constant invariant state $\pi_{\lambda_t}=\pi$ like, for instance, in a relaxation in the absence of driving. In that case, all the quantities in Eqs.~(\ref{dynamical-sl1}-\ref{dynamical-sl2}) are exact differentials. In particular, the non-adiabatic entropy production when the system relaxes from $\rho_0$ to $\rho_t$ is given by
\begin{equation}\label{sna_nodriving}
\Delta S_{\rm na}=-S(\rho_\tau||\pi)+S(\rho_0||\pi)\geq 0
\end{equation}
which equals $\Delta S_{\rm na}=S(\rho_0||\pi)$ for a full relaxation to $\rho_\tau=\pi$. The later coincides with the entropy production introduced by Spohn \cite{Spohn}.

\section{Examples} \label{S-Examples}

We illustrate our findings with three paradigmatic examples. In the first one, we consider a two-qubit CNOT gate as a simple process with a finite size environment
to illustrate the differences between the inclusive and non-inclusive entropy production introduced in  section \ref{sec:entropy}. 
The second and third examples correspond to two representative examples of non-equilibrium quantum Markov systems.
The second example is an autonomous system coupled to several thermal baths. In this case, the non-adiabatic entropy production is zero except during the transient relaxation to the steady state. 
However, it provides an intuitive picture of how entropy is produced in non-equilibrium setups. The third example is a driven system that does not fulfill condition (\ref{condition}) and, 
consequently, does not admit the splitting of the entropy into adiabatic and non-adiabatic contributions with positive averages. 

\subsection{A two-qubit CNOT gate}
\label{sec:cnot}

The difference between the inclusive and non-inclusive entropy production introduced in section \ref{sec:entropy} becomes especially relevant for processes where the system of interest repeatedly interacts with a finite-size reservoir.
As an extreme case we consider both system and environment to be qubits with the same energy spacing $\epsilon$. 
Their Hamiltonians are given by $H_S = \epsilon \ket{1}\bra{1}_S$ and $H_E = \epsilon \ket{1}\bra{1}_E$. We assume the initial state of the system to be partially coherent, $\rho_S = (\mathbb{I}+\alpha\,\sigma_x)/2$ with $0 \leq \alpha \leq 1$, 
and the environmental qubit starting in a thermal state $\rho_E = e^{- \beta H_E}/Z_E$ at inverse temperature $\beta = 1/k_B T \geq 0$,  $Z_E=1+e^{-\beta\epsilon}$ being the partition function. The initial state can be written as
\begin{equation}
\rho_{SE}=\rho_S \otimes \rho_E =\frac{1}{4}\left(\mathbb{I}+\alpha\,\sigma_x\right)\otimes \left(\mathbb{I}+\kappa\,\sigma_z\right).
\end{equation}
where $\kappa \equiv \tanh(\beta \epsilon/2)$, $\sigma_j$ with $j=x,y,z$ are the Pauli matrices, and we take the standard qubit basis  $\{\ket{0},\ket{1}\}$ for both the system and the environment. The eigenbasis of $\rho_{SE}$ determines the projectors of the initial measurements $\{ \mathcal{P}_n, \mathcal{Q}_\nu\}$, which are in this case  $\mathcal{P}_\pm = \ket{\psi_\pm}\bra{\psi_\pm}$ with $\ket{\psi_\pm}=(\ket{0}\pm \ket{1})/\sqrt{2}$ and $\mathcal{Q}_\nu = \ket{\nu}\bra{\nu}$ with $\nu=0,1$.

System and environment interact through a CNOT gate, $U_\mathrm{CNOT}$, where the system  acts as the control qubit \cite{NielsenChuang}. The interaction leads to the following global system-environment state
\begin{align}\label{rhosep}
 \rho_{SE}^\prime &= U_\mathrm{CNOT} \left( \rho_S \otimes \rho_E \right) U_\mathrm{CNOT}^\dagger  \\ 
 &= \frac{1}{4}\left(\mathbb{I}+  \alpha \, \sigma_x \otimes \sigma_x - \alpha \kappa \,\sigma_y \otimes \sigma_y + \kappa\, \sigma_z \otimes \sigma_z \right). \nonumber
\end{align}
Notice that $\rho_{SE}^\prime$ has maximally mixed reduced states both in system and environment. 
As a consequence, for any choice of the final projectors $\{ \mathcal{P}_m^\ast, \mathcal{Q}_\mu^\ast\}$, we have $\rho_S^\prime = \rho_S^\ast =\rho_E^\prime = \rho_E^\ast= \mathbb{I}/2$. In contrast, the global state $\rho_{SE}^\ast$ depends on the 
final projectors. 
The average work done during the interaction is $W = \tr[(H_S + H_E) (\rho_{SE}^\prime- \rho_{SE})] = \epsilon (1/2 - e^{-\beta \epsilon}/Z_E)>0$, while there is no further energy contributions from local measurements.

The inclusive entropy production in Eq.~\eqref{sq2} is just given by the erasure of quantum correlations in the final measurements, $\Delta_\mathrm{i} S_\mathrm{inc} = \mathcal{I}(\rho_{SE}^\prime) - \mathcal{I}(\rho_{SE}^\ast)$. This is the so-called mutual induced disturbance introduced by Luo \cite{Luo08}. Moreover if, following Refs. \cite{Simdisc1,Simdisc2}, we maximize $\mathcal{I}(\rho_{SE}^\ast)$ over $\{ \mathcal{P}_m^\ast, \mathcal{Q}_\mu^\ast\}$, 
then the inclusive entropy production is equal to the (symmetric) quantum discord \cite{ZurekDiscord, ModiRev} of the state $\rho_{SE}^\prime$.
On the other hand, the non-inclusive entropy production in Eq.~\eqref{sprod2} is given 
by the total correlations in state $\rho_{SE}^\prime$, that is, $\Delta_\mathrm{i} S = \Delta_\mathrm{i} S_{\rm inc} + \mathcal{I}(\rho_{SE}^\ast) = \mathcal{I}(\rho_{SE}^\prime)$, and it is independent of the choice of the local projectors of the final measurements $\{ \mathcal{P}_m^\ast, \mathcal{Q}_\mu^\ast\}$.

The entropy production per trajectory $\Delta_{\rm i} s_\gamma$ can be calculated as explained in Sec.~\ref{S-TotalEP}. Recall that we may obtain both the inclusive or non-inclusive entropy production
depending on our choice for the initial state of the backward process, and that the two quantities verify the integral fluctuation theorem \eqref{inttotalFT}.

\begin{figure}[t]
\includegraphics[width = 1.0 \linewidth]{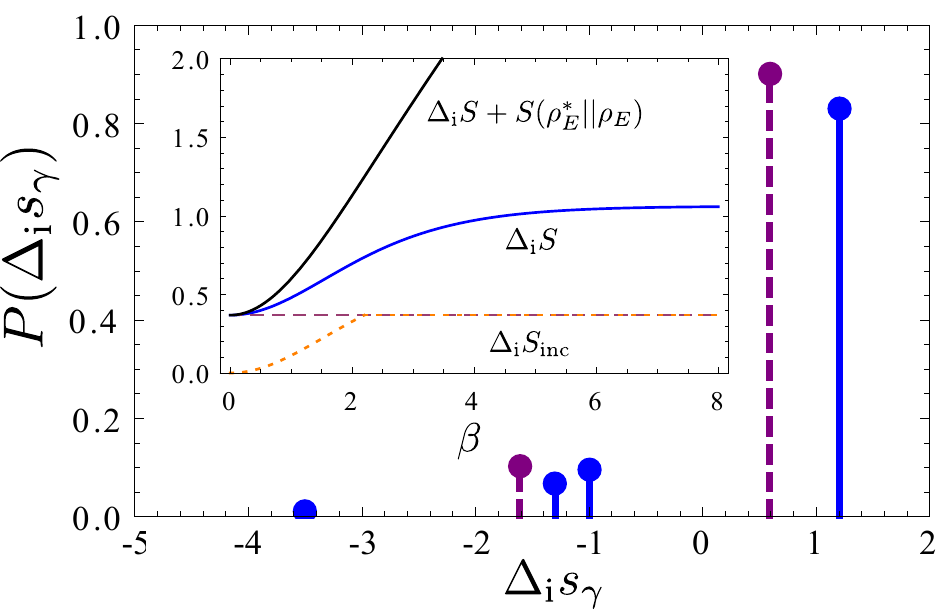}
\caption{Probability distribution $P(\Delta_\mathrm{i} s_\gamma)$ of the entropy production per trajectory  for non-inclusive (blue solid) and inclusive (purple dashed) cases. Initial states of system and environment correspond to parameters 
$\alpha = 0.8$ and $\beta \epsilon= 2.5$. Inset: plot of the different versions of the average entropy production as a function of $\beta$ ($\epsilon = 1$ and $\alpha=0.8$).} 
\label{F-qubits}
\end{figure}

In Fig.~\ref{F-qubits} we show the  probability distribution of the entropy production $P(\Delta_\mathrm{i} s_\gamma)$ for $\beta\epsilon = 2.5 $ and $\alpha = 0.8$. Blue solid bars correspond to the non-inclusive version and purple dashed bars correspond to the inclusive one. The latter does depend on the final measurements. Here we have taken as final projectors the local energy eigenbasis,  $\{ \mathcal{P}_m^\ast = \ket{m}\bra{m}_S, \mathcal{Q}_\mu^\ast = \ket{\mu}\bra{\mu}_E\}$ for $m,\mu = 0,1$.
The different types of average entropy production are plotted in the inset figure as functions of $\beta$ for the same value of $\alpha=0.8$.
There, black and blue solid lines correspond to the average non-inclusive entropy production with and without including the term $S(\rho_{E}^\ast || \rho_E)$ due to local disturbance of the  environment [see Eq.~\eqref{eptrajectory-reset}], respectively. 
Dashed and dotted lines show
to the average inclusive entropy production  for different choices of the local projectors in the final measurement $\{ \mathcal{P}_m^\ast, \mathcal{Q}_\mu^\ast\}$. The purple dashed line is obtained when the final projectors are given by the local energy eigenbasis $\{ \mathcal{P}_m^\ast = \ket{m}\bra{m}_S, \mathcal{Q}_\mu^\ast = \ket{\mu}\bra{\mu}_E\}$ for $m,\mu = 0,1$. The orange dotted line is the symmetric quantum discord, obtained when 
maximizing $\mathcal{I}(\rho_{SE}^\ast)$.

As mentioned in section \ref{sec:entropy}, inclusive and non-inclusive entropy production apply to different physical situations depending on how the system and the environment are manipulated after the process. If system and environment are separated and every further manipulation is local, then  we do not make use of the classical correlations given by the mutual information  $\mathcal{I}(\rho_{SE}^\ast)$; in this case the non-inclusive entropy production is the magnitude that adequately describes the increase of entropy. On the other hand,  global operations on the whole system+environment can make use of those correlations and, for instance, extract more energy from a thermal bath. We illustrate this possibility in our simple example by considering a second CNOT interaction after the final local measurements. For simplicity we perform the final measurements in the local energy basis, 
$\{ \mathcal{P}_m^\ast = \ket{m}\bra{m}_S, \mathcal{Q}_\mu^\ast = \ket{\mu}\bra{\mu}_E\}$ for $m,\mu = 0,1$. Applying these projectors to state $\rho_{SE}^\prime$ in Eq.~\eqref{rhosep}, one obtains the final global state
\begin{equation}
 \rho_{SE}^\ast = \frac{1}{4}\left(\mathbb{I} + \kappa\, \sigma_z \otimes \sigma_z\right).
\end{equation}
Applying the second CNOT to this state, one gets 
\begin{equation} \label{primeprime}
\rho_{SE}^{\prime \prime} = U_\mathrm{CNOT} ~\rho_{SE}^\ast~ U_\mathrm{CNOT}^\dagger = \rho_S^\ast \otimes \rho_E,
\end{equation}
where $\rho_E$ is the initial thermal state of the environment. As we can see, in this second process system and environment become completely 
decorrelated after interaction, while a work $W_{\rm ext}=\tr[(H_S + H_E)( \rho_{SE}^{\prime\prime} - \rho_{SE}^\ast)] =  \epsilon (1/2 - e^{-\beta \epsilon}/Z_E)$ is extracted when performing the second gate. Notice that the extracted work equals the work performed in the first gate. This work extraction is impossible if we only have local operations at our disposal, for which the final state $\rho^\ast_{SE}$ is completely equivalent to the uncorrelated state $\rho_S^\ast \otimes \rho_E^\ast$. 

This simple example highlights the importance of distinguishing between inclusive and non-inclusive entropy production in a small finite-size environment. 
Similar conclusions can be applied for the term $S(\rho_E^\ast || \rho_E)$.

\subsection{Autonomous thermal machine}

\begin{figure}[t]
\begin{center}
\includegraphics[width = \linewidth]{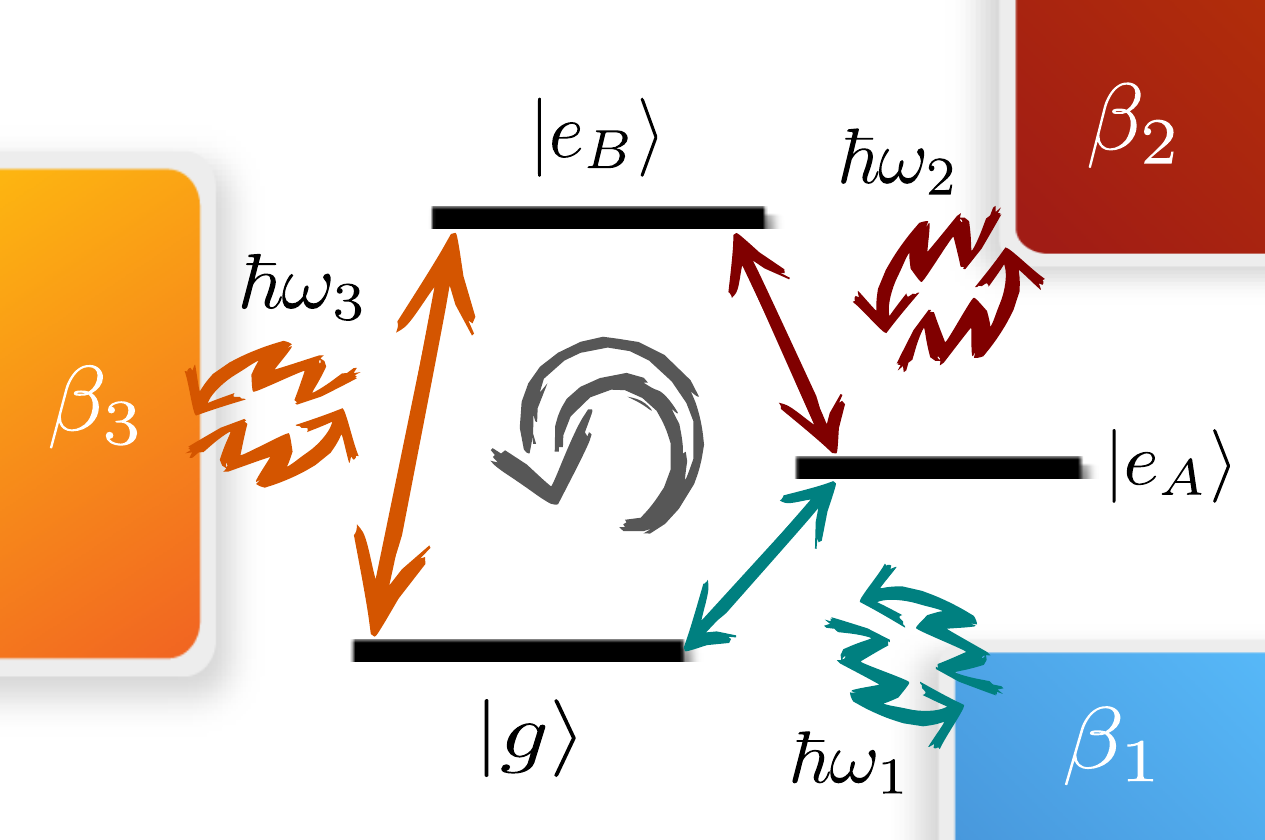}
\caption{Schematic diagram of a three-level thermal machine acting as a refrigerator. The three transitions of the machine are weakly coupled to 
thermal reservoirs at temperatures $\beta_1$, $\beta_2$ and $\beta_3$, inducing jumps between the machine energy levels (double arrows). 
In a refrigeration cycle the machine performs a sequence of three jumps $\ket{g} \rightarrow \ket{e_A} \rightarrow \ket{e_B} \rightarrow \ket{g}$, 
where it absorbs a quantum of energy $\hbar \omega_1$ from the cold reservoir, together with a quantum $\hbar \omega_2$ from the hot one, while 
emitting a quantum $\hbar \omega_3$ into the reservoir at intermediate temperature.} 
\label{F-Threelevel}
\end{center}
\end{figure}

Consider an  autonomous three-level thermal machine powered by three thermal reservoirs at different 
temperatures, as depicted in Fig.~\ref{F-Threelevel} \cite{Scovil,Geusic, CorreaSqz, Palao, Kosloff}. Each bath mediates a different transition between the energy levels, $\{\ket{g}, \ket{e_A}, \ket{e_B}\}$. 
The Hamiltonian of the system is
\begin{equation}
H_S = \hbar \omega_1 \ket{e_A}\bra{e_A}  +  \hbar (\omega_1 + \omega_2) \ket{e_B}\bra{e_B},
\end{equation}
that is, the three possible transitions $g \leftrightarrow e_A$, $e_A \leftrightarrow e_B$ and $g \leftrightarrow e_B$ have frequency gaps 
$ \omega_1$, $\omega_2$, and $\omega_3 \equiv \omega_1 + \omega_2$, respectively. Each transition is weakly coupled 
to a bosonic thermal reservoir in equilibrium at inverse temperature $\beta_r = 1/ k T_r$ with $r= 1, 2, 3$, where we assume 
$\beta_1 \geq \beta_3 \geq \beta_2$ for concreteness. 

The dynamics of the three-level thermal machine  can be described by a Lindblad master equation obtained in the weak 
coupling limit by standard techniques from open quantum systems theory \cite{BreuerBook, Wiseman, Rivas}. It reads
\begin{equation}\label{Meq}
\dot{\rho}_t = -\frac{i}{\hbar}[H_S, \rho_t] + \mathcal{L}_1 (\rho_t) + \mathcal{L}_2 (\rho_t) + \mathcal{L}_3 (\rho_t),
\end{equation}
where $\rho_t$ is the density operator of the three level system and Lamb-Stark shifts have been neglected. The three dissipative terms in the above 
equation describe the irreversible dynamical contributions induced by each of the three thermal reservoirs: 
\begin{align}\label{Lindbladian}
\mathcal{L}_r (\rho_t) & = \Gamma_\downarrow^{(r)}\left(a_r \rho_t a_r^\dagger - \frac{1}{2}\{a_r^\dagger a_r, \rho_t\} \right) \nonumber \\ 
& +~ \Gamma_\uparrow^{(r)} \left(a_r^\dagger \rho_t a_r - \frac{1}{2}\{a_r  a_r^\dagger, \rho_t \} \right),
\end{align}
$r = 1,2, 3$ where $a_1 = \ket{g}\bra{e_A}$, $a_2 = \ket{e_A}\bra{e_B}$ and $a_3 = \ket{g}\bra{e_B}$ are the ladder operators of the three-level 
system. Equation~(\ref{Lindbladian}) describes the emission and absorption of excitations of energy $\hbar \omega_r$ to or from the reservoir $r$, at rates $\Gamma_{\downarrow}^{(r)} = \gamma_r (n_r^{\rm th} +1)$ and $\Gamma_{\uparrow}^{(r)} = \gamma_r n_r^{\rm th}$, 
fulfilling detailed balance $\Gamma_{\downarrow}^{(r)} = e^{\beta_r \hbar \omega_r} \Gamma_{\uparrow}^{(r)}$. Here $n_r^{\rm th} = (e^{\beta_r \hbar \omega_r} - 1)^{-1}$ is the mean number of 
excitations of energy $\hbar \omega_r$ in reservoir $r$, and $\gamma_r \ll \omega_{r'} ~\forall r,r'=1,2,3$ are the spontaneous emission decay rates associated to each transition. 
The heat fluxes entering from the reservoirs associated to the imbalance in emission and absorption can be defined as $\dot{Q}_r = \tr[H_S \mathcal{L}_r(\rho_t)]$ for $r=1,2,3$, and the 
first law of thermodynamics reads $\dot{U} = \tr[H_S \dot{\rho}_S] = \dot{Q}_1 + \dot{Q}_2 + \dot{Q}_3$.

Therefore, in our example we have six Lindblad operators ($r=1,2,3$): 
\begin{equation}
L_{\downarrow}^{(r)} = \sqrt{\Gamma_\downarrow^{(r)}}a_r,  \qquad 
L_{\uparrow}^{(r)} = \sqrt{\Gamma_\uparrow^{(r)}}a^\dagger_r \qquad 
\end{equation}
that define the infinitesimal CPTP map (\ref{inf-time-ev}) with the Kraus representation given by  Eqs.~(\ref{Inf_k_op_0}-\ref{Inf_k_op}). In particular:
\begin{align} \label{L_operators2}
M_{\downarrow}^{(r)} & = \sqrt{dt} L_{\downarrow}^{(r)} = \sqrt{dt ~\Gamma_\downarrow^{(r)}} a_r,  \\ 
M_{\uparrow}^{(r)} & = \sqrt{dt} L_{\uparrow}^{(r)} = \sqrt{dt ~\Gamma_\uparrow^{(r)}} a_r^\dagger. 
\end{align}
Here the stochastic jumps during the evolution correspond to simple transitions between 
the energy levels $\{ \ket{g}, \ket{e_A}, \ket{e_B}\}$.
Therefore, the stochastic dynamics is completely equivalent to a classical Markov process if the initial state of the machine is diagonal in the Hamiltonian eigenbasis. In particular, the stationary state reads:
\begin{equation}\label{ss}
\pi = \pi_g \ket{g}\bra{g} + \pi_{A} \ket{e_A}\bra{e_A} + \pi_{B} \ket{e_B}\bra{e_B}, 
\end{equation}
In appendix \ref{appD}, we explicitly calculate the occupation probabilities $\pi_g$, $\pi_{A}$, and $\pi_{B}$.
Nevertheless, the transient dynamics could exhibit some quantum effects when the initial state exhibit coherences in the Hamiltonian eigenbasis. For instance it has been recently pointed that initial coherence can be used to reach lower temperatures during the transient dynamics \cite{Mitchison, Brask}.

The backward trajectory $\tilde{\gamma} = \{ m, (k_N, t_N),\dots,$ $(k_1, t_1), n\}$ is defined by the inverse sequence of events with respect to $\gamma$, occurring in 
the backward process. We consider the initial state of the backward process the inverted final state of the forward process, $\Theta_S \rho_{t_{\rm f}} \Theta_S^\dagger$, 
while the thermal reservoirs have the same state as in the forward process. We further assume the simplest form for the time inversion operator $\Theta_S$, namely, the complex conjugation, i.e., $\Theta_S\psi=\psi^*$, which commutes with any matrix with real entries, as the Hamiltonian and the jump operators $a_r$, $a^\dagger_r$. 

The Lindblad operators in this case come in pairs $L_{\downarrow}^{(r)} = ~e^{\beta_r \hbar \omega_r/ 2} 
L_{\uparrow}^{(r) \dagger}$. Hence the stochastic entropy change in the environment $\sigma^E_k$ for each operator $L_k$ is given by Eq.~(\ref{entpairs}), where the label $k$ takes on the six possible values $k=(\uparrow, r)$ and $k=(\downarrow,r)$ with $r=1,2,3$:
\begin{equation}\label{sigma-r}
\sigma^{(r)}_{\downarrow} =  \beta_r \hbar \omega_r, ~~~~ \sigma^{(r)}_{\uparrow} =  -\beta_r \hbar \omega_r.
\end{equation}
This is as expected, since the upward jump $r$ induced by the operator $L^{(r)}_{\downarrow}$ in the forward trajectory $\gamma$  dissipates a heat $\hbar\omega_r$ to  the reservoir at inverse temperature $\beta_r$. Equivalently, in the downward jump  $r$ a heat $\hbar\omega_r$ is extracted from the thermal bath reducing its entropy by an amount $\beta_r\hbar\omega_r$. 
The Kraus operators of the backward map are given by Eq.~(\ref{mpairs}): 
$\tilde{M}_\downarrow^{(r)} =  M_\uparrow^{(r)}$, and $\tilde{M}_\uparrow^{(r)} =   M_\downarrow^{(r)}$,
and $\tilde{M}_0 = \Theta_S M_0 \Theta_S^\dagger$ for the no-jump evolution. Indeed, by virtue of Eq.~(\ref{drift-sym}), we obtain 
$\tilde{U}_{\rm eff} = \Theta_S U_{\rm eff}^\dagger \Theta_S^\dagger = U_{\rm eff}$ for the effective evolution operator describing the dynamics between 
jumps in the backward process. From the above equations we see that the backward map $\tilde{\E}$ is obtained from the forward map  $\E$ inverting the jumps. We also notice that, consequently, 
the backward map $\tilde{\E}$ admits the time-reversed steady state $\tilde{\pi} = \Theta_S \pi \Theta_S^\dagger = \pi$ as an invariant state.

\begin{figure}[t]
\includegraphics[width = \linewidth]{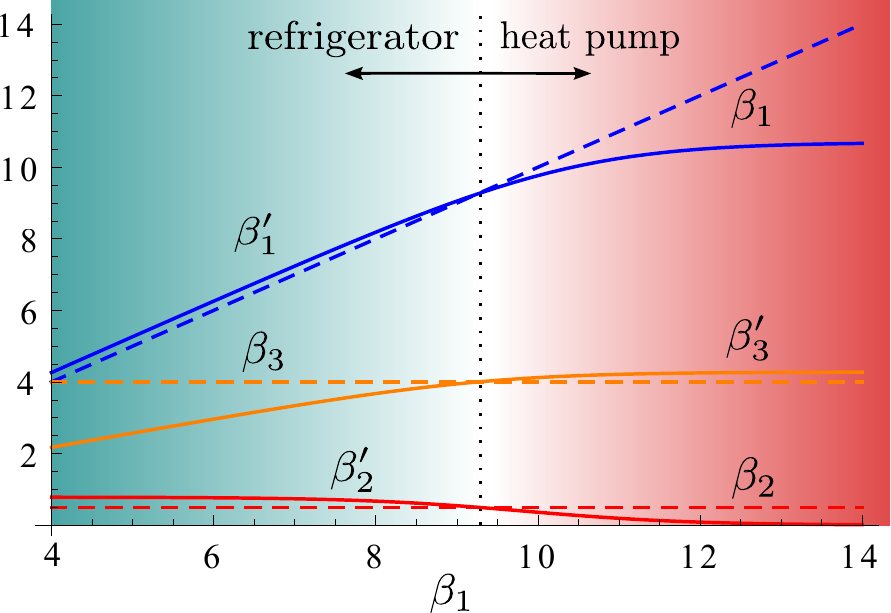}
\caption{Comparison between the inverse effective (or virtual) temperatures $\beta_r^\prime$ (solid lines) and the real inverse temperatures of the reservoirs $\beta_r$ (dashed lines) 
for $r=1,2,3$ (blue, red, orange), as a function of $\beta_1$ when $\hbar\omega_1 = 1$, and $\hbar\omega_2 = 1.5$. In the refrigerator regime, the transition 
$g \leftrightarrow e_A$ is at an effective temperature colder than the coldest reservoir, $\beta_1^\prime \geq \beta_1$, inducing heat extraction from it, while the other transitions 
induce dissipation of heat to the reservoir at intermediate temperature, $\beta_2 \geq \beta_2^\prime$, and absorption of heat in the hotter one $\beta_2^\prime \geq \beta_2$. In 
the heat pump regime the three heat flows change its directions as the previous inequalities become inverted.} 
\label{F-Regimes}
\end{figure}

We next construct the dual and dual-reverse processes for the model. The condition for the Lindblad operators 
to be of the form in Eq.~(\ref{condition}) is fulfilled here. Indeed, the non-equilibrium potential, $\HAT{\Phi} = - \ln \pi$, 
obeys $[\HAT{\Phi}, H_S] = 0$ and
\begin{equation}\label{condexample}
 [\HAT{\Phi}, L_k^{(r)}] = \Delta \phi_k^{(r)} L_k^{(r)}, ~~~~ [\HAT{\Phi}, L_k^{(r) \dagger}] = - \Delta \phi_k^{(r)} L_k^{(r) \dagger},
\end{equation}
where the nonequilibrium potential changes associated to each jump in the trajectory read
\begin{equation}\label{nep}
 \Delta \phi_0 = 0, ~~~~ \Delta \phi_{r \downarrow} = - \beta_r' \hbar \omega_r, ~~~~ \Delta \phi_{r \uparrow} = \beta_r' \hbar \omega_r.
\end{equation}
Here we have introduced the quantities $\beta_1'= \ln(\frac{\pi_0}{\pi_1}) /\hbar \omega_1$, $\beta_2'= \ln(\frac{\pi_1}{\pi_2}) /\hbar \omega_2$ and 
$\beta_3'= \ln(\frac{\pi_1}{\pi_2}) /\hbar \omega_3$, which can be seen as the local inverse temperature (or virtual temperature \cite{VirtualQubits, VirtualTemperatures,Ral16})  of each transition in the steady state $\pi$. As shown in appendix \ref{appD}, they determine the direction of the heat flow in the stationary regime, i.e., if $\beta'_r>\beta_r$, then the temperature of reservoir $r$ is higher than the local temperature of the machine and the heat $\dot Q_r$ is positive (energy flows from the reservoir to the machine), and viceversa. Moreover, for $\beta'_r\simeq \beta_r$ the heat flow is proportional to  $\beta'_r -\beta_r$; therefore, the difference $\beta'_r -\beta_r$ can be considered as a thermodynamic force for the heat flow between the reservoir $r$ and the system. In Fig.~\ref{F-Regimes} we plot the local inverse temperatures $\beta_r'$ compared to the reservoir temperatures $\beta_r$ for a specific choice of $\beta_2=0.5$ and $\beta_3=4$ and as a function of $\beta_1$, the inverse temperature of the coldest bath (we use units of $(\hbar \omega_1)^{-1}$). There is a point, around $\beta_1=9.3$, where $\beta'_r=\beta_r$ and all the heat flows in the stationary regime vanish. Below that point, the steady heat flow from the coldest reservoir at inverse temperature $\beta_1$ is positive, i.e., the machine acts as a refrigerator that extract energy from the coldest bath 1. On the other hand, for $\beta_1>9.3$, heat flows from the machine to the hottest bah at inverse temperature $\beta_2$, so  the machine acts as a heat pump capable to heat up the hottest reservoir 2.

The Kraus operators for dual and dual-reverse maps, $\D$ and $\tilde{\D}$, can be obtained from Eqs.~(\ref{dual_k2}) and (\ref{dualr_k2}) respectively, 
by using Eqs.~(\ref{sigma-r}) and (\ref{nep}). They read:
\begin{eqnarray}
 D_\downarrow^{(r)} &= \sqrt{dt} ~ e^{(\beta_r' - \beta_r) \hbar \omega_r} L_\downarrow^{(r)},   \\ 
 D_\uparrow^{(r)} &= \sqrt{dt} ~ e^{-(\beta_r' - \beta_r) \hbar \omega_r} L_\uparrow^{(r)},  \\
\tilde{D}_\downarrow^{(r)} &= \sqrt{dt} ~ e^{-\beta_r' \hbar \omega_r} L_\downarrow^{(r)},  \\ 
\tilde{D}_\uparrow^{(r)} &= \sqrt{dt} ~ e^{\beta_r' \hbar \omega_r} L_\uparrow^{(r)}. 
\end{eqnarray}
We see that the dual and dual-reversed dynamics induce similar jumps in the three-level system, 
but with modified rates depending on the differences $\beta_r^\prime - \beta_r$. Only when $\beta_r^\prime = \beta_r$ for each $r$, the dual process becomes equal to the forward process, 
and hence the dual-reverse process equals the backward process (see Fig. \ref{F-Regimes}).

Notice that Eq.~(\ref{condexample}), together with the backward map having $\tilde{\pi}=\pi$ as an invariant state, are sufficient conditions to ensure the existence of the three fluctuation theorems 
for the adiabatic, non-adiabatic and total entropy productions during trajectory $\gamma$. They explicitly read:
\begin{align}\label{stochasticversion1}
\Delta_{\rm i} s^{\rm a}_{\gamma} & = \sum_{r=1}^3 (\beta_r' - \beta_r) q_\gamma^{(r)}, \\
\Delta_{\rm i} s^{\rm na}_\gamma & = \sigma_{n m}^S - \sum_{r=1}^3 \beta_{r}'q_\gamma^{(r)}, \\  \label{stochasticversion3}
\Delta_{\rm i} s_\gamma & = \sigma_{n m}^S - \sum_{r=1}^3 \beta_r q_\gamma^{(r)},  
\end{align}
where  $\sigma_{n m}^S$ is the stochastic entropy increase in the system, and $q_\gamma^{(r)} = \hbar \omega_r (n_{\uparrow}^{(r)} - n_{\downarrow}^{(r)})$
is the stochastic heat entering the system from reservoir $r$, $n_{\uparrow \downarrow }^{(r)}$ being  the total number of upward or downward jumps in transition $r$. The expression for the adiabatic entropy production is particularly interesting, since it is equal to the entropy generated by the heat transfer between reservoirs at inverse temperatures $\beta_r$ and $\beta'_r$. In particular, the adiabatic entropy production is identically zero when $\beta_r=\beta_r'$ even though it is possible to have transient flows of heat.

We can now calculate the average rates of nonequilibrium potential and reservoirs entropy changes:
\begin{eqnarray}\label{dotS}
\langle \dot{\sigma}^{(r)} \rangle &=& \sum_{k =\uparrow,\downarrow} \tr[L_{k}^{(r) \dagger} L_{k}^{(r)} \rho_t] \sigma^{(r)}_k = - \beta_r \dot{Q}_r, \\
\dot{\Phi}_r &=& \sum_{k =\uparrow,\downarrow} \tr[L_{k}^{(r) \dagger} L_{k}^{(r)} \rho_t] \Delta \phi_k^{(r)} = \beta_r' \dot{Q}_r,
\end{eqnarray}
where we split in three parts the nonequilibrium potential flow $\dot{\Phi} = \dot{\Phi}_1 + \dot{\Phi}_2 + \dot{\Phi}_3
= -\tr[\dot{\rho}_S \ln \pi]$. The entropy production rates hence read:
\begin{align}
\dot{S}_{\rm a} &= \sum_r (\beta_r' - \beta_r) \dot{Q}_r \geq 0, \\
\dot{S}_{\rm na} &= \dot{S} - \sum_r \beta_r' \dot{Q}_r \geq 0, \\ 
\dot{S}_{\rm i} &= \dot{S} - \sum_r \beta_r \dot{Q}_r \geq 0, 
\end{align}
showing the same structure as the trajectory entropies in Eqs.~(\ref{stochasticversion1}-\ref{stochasticversion3}). Since there is no driving in this example, the non-adiabatic entropy production reads  as in Eq.~(\ref{sna_nodriving}), and equals $\Delta S_{\rm na}=S(\rho_0||\pi)$ for a full relaxation to the steady state $\pi$.

In the steady state we have $\dot{S}_{\rm na} = 0$, and the first law becomes $\sum_r \dot{Q}_r^{\rm ss} = 0$. 
This implies that the only contribution to the entropy production rate is the adiabatic one, which can be written as:
\begin{equation}\label{SL_steady}
\dot{S}_{\rm a} = \dot{S}_{\rm i} = (\beta_3 - \beta_2)\dot{Q}_2^{\rm ss} - (\beta_1 - \beta_3) \dot{Q}_1^{\rm ss} \geq 0.
\end{equation}
This equation can be used to bound the efficiency of the machine in the different regimes of operation. For instance, the efficiency of the machine acting as a refrigerator  is given by
\begin{equation}\label{efbound}
\epsilon = \frac{\dot{Q}_1^{\rm ss}}{\dot{Q}_2^{\rm ss}} \leq \frac{\beta_3 - \beta_2}{\beta_1 - \beta_3} \equiv \epsilon_C,
\end{equation}
where $\epsilon_C$ is the so-called Carnot efficiency  of a refrigerator \cite{VirtualQubits}.

\subsection{Periodically driven cavity mode} \label{S-cavity}

\begin{figure}[t]
\begin{center}
\includegraphics[width= \linewidth]{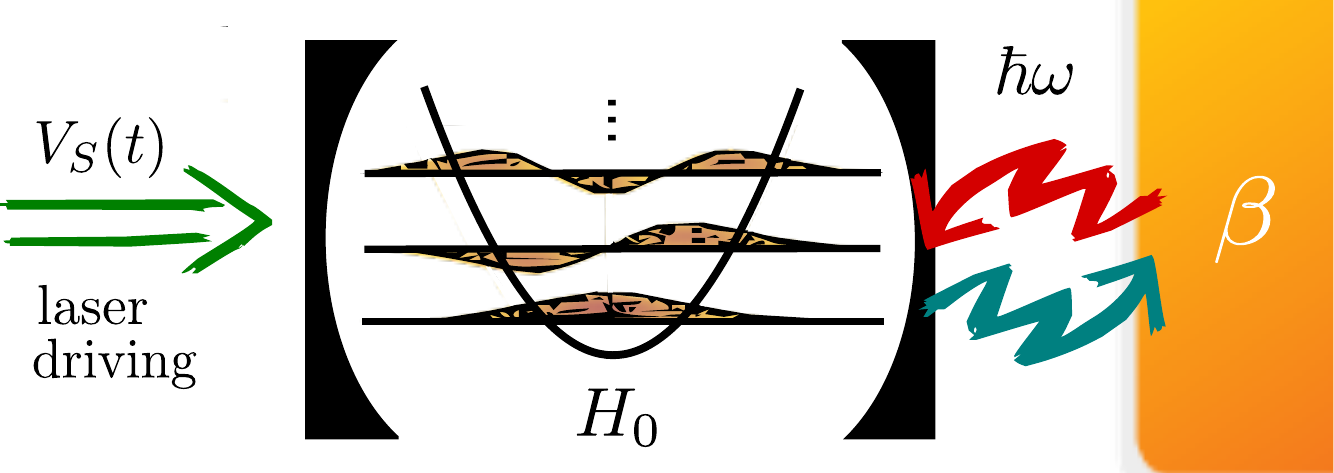}
\caption{Schematic picture of the setup. The intracavity mode $H_0$ is externally driven by a resonant laser field $V_S(t)$, 
while in weak contact with the environment at inverse temperature $\beta$, producing the emission and absorption of photons.} 
\label{F-cavity}
\end{center}
\end{figure}

Our third example consists of a single electromagnetic field mode with frequency $\omega$ in a microwave cavity with slight losses in one of the two mirrors. 
The losses of the cavity are produced by the weak coupling of the cavity mode to a bosonic thermal reservoir in equilibrium at inverse temperature 
$\beta = 1/k T$. In addition, an external laser of the same frequency $\omega$ and weak intensity drives 
the cavity mode producing excitations. The Hamiltonian of the system can be expressed as $H_S(t) = H_0 + V_S(t)$ consisting of two terms: 
the first one is the Hamiltonian of the undriven mode $H_0 = \hbar \omega a^\dagger a$ and 
\begin{equation}
V_S(t) = i \hbar (\epsilon a^\dagger e^{-i\omega t} - \epsilon^{\ast} a e^{i \omega t}),
\end{equation}
describes the effect of the classical resonant laser field with complex amplitude $\epsilon = |\epsilon| e^{i \varphi}$. Here the subscript $S$ stands for the Schr\"odinger picture, whereas operators and density matrices without any subscript will correspond to the interaction picture with respect to $H_0$ ($H_0$ is of course the same in the two pictures). Figure \ref{F-cavity} shows a schematic picture of the setup.

The reduced evolution of the cavity mode can be described by the following Lindblad master equation  \cite{Wiseman}
\begin{equation}\label{mastereqS}
\dot{\rho}_S(t) = - \frac{i}{\hbar}[H_S(t), \rho_S(t) ] + \mathcal{L}(\rho_S(t)).
\end{equation}
with the dissipative part
\begin{align} \label{dissipator}
\mathcal{L}(\rho) =~ & \Gamma_\downarrow \left(a \rho a^\dagger - \frac{1}{2} \{a^\dagger a, \rho \}\right) \nonumber \\ 
 +~ & \Gamma_\uparrow \left(a^\dagger \rho a - \frac{1}{2} \{a a^\dagger, \rho \} \right).
\end{align}  
This term accounts for emission and absorption of photons by the cavity mode from the equilibrium reservoir at respective rates $\Gamma_\downarrow = \gamma_0 (n^{\rm th} + 1)$ and 
$\Gamma_\uparrow = \gamma_0 ~n^{\rm th}$. Here again $n^{\rm th} = (e^{-\beta \hbar \omega}- 1)^{-1}$, and $\gamma_0$ is the spontaneous emission decay rate 
in absence of driving. The dissipative term ${\cal L}(\rho)$ does not depend on the driving: it induces jumps in the eigenbasis of $H_0$ and is also invariant under the change of picture. Notice that this is an approximation. For slow driving, for instance, the bath induces jumps between the instantaneous eigenstates of the Hamiltonian $H_S(t)$. The dissipator \eqref{dissipator} is valid for weak driving and weak coupling with the thermal bath, that is $\gamma_0 \sim |\epsilon| \ll \omega$ \cite{rivas2010}. 

In the interaction picture with respect to $H_0$,  the Lindblad equation (\ref{mastereqS}) reads  \cite{Wiseman}
\begin{equation}\label{mastereq}
\dot{\rho}(t) = - \frac{i}{\hbar}[V, \rho(t) ] + \mathcal{L}(\rho(t)),
\end{equation}
where $V = i \hbar(\epsilon a^\dagger - \epsilon^{\ast} a)$ is the driving Hamiltonian in the interaction picture, which turns 
to be constant.

Before discussing the FT applied to this example, let us calculate the energetics of the system from the Lindblad equation. For this purpose, it is more convenient to express the internal energy in the Schr\"odinger picture: $U(t)=\tr[H_S(t)\rho_S(t)]$. The first law reads $\dot U(t)=\dot W(t)+\dot Q(t)$, where the average work is given by 
\begin{align}
\dot{W}(t) &= \tr[\dot{H}_S(t) \rho_S(t)] \nonumber \\ 
 & = \hbar\omega \, \tr\left[ \left(\epsilon  a^\dagger e^{-i\omega t} + \epsilon^*ae^{i\omega t} \right)\rho_S(t)\right] \nonumber \\
& = \hbar\omega\, \tr\left[ \left(\epsilon  a^\dagger  + \epsilon^*a \right)\rho(t)\right]
\end{align}
and the average energy change not accounted for by work we denote
\begin{align}
\dot{Q}(t) & =  \tr[H_S(t) \dot\rho_S(t)]= \tr\left[H_S(t){\cal L}(\rho_S(t))\right] \nonumber \\
& = \tr\left[(H_0+V){\cal L}(\rho(t))\right],\label{q3}
\end{align}
although it is not necessarily equal to the heat, i.e., the energy reversibly exchanged with a thermal reservoir that accounts for the reservoir's entropy change~\cite{Horowitz2016}.
Below we discuss in detail the physical nature of this energy transfer. 

The steady state of the dynamics (\ref{mastereq}) obeys $- \frac{i}{\hbar}[V, \pi ] + \mathcal{L}(\pi) = 0$. This equation can be solved by noticing that the term $[V,\rho]$ in (\ref{mastereq}) 
cancels under the transformation $a \rightarrow a - \alpha$, where $\alpha = 2 \epsilon/\gamma_0$. The resulting steady state is
\begin{equation} \label{ssi}
\pi = \mathcal{D}(\alpha) ~\frac{e^{- \beta H_0}}{Z_0} ~\mathcal{D}^\dagger(\alpha), 
\end{equation}
where $Z_0 = \tr[\exp(-\beta H_0)]$, and $\mathcal{D}(\alpha) = \exp(\alpha a^\dagger - \alpha^\ast a)$ is the unitary displacement operator in optical phase space, fulfilling 
$\mathcal{D}(\alpha) a \mathcal{D}^\dagger(\alpha) = a - \alpha$, $\mathcal{D}^\dagger(\alpha) = \mathcal{D}(-\alpha)$. 
In contrast to the undriven case, here the cavity does not reach equilibrium 
with the reservoir: coherences in the energy basis do not decay to zero due to the work performed by the external laser. Notice also that the state $\pi$ 
defines a limit cycle (unitary orbit) in the Schr\"odinger picture. In the stationary regime $\pi_S(t) =e^{-i H_0 t/	\hbar}~\pi~ e^{i H_0 t/\hbar}$, i.e., the mode rotates in optical phase space, according to the free evolution 
$\dot{\pi}_S = (-i/\hbar)[H_0, \pi_S]$.

The energetics in this stationary regime is rather simple. The internal energy is constant, even though the state $\pi_S(t)$ depends on time:   $U_{\rm ss}=\tr [H_S\pi_S]=\tr [(H_0+V)\pi]=\tr [H_0\pi]= \hbar \omega (n^{\rm th}+|\alpha|^2)$, bigger than the thermal average energy $\hbar\omega n^{\rm th}$. The laser introduces energy at a rate:
\begin{equation}\label{sspower}
\dot{W}_{\rm ss} = \hbar \omega \tr[(\epsilon a^\dagger + \epsilon^\ast a) \pi] = \hbar\omega{\gamma_0}{|\alpha|^2} \geq 0,
\end{equation}
which is dissipated as heat to the thermal bath  $\dot Q_{\rm ss}=-\dot W_{\rm ss}$.

The FT can be applied both to the Srchr\"odinger or the interaction picture. Here it is is more convenient to determine the forward and backward processes in the interaction picture, where there is no driving.
The Kraus operators for the map $\E$ in Eq.~(\ref{inf-time-ev}) read in this case:
\begin{equation}
M_0 =  \mathbb{I} - dt  \left(\frac{i}{\hbar} V + \frac{1}{2} \sum_{k=\downarrow, \uparrow} L_{k}^\dagger L_{k}\right), \nonumber
\end{equation}
for the no-jump evolution, and 
\begin{align}\label{Lindop}
M_\downarrow = \sqrt{dt} L_\downarrow = \sqrt{dt \Gamma_\downarrow } a, \nonumber \\ 
M_\uparrow = \sqrt{dt} L_\uparrow = \sqrt{dt \Gamma_\uparrow } a^\dagger, \nonumber
\end{align}
for the downward and upward jumps corresponding to emission and absorption of photons.

The trajectory $\gamma = \{ n, (k_1, t_1), ... ,(k_N, t_N), m \}$ is then constructed as in the previous example by counting the jumps induced by the reservoir and registering the times at which they occur.

Since the forward dynamics is governed by a single pair of Lindblad operators $\{L_{\downarrow} = \sqrt{\Gamma_\downarrow} ~a,  L_{\uparrow} = \sqrt{\Gamma_\uparrow} ~a^\dagger\}$,
condition (\ref{trace-pre}) allows us to obtain the stochastic entropy changes in the reservoir:
\begin{equation}\label{er2}
 \sigma_0^E = 0, ~~~~~ \sigma_{\downarrow}^E = \beta \hbar \omega, ~~~~~ \sigma_{\uparrow}^E = - \beta \hbar \omega.
\end{equation}
That is, in a downward (upward) jump, the entropy in the environment increases (decreases) by $\beta \hbar \omega$, corresponding to a transfer of energy  $\hbar \omega$. 
In average, this transfer of energy equals $\tr[H_0 \mathcal{L}(\rho(t))]$, whereas the energy not accounted for by work is given by Eq.~\eqref{q3}, i.e., by $\dot{Q} (t)= \tr[(H_0+V) \mathcal{L}(\rho(t))]$. The origin of this discrepancy is the choice of a dissipator (\ref{dissipator})  independent of the driving. As already mentioned, the dissipator is valid for weak driving \cite{rivas2010}, when the term  $\tr[V \mathcal{L}(\rho(t))]\sim O(\gamma|\epsilon_0|)$ is negligible.

However, it is worth it to notice that our approach does not depend on the physical nature of the environment and its interaction with the system. As shown in section \ref{S-MarkovianME}, once a Lindblad equation like \eqref{mastereqS} with a given set of Lindblad operators for its unraveling has been specified, no matter how it has been derived, induces an entropy change in the environment given by Eq. \eqref{er2}. This is a direct consequence of micro-reversibility that yields condition \eqref{trace-pre} on the Lindblad operators. When these operators come in pairs, as it is the case in our example, the condition determines the entropy change in the environment [see Eq.~\eqref{entpairs}]. 

Therefore, if one could conceive physical situations where the Lindblad equation \eqref{mastereqS} is exact, then the entropy production would be given by Eq. \eqref{er2} and the energy transfer $\tr[V \mathcal{L}(\rho(t))]$ would not contribute to the entropy of the environment. A clue on the nature  of this energy transfer is provided by  Ref.~\cite{Auffeves2}. In that paper, Elouard \emph{et. al.} introduce  a driven two-level system in an engineered thermal bath where  excitations occur through a third level with a very short lifetime and transitions are monitored by measuring emitted photons. The resulting Lindblad equation is the analog of Eq. \eqref{mastereqS} for a two-level system and the entropy change in the environment is precisely \eqref{er2}. They show that the energy transfer $\tr[V \mathcal{L}(\rho(t))]$ is due to the collapse of a coherent state induced by the photon detection. This energy transfer  does not change the entropy of the universe and has been categorized either as ``measurement work'' \cite{JordanParrondo,JordanPRA} or as ``quantum heat'' \cite{Auffeves, Auffeves2} due to measurement. 

The Kraus operators of the backward map are given by  Eq.~(\ref{mpairs}):
\begin{align}
\tilde{M}_0 &=~ \Theta M_0 \Theta^\dagger = M_0,  \\
\tilde{M}_\downarrow &= \sqrt{dt} \tilde{L}_\downarrow = \sqrt{dt} \Theta L_\uparrow \Theta^\dagger = M_\uparrow,  \\ 
\tilde{M}_\uparrow &= \sqrt{dt} \tilde{L}_\uparrow = \sqrt{dt} \Theta L_\downarrow \Theta^\dagger = M_\downarrow, 
\end{align} 
implying again  that the forward and the backward map are equivalent, i.e., the jumps up (down) in the forward process are transformed in jumps down (up) in the backward process.

The main feature of this example is that the key condition (\ref{condition}) is not fulfilled. Recall that this condition is needed to define the dual and dual-reverse dynamics as well as the stochastic adiabatic and non-adiabatic entropy production at the trajectory level. Using the expression for the stationary state  Eq.~(\ref{ssi}), we can calculate the nonequilibrium potential in the interaction picture
\begin{align} \label{nonequ-cavity}
\HAT{\Phi} &= -\ln\pi= \beta \mathcal{D}(\alpha) H_0 \mathcal{D}^\dagger(\alpha) + \ln Z_0  \nonumber \\ 
&=  \beta H_0 - \beta \hbar \omega |\alpha|( x_{\varphi}-|\alpha|)  + \ln Z_0,
\end{align}
where we have introduced 
the field quadrature 
\begin{equation}
 x_{\varphi}  =  a^\dagger e^{i \varphi} +  a^{~}e^{-i \varphi}.
 \end{equation}

The nonequilibrium potential $\Phi$ in Eq. \eqref{nonequ-cavity} does not obey Eq.~(\ref{conditionrelations}), because the Lindblad operators appearing in the dynamics (\ref{mastereq}) promote jumps among the  eigenstates of the unperturbed Hamiltonian $H_0$, instead of the eigenstates of the steady density matrix $\pi$. This implies that we cannot associate a single change in the nonequilibrium potential to each of the Lindblad jump operators, 
nor to $M_0$. As a consequence, the entropy production per trajectory  cannot be decomposed in an adiabatic and  a non-adiabatic contributions and the corresponding fluctuation theorems do not apply. 
However, the conditions in Eq.~(\ref{conditionrelations}) can be recovered in some cases by properly including the effect of the driving on the Lindblad operators~\cite{JordanPRA}.

\begin{figure*}[t]
\begin{center}
\includegraphics[width = \linewidth]{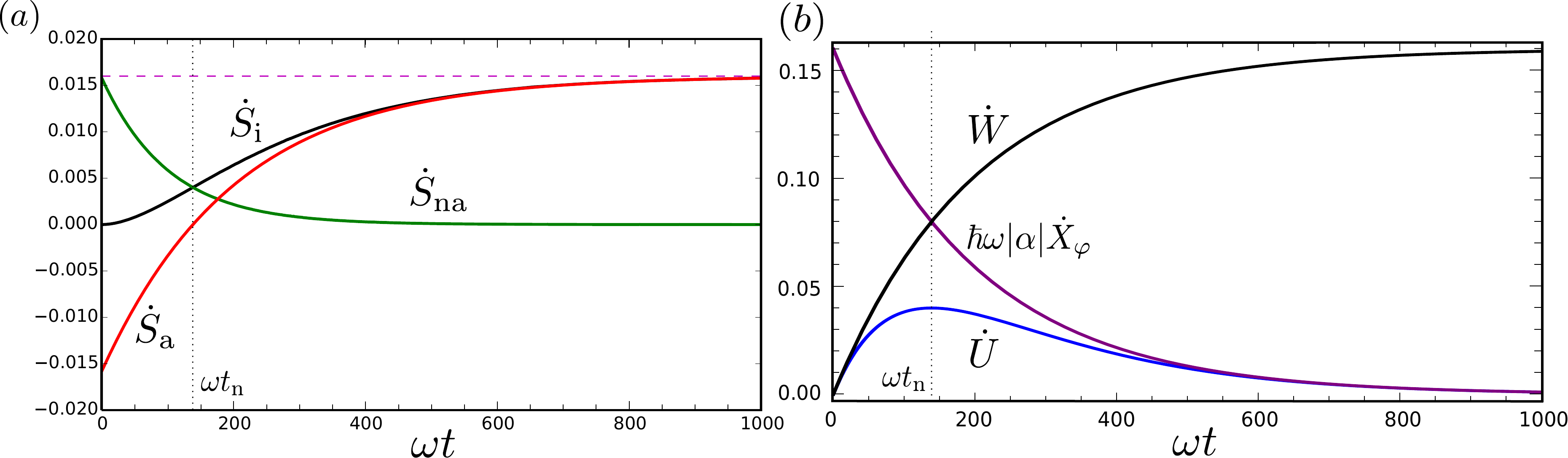}
\caption{Time evolution of $(a)$ adiabatic ($\dot{S}_{\mathrm a}$), non-adiabatic ($\dot{S}_{\mathrm{na}}$), and total ($\dot{S}_{\mathrm i}$) entropy production rates represented by 
color solid lines, and $(b)$ input power ($\dot{W}$), rate at which the cavity mode absorbs energy ($\dot{U}$), and displacement rate ($\dot{X}_{\varphi}$). The cavity mode starts in 
equilibrium with the thermal reservoir, $\rho_{0}= e^{- \beta \hat{H}_0}{Z}$, and the laser driving is switched on at $t=0$. The dashed line in $(a)$ corresponds to $\beta \dot{W}_{\rm ss}$, and 
we used vertical dotted lines to highlight the instant of time at which the adiabatic entropy production rate changes its sign ($t_\mathrm{n}$). We used parameters 
$\epsilon = 0.02 \omega$, $\gamma_0 = 0.01 \omega$, and temperature $kT = 10  \hbar \omega$, for $\hbar \omega = 1$ units.} 
\label{F-cavityentropy}
\end{center}
\end{figure*}

Even though the adiabatic and non-adiabatic entropy production cannot be defined at the trajectory level, we can calculate their averages   \cite{Spohn} using, for instance, Eqs.~(\ref{dynamical-sl2}):
\begin{eqnarray}\label{nonadeprate}
 \dot{S}_{\rm na} &= &-\dot S(\rho(t)||\pi)= \dot{S} - \beta \left( \dot{U} - \hbar\omega|\alpha| \dot{X}_{\varphi} \right)   \\
 \dot{S}_{\rm a} &=& \dot{S}_{\rm i} - \dot{S}_{\rm na} = \beta(\dot{W} - \hbar\omega|\alpha| \dot{X}_\mathcal{\varphi}).
\end{eqnarray}
Here we have used $\dot{S}_{\rm i}=\dot S-\beta\dot Q$ and $\dot U(t)=\langle \dot H_0\rangle =\dot Q+\dot W$, and introduced $\dot{X}_{\varphi} \equiv \tr[x_{\varphi} \dot{\rho}(t)]$, the rate at which the cavity field mode is displaced by the laser (with phase $\varphi$) until the 
steady state is reached. 
Since there are no fluctuation theorems for these quantities, in principle they could be negative. The non-adiabatic entropy production, however, still obeys (\ref{non-adiab_lind}) and, since the steady state $\pi$ is constant in the interaction picture, it can be written as the change 
of the relative entropy between the state $\rho(t)$ and $\pi$, which is always positive: $\dot S_{\rm na}=-\dot S(\rho(t)||\pi)\geq 0$. This is not the case of the adiabatic entropy production $\dot S_{\rm a}$, which indeed can take on negative values. The expression for $\dot S_{\rm a}$ 
in Eqs.~(\ref{nonadeprate}) does equal the entropy production in the steady state, $\dot{S}_\mathrm{a} \rightarrow \beta W_\mathrm{ss}=- \beta Q_\mathrm{ss} \geq 0$. However, it can be negative in the transient regime, as  shown in Fig. \ref{F-cavityentropy} (see also appendix \ref{appE}). 
In Fig.\ref{F-cavityentropy}(a) we depict the evolution of the three entropy production rates when the cavity mode starts in a Gibbs thermal state in equilibrium with the reservoir temperature, $\rho_0 = \exp(-\beta \hat{H}_0)/Z_0$. We find that the entropy of the mode is kept constant during 
the evolution, $\dot{S} = 0~ \forall t$, which implies $\dot{S}_{\mathrm i}= - \beta \dot{Q} \geq 0$, and $\dot{S}_{\mathrm{na}} = \beta(\hbar\omega|\alpha| \dot{X}_{\varphi} - \dot{U}) \geq 0$. On the other hand, the adiabatic entropy production rate $\dot{S}_{\mathrm a} = \beta(\dot{W} - \mu \dot{X}_{\varphi})$ 
is negative for times $t<t_{\rm n}\equiv 2\ln2 /\gamma_0$. It is worth mentioning that for this initial condition the term $\tr[V{\cal L}(\rho(t))]$  in the energetics vanishes at any time $t$.

To explore the origin of this purely quantum effect, we plot the energetics of the relaxation process  Fig.~\ref{F-cavityentropy}(b). The cavity mode absorbs energy at constant entropy from the external laser 
until the periodic steady state is reached, $\dot{U}= \dot{W}e^{- \gamma_0 t /2}$, where $\dot{W} = \dot{W}_{\mathrm{ss}}(1 - e^{-\gamma_0 t/2})\geq 0$ is the input power and, consequently,  heat is dissipated 
at a rate $ - \dot{Q} = \dot{W} (1- e^{-\gamma_0 t/2}) \geq 0$. When the relaxation is completed, the input laser power is fully dissipated into the reservoir, i.e. $\dot{Q}_{\mathrm{ss}} = - \dot{W}_{\mathrm{ss}}$. 
The energy absorbed by the cavity mode during the evolution is fully employed to generate the unitary displacement $\alpha$, that is, $\Delta U = \hbar\omega|\alpha| \Delta X_{\varphi} = \hbar \omega |\alpha|^2$. 
However, the transient dynamics ruling this process is far from trivial. The cavity mode is always displaced, i.e. gaining coherence, at a higher rate than energy, $\dot{U} = \hbar \omega |\alpha| \dot{X}_{\mathcal{S}} (1- e^{-\gamma_0 t/2})$, 
in accordance with the positive non-adiabatic entropy production rate. In addition, by comparing Figs. \ref{F-cavityentropy}(a) and \ref{F-cavityentropy}(b) the energetic meaning of the adiabatic 
entropy production rate can be clarified. In the initial transient where $\dot{S}_{\mathrm{a}} < 0$ the coherence gain surpass the input power, i.e. 
$\hbar \omega |\alpha| \dot{X}_{\varphi} > \dot{W}$, which in turn implies that the rate at which the cavity mode gains energy speeds up in this period $\ddot{U} > 0$. 
At time $t_{\rm n}$, when $\dot{S}_{\mathrm a}=0$, we have $\dot{W} = \hbar \omega |\alpha| \dot{X}_{\varphi} = \dot{W}_{\mathrm{ss}}/2$, and $\dot{U}$ peaks at its maximum. 
After this time, the adiabatic entropy production rate is positive $\dot{S}_{\mathrm a} >0$, implying $\hbar \omega |\alpha| \dot{X}_{\varphi} < \dot{W}$, and $\dot{U}$ 
decreases until it becomes zero in the long time run, when the periodic steady state is reached. In conclusion, we obtained that the sign of the adiabatic entropy 
production rate spotlights the acceleration in the internal energy changes of the cavity mode.

\section{Conclusions}  \label{S-Conclusions}

In this paper we have analyzed the production of entropy in general processes embedded in a two measurement protocol, with local measurements performed in both system and environment. 
Our first main result is the fluctuation theorem (\ref{m_logratio}), which compares the probability of forward and backward trajectories. 
Particularizing this expression to certain initial conditions of the backward process, one can obtain FT's for the change of inclusive (\ref{sq2}) and exclusive (\ref{sprod2}) entropy production, 
i.e., for the entropy production that results from keeping or neglecting the classical correlations generated between the system and the environment during the process. 
These differences have been illustrated for the case of two qubits interacting through a CNOT quantum gate.

We have also explored whether it is possible to split the total entropy production into adiabatic and non-adiabatic contributions, as it is customary in classical systems far from equilibrium  \cite{EspositoFacesI, EspositoFacesII}. 
For that purpose, we have introduced a dual dynamics for the reduced evolution of the system, which in turn  allowed us to clarify the interpretation of previous FT's derived for quantum CPTP maps \cite{MHP}. 
We have shown that the aforementioned decomposition is possible only if the reduced dynamics satisfies certain condition, namely Eq.~(\ref{condition}). 
In fact, we give an explicit example where that condition is not fulfilled and the adiabatic entropy production takes on negative values. 
This is a pure quantum feature whose consequences, we believe, are worth to be further explored.

Our results can be applied to a broad range of quantum processes including multipartite environments and concatenations of CPTP maps. In particular, we developed in detail the application to processes described by 
Lindblad master equations. We have introduced a general method to identify the environmental entropy change during the trajectories induced by quantum jumps [see Eq.~(\ref{trace-pre}) and below], which 
allows us to recover the FT's. The meaning of the terms adiabatic and non-adiabatic become clear in this situation, since the non-adiabatic contribution tends to zero for quasi-static driving.

We have finally studied the decomposition of the total entropy production in two specific situations of interest: an autonomous three-level thermal machine and  a dissipative cavity mode resonantly driven by a classical field.  

Summarizing, our results provide an exhaustive characterization of the entropy production in open quantum systems undergoing arbitrary processes. This includes: systems in contact with non-thermal or finite-size reservoirs, 
configurations with several equilibrium baths with different temperatures or chemical potentials, driven systems, etc. In all those cases, one should be able to assess the entropy production and characterize its fluctuations 
within the theoretical framework presented in this paper. Therefore, our results clarify the origin of entropy production from coarse-graining and its link 
to thermodynamical notions when particular choices for the environment are made.

\acknowledgments{
We thank Gili Bisker for comments. The authors acknowledge funding from MINECO (Proyecto TerMic, FIS2014-52486-R). G. M. acknowledges support from FPI grant no. BES-2012-054025. 
This work has been partially supported by COST Action MP1209 ``Thermodynamics in the quantum regime''.
}

\appendix

\section{Infinitesimal changes in the state of the reservoir} \label{appA}

In this appendix we show that the term $S(\rho_E^\ast || \rho_E)$ appearing in Eq.~(\ref{eptrajectory-reset}) is negligible for infinitesimal changes in the environment density matrix. 
Let us assume the change in the environment density operator in the following general form: 
\begin{equation}
 \rho_E^\ast = \rho_E + \epsilon \Delta \rho_E,
\end{equation}
where $\tr[\Delta \rho_E] = 0$, and $\epsilon \geq 0$ is a small real number. Using the definition of the quantum relative entropy, we can then write:
\begin{equation} \label{rel-ent}
 S(\rho_E^\ast || \rho_E) =  S(\rho_E) - S(\rho_E^\ast) - \epsilon \tr[\Delta \rho_E \ln \rho_E].
\end{equation}
In the following we show that if $\epsilon \ll 1$, then $S(\rho_E^\ast) - S(\rho_E) \simeq - \epsilon \tr[\Delta \rho_E \ln \rho_E]$, and consequently $S(\rho_E^\ast || \rho_E) \rightarrow 0$.
This can be done by applying perturbation theory. Let the eigenvalues and eigenstates of $\rho_E^\ast$, the set $\{q_\mu^\ast, \ket{\phi_\mu^\ast}\}$, be expanded to second order in $\epsilon$:
\begin{align}
 q_\mu^\ast  &\simeq q_\mu + \epsilon q_\mu^{(1)} + \epsilon^2 q_\mu^{(2)},   \\
 \ket{\phi_\mu^\ast} &\simeq \ket{\phi_\mu} + \epsilon \ket{\phi_\mu^{(1)}} + \epsilon^2 \ket{\phi_\mu^{(2)}}, 
\end{align}
where the zeroth order contributions obey $\rho_E \ket{\phi_\mu} = q_\mu \ket{\phi_\mu}$, and we have for the first order terms:
\begin{align}
 q_\mu^{(1)} &= \bra{\phi_\mu} \Delta \rho \ket{\phi_\mu},  \\
 \ket{\phi_\mu^{(1)}} &= \sum_{\nu \neq \mu} \frac{\bra{\phi_\nu} \Delta \rho \ket{\phi_\mu}}{q_\mu - q_\nu} \ket{\phi_\nu}. 
\end{align}

We now calculate the entropy change up to second order in $\epsilon$:
\begin{eqnarray} \label{echange}
  & S(\rho_E^\ast) - S(\rho_E)  = - \sum_\mu q_\mu^\ast \ln q_\mu^\ast + \sum_\nu q_\nu \ln q_\nu \nonumber \\ 
  & \simeq - \epsilon \sum_\mu q_\mu^{(1)} \ln q_\mu - \epsilon^2 \big(q_\mu^{(2)} \ln q_\mu + \sum_\mu \frac{q_\mu^{(1)~2}}{2 q_\mu} \big), 
\end{eqnarray}
to be compared with:
\begin{align} \label{eflow}
- \epsilon \tr[\Delta \rho_E \ln \rho_E] & = - \epsilon \sum_\mu \bra{\phi_\mu} \Delta \rho \ket{\phi_\mu} \ln q_\mu \nonumber \\ 
& = - \epsilon \sum_\mu q_\mu^{(1)} \ln q_\mu.
\end{align}
The above equations (\ref{echange}) and (\ref{eflow}) are equal up to first order, differing in $\mathcal{O}(\epsilon^2)$. Therefore, using Eq. (\ref{rel-ent}), we conclude that:
\begin{equation}
 S(\rho_E^\ast || \rho_E) \simeq - \epsilon^2 \big(q_\mu^{(2)} \ln q_\mu + \sum_\mu \frac{q_\mu^{(1)~2}}{2 q_\mu} \big) = \mathcal{O}(\epsilon^2), 
\end{equation}
and when $\epsilon \ll 1$ we can consider $S(\rho_E^\ast || \rho_E) \rightarrow 0$ up to first order, in contrast to the change in entropy [Eq. (\ref{echange})].

\section{Multipartite environments} \label{appB}

Recall that we assume $R$ ancillary systems in an uncorrelated state, $\rho_E = \rho_1 \otimes ... \otimes \rho_R$, and local measurements in each separate environmental ancilla. 
We denote the local density operators of the environmental ancilla $r$ at the beginning and at the end of the (forward) process as
\begin{equation}
\rho_r = \sum_{\nu} q^{(r)}_{\nu} \HAT{\mathcal{Q}}^{(r)}_{\nu}, ~~~~~~~ \rho_r^\ast = \sum_{\mu} q^{(r) \ast}_{\mu} \HAT{\mathcal{Q}}^{(r) \ast}_{\mu}, 
\end{equation}
with eigenvalues $q^{(r)}_{\nu}$ and $q^{(r) \ast}_{\mu}$, and orthogonal projectors onto its eigenstates $\HAT{\mathcal{Q}}^{(r)}_{\nu} = \ket{\chi^{(r)}_{\nu}}\bra{\chi^{(r)}_{\nu}}_E$ and 
$\HAT{\mathcal{Q}}^{(r) \ast}_{\mu} = \ket{\chi^{(r) \ast}_{\mu}}\bra{\chi^{(r) \ast}_{\mu}}_E$.

The generalization of the results is then straightforward by considering the same steps and assumptions as before. The reduced system dynamics is 
again given by Eq.~(\ref{f_map}), but the operators $\HAT{M}_{\mu \nu}$ now using collective indices 
\begin{equation}
(\mu, \nu) = \{(\nu^{(1)}, \mu^{(1)}), ... , (\nu^{(R)}, \mu^{(R)})\},
\end{equation} 
representing the set of transitions obtained in the projective measurements of all environmental ancillas:
\begin{equation}\label{e_transition}
\ket{\chi^{(r)}_{\nu^{(r)}}}_E \rightarrow \ket{\chi^{(r) \ast}_{\mu^{(r)}}}_E ~~~~~\mathrm{ for}~ r= 1, ..., R.
\end{equation}
That is, the Kraus operators of the forward process are given by
\begin{equation}\label{k_ops_multi}
\HAT{M}_{\mu \nu} = \left( \prod_{r=1}^R \sqrt{q^{(r)}_{\nu^{(r)}}} \right) \bra{\chi^{(1) \ast}_{\mu^{(1)}} ... ~\chi^{(R) \ast}_{\mu^{(R)}}}_{E} \HAT{U}_{\Lambda} \ket{\chi^{(1)}_{\nu^{(1)}} ...~ \chi^{(R)}_{\nu^{(R)}}}_{E}, 
\end{equation}
and analogously for the Kraus operators of the backward process (\ref{b_kraus}) we have
\begin{align}\label{b_k_ops_multi} 
\HAT{\tilde{M}}_{\nu \mu} = & \left( \prod_{r=1}^R \sqrt{\tilde{q}^{(r)}_{\mu^{(r)}}} \right) \\  
& \times \bra{\chi^{(1)}_{\nu^{(1)}} ... ~\chi^{(R)}_{\nu^{(R)}}}_{E} \HAT{\Theta}_E^\dagger \HAT{U}_{\tilde{\Lambda}} \HAT{\Theta}_E \ket{\chi^{(1) \ast}_{\mu^{(1)}} ...~ \chi^{(R) \ast}_{\mu^{(R)}}}_{E}. \nonumber
\end{align}
The key relation (\ref{b_kraus2}) necessary to obtain the fluctuation theorem for the total entropy production (\ref{m_logratio}) hence follows as well in this case, 
with a decomposition of the environment boundary term
\begin{equation}
 \sigma_{\mu \nu}^E = \sum_{r=1}^R \sigma_{\mu^{(r)} \nu^{(r)}}^{(r)},
\end{equation}
being $\sigma_{\mu^{(r)} \nu^{(r)}}^{(r)} \equiv -\ln \tilde{q}^{(r)}_{\mu^{(r)}} + \ln q^{(r)}_{\nu^{(r)}}$.

The application of the above formalism introducing the dual and dual-reverse processes follows immediately in the same manner, leading to the fluctuation theorems 
for the adiabatic and non-adiabatic entropy production in detailed and integral versions, Eqs. (\ref{dualr_dft}), (\ref{sa2}) and (\ref{i-FT}).
The adiabatic entropy production per trajectory and its average then read in this case:
\begin{eqnarray}
\Delta_\mathrm{i} s^{\mathrm{a}}_{\mu \nu} &= \sum_{r=1}^R \sigma^r_{\mu^{(r)} \nu^{(r)}} + ~\Delta \phi_{\mu \nu},  \\
\Delta_\mathrm{i} S_\mathrm{ a} &=  \sum_{r=1}^R  S(\rho_r^\ast) - S(\rho_r)  + \langle \Delta \phi \rangle \geq 0,
\end{eqnarray}
where in the averaged version we set again (uncorrelated) reversible boundaries, $\tilde{\rho}_{\mathcal{S} E} = \HAT{\Theta} (\rho_{\mathcal{S}}^\ast \otimes \rho_1^\ast \otimes ... ~\otimes \rho_R^{\ast}) \HAT{\Theta}^\dagger$.

\section{Concatenations of CPTP maps} \label{appC}

In the following we focus in the derivation of FT's for concatenations of CPTP maps reported in Sec. \ref{S-Concatenations}. Here we assume the environment is a single reservoir or ancilla. 
However, the extension to multiple reservoirs follows in the same manner than in the one map case (see Appendix \ref{appB}).

Consider the maps concatenation $\Omega$ in Eq.~(\ref{Omega}). For any map $\E^{(l)}$ in the sequence the environmental ancilla starts in a generic state
\begin{equation}\label{rhoEcon}
 \rho_E^{(l)} \equiv \sum_{\alpha} q_{\alpha}^{(l)} \HAT{\mathcal{Q}}_{\alpha}^{(l)},
\end{equation} 
and it is measured at the beginning and at the end of the interaction with the system, generating outcomes labeled as $\nu_l$ and $\mu_l$, respectively. 
The measurements are specified by the rank-one projective 
operators $\{ \HAT{\mathcal{Q}}_{\nu_k}^{(l)}\} \equiv \{ \ket{\phi_{\nu_l}^{(l)}} \bra{\phi_{\nu_l}^{(l)}}\}$ for the initial measurement 
and $\{\HAT{\mathcal{Q}}_{\mu_l}^{(l) \ast} \equiv \ket{\phi_{\mu_l}^{(l) \ast}} \bra{\phi_{\mu_l}^{(l) \ast}} \}$ for the final one. 
Under this conditions, each map in the concatenation can be written as: 
\begin{equation} \label{k_op_c}
 \E^{(l)}(\cdot) = \sum_{\mu_l, \nu_l} M_{\mu_l, \nu_l}^{(l)} (\cdot )~ M_{\mu_l \nu_l}^{(l) \dagger},
\end{equation}
with $M_{\mu_l \nu_l}^{(l)} \equiv \sqrt{q_{\nu_l}^{(l)}} \bra{\phi_{\mu_l}^{(l) \ast}} \HAT{U}_\Lambda^{(l)} \ket{\phi_{\nu_l}^{(l)}}$, 
where the unitary evolution $\HAT{U}_\Lambda^{(l)}$ is as in Eq.~(\ref{unitary}). 
Here we consider always the same total time-dependent Hamiltonian $\HAT{H}(t)$, following an arbitrary driving protocol $\Lambda = \{\lambda_t |~ 0 \leq t \leq N\tau \}$. 
For convenience the latter can be also split into $N$ intervals; hence the partial protocol $\Lambda_l = \{ \lambda_t |~ t_{l-1} \leq t \leq t_l \}$ generates the unitary 
operator $\HAT{U}_\Lambda^{(l)}$.

A quantum trajectory in this context is defined as follows. At time $t=0$ we start with our system in $\rho_S$, which is measured with eigenprojectors $\{ \HAT{\mathcal{P}}_n\}$, 
obtaining outcome $n$. Then the sequence of maps $\HAT{\Omega}$ defined in Eq.~(\ref{Omega}) is applied, obtaining outcomes $\{ \mu_l, \nu_l \}$ from each of the $l = 1, ..., N$ 
pairs of measurements in the environment. Finally at time $t = N \tau$ the system is measured again with arbitrary (rank-one) projectors $\{ \HAT{\mathcal{P}}_m^\ast \}$ giving outcome $m$. 
A quantum trajectory is now completely specified by the set of outcomes, $\gamma = \{n , (\nu_1, \mu_1), ..., (\nu_N, \mu_N), m \}$, and 
occurs with probability
\begin{equation}\label{fp_c}
P (\gamma) = p_n ~{\rm Tr}[\HAT{\mathcal{P}}_m^\ast~ \E^{(N)}_{\mu_N \nu_N} \circ ... \circ \E^{(1)}_{\mu_1 \nu_1} (\HAT{\mathcal{P}}_n)].
\end{equation}

Now we can apply the same arguments in previous sections to construct the three different processes used to state the FT's. 
For the initial state of the backward process, we consider again an arbitrary initial state of the system $\tilde{\rho}_S = \sum_m \tilde{p}_m \HAT\Theta \HAT{\mathcal{P}}_m^\ast \HAT\Theta^\dagger$, uncorrelated from the environment initial states $\tilde{\rho}_E^{(l)} = \sum_\alpha \tilde{q}_\alpha \HAT\Theta \HAT{\mathcal{Q}}_\alpha^\ast \HAT\Theta^\dagger$, 
and apply the sequence of maps $\tilde{\Omega} = \tilde{\E}^{(1)} \circ ... \circ~\tilde{\E}^{(l)} \circ ...\circ ~\tilde{\E}^{(N)}$, generating a trajectory $\tilde{\gamma} = \{m , (\mu_1, \nu_1), ..., (\mu_N, \nu_N), n \}$ with probability:
\begin{equation}
\tilde{P} (\tilde{\gamma}) = \tilde{p}_m ~\tr[\HAT\Theta_S \HAT{\mathcal{P}}_n \HAT\Theta_S^\dagger ~\tilde{\E}^{(1)}_{\nu_1 \mu_1} \circ ... \circ ~\tilde{\E}^{(N)}_{\nu_N \mu_N} (\HAT\Theta_S \HAT{\mathcal{P}}_m^\ast \HAT\Theta_S^\dagger)]. 
\end{equation}
Here the backward maps, $\tilde{\E}^{(l)}$, and their corresponding operations, are defined from each map $\E^{(l)}$ in the concatenation $\HAT{\Omega}$ by applying Eqs.~(\ref{b_ops}) and (\ref{b_kraus}). 

Dual and dual-reverse maps and operations also follow from its definitions in Sec.~\ref{S-DecompositionFTs} when conditions (\ref{condition}) and $\tilde{\E}^{(l)}(\tilde{\pi}^{(l)}) = \tilde{\pi}^{(l)}$ are met for each map in the sequence. 
The corresponding probabilities for trajectory $\gamma$ in the dual process, and trajectory $\tilde{\gamma}$ in the dual-reverse are:
\begin{align}
P_D (\gamma) &= p_n ~\tr[\HAT{\mathcal{P}}_m^\ast ~\D^{(N)}_{\mu_N \nu_N} \circ ... \circ \D^{(1)}_{\mu_1 \nu_1} (\HAT{\mathcal{P}}_n)],  \\
\tilde{P}_D (\tilde{\gamma}) &= \tilde{p}_m ~\tr[\HAT\Theta \HAT{\mathcal{P}}_n \HAT\Theta^\dagger ~\tilde{\D}^{(1)}_{\nu_1, \mu_1} \circ ... \circ \tilde{\D}^{(N)}_{\nu_N \mu_N} (\HAT\Theta \HAT{\mathcal{P}}_m^\ast \HAT\Theta^\dagger)], 
\end{align}
where in the dual-reverse trajectories we took again the sequence of maps in inverted order, that is, we applied $\tilde{\D}^{(1)} \circ ... \circ \tilde{\D}^{(N)}$ over the initial state $\tilde{\rho}_S$.  

Again, the Kraus operators for the backward, dual, and dual-reverse trajectories, fulfill the set of operator detailed-balance relations:
\begin{eqnarray}
\HAT\Theta^\dagger \HAT{\tilde{M}}_{\nu \mu}^{(l)} \HAT\Theta &=& e^{-\sigma^E_{\mu_{l}, \nu_{l}}/2}  \HAT{M}_{\mu \nu}^{(l) \dagger},  \label{d_b_b_c} \\ 
\HAT\Theta^\dagger \HAT{\tilde{D}}_{\nu \mu}^{(l)} \HAT\Theta &=& e^{~\Delta \phi_{\mu \nu}^{(l)} /2} ~ \HAT{M}_{\mu \nu}^{(l) \dagger},  \label{d_b_b_c2}\\  
\HAT{D}_{\mu \nu}^{(l)} ~&=& e^{-(\sigma^E_{\mu_{l}, v_{l}} + \Delta \phi_{\mu \nu}^{(l)})/2} ~\HAT{M}_{\mu \nu}^{(l)}, \label{d_b_b_c3}
\end{eqnarray}
where the nonequilibrium potential changes are defined with respect to the invariant state $\pi^{(l)}$ of each map $\E^{(l)}$ as in the single map case: $\Delta \phi^{(l)}_{\mu \nu} = - \ln \pi_{\mu}^{(l)} + \ln \pi_{\nu}^{(l)}$.

The set of equations (\ref{d_b_b_c}-\ref{d_b_b_c3}) immediately implies the detailed FT's for concatenations in Eqs.~(\ref{dualr_dft_c}-\ref{total_dft_c}).
Its corresponding integral versions and second-law-like inequalities follow immediately as a corollary.

Finally, it is interesting to consider the expression of the average nonequilibrium potential change during the whole sequence. By denoting $\rho_S(t_l)$ the reduced state of the system at time $t_l$, we have:
\begin{align} \label{neqtp_c}
 \Delta \Phi &= \sum_{l=1}^N \tr[\E_{\mu_l \nu_l}^{(l)} (\rho_S(t_{l-1}))] \Delta \phi^{(l)}_{\mu_l \nu_l} \nonumber \\
 & = \sum_{l=1}^N \tr[(\rho_S(t_{l}) - \rho_S(t_{l-1})) \HAT{\Phi}_l],
\end{align}
where $\HAT{\Phi}_l = - \ln \pi^{(l)}$. 
The above expression can be decomposed into the following {\it boundary} and {\it path} contributions:
\begin{align}
 \Delta \Phi_{\rm b} &= {\rm Tr}[\rho_S' \HAT{\Phi}_N] - {\rm Tr}[\rho_S \HAT{\Phi}_1], \\ 
 \Delta \Phi_{\rm p} &= - \sum_{l=1}^{N-1} {\rm Tr}[\rho_S(t_l) (\HAT{\Phi}_{l+1} - \HAT{\Phi}_l)].
\end{align}
When all the maps in the concatenation have the same invariant state, $\HAT{\Phi}_{l+1} = \HAT{\Phi}_l \equiv \HAT{\Phi} ~ \forall l$, 
we obtain $\Delta \Phi_{\rm p} = 0$, while $\Delta \Phi_{\rm b} = {\rm Tr}[(\rho_S' - \rho_S) \HAT{\Phi}]$ and we recover the expression for the single map case, c.f. Eq.~(\ref{deltaphi4}). 
In the other hand the boundary term only vanishes for cyclic processes, such that $\rho_S' = \rho_S$, implemented by cyclic concatenations with $\HAT{\Phi}_N = \HAT{\Phi}_1$. 
In this case $\Delta \Phi_{\rm b} = 0$ while $\Delta \Phi_{\rm p}$ gives in general a non-zero contribution.

The dynamical versions of these boundary and path terms read:
\begin{eqnarray}
\dot{\Phi}_{\rm b} = \frac{d}{dt} \left( \tr[\rho_t \HAT{\Phi}(\lambda_t)] \right), ~~
\dot{\Phi}_{\rm p} = -\tr[\rho_t \dot{\HAT{\Phi}}(\lambda_t)], ~~
\end{eqnarray}
which are also analogous to their classical counterparts \cite{EspositoFaces, EspositoFacesI,EspositoFacesII}.

\section{Autonomous quantum thermal machines details} \label{appD}

The setup presented in Sec. \ref{S-Examples} constitutes the simplest model of an ideal quantum absorption heat pump and refrigerator, usually considered to operate at 
steady-state conditions \cite{CorreaSqz,Palao,Kosloff}. We now focus in the heat pump configuration, but similar conclusions follows as well in the heat pump mode of operation. 
The cooling mechanism exploit the average heat flow entering from the reservoir at the hottest temperature, $\dot{Q}_2 > 0$, to continuously extract heat from the reservoir at the 
lowest temperature, $\dot{Q}_1 > 0$, while draining $\dot{Q}_3 < 0$ to the reservoir at the intermediate (inverse) temperature, $\beta_3$ (see Fig. \ref{F-Threelevel}).

The three average heat fluxes entering from the reservoirs associated to the imbalance in emission and absorption processes, $\dot{Q}_r = \tr[H_S \mathcal{L}_r(\rho_t)]$, read:
\begin{eqnarray}
 \dot{Q}_1 &=  \hbar \omega_1 \left( \Gamma_\uparrow^{(1)} p_g(t) - \Gamma_\downarrow^{(1)} p_{A}(t) \right),  \\
 \dot{Q}_2 &=  \hbar \omega_2 \left( \Gamma_\uparrow^{(2)} p_{A}(t) - \Gamma_\downarrow^{(2)} p_{B}(t) \right),  \\
 \dot{Q}_3 &=  \hbar \omega_3 \left( \Gamma_\uparrow^{(3)} p_{g}(t) - \Gamma_\downarrow^{(3)} p_{B}(t) \right), 
\end{eqnarray}
where $p_i(t)$  are the instantaneous populations of the machine 
energy levels $\ket{g}$, $\ket{e_A}$, $\ket{e_B}$, and $\sum_i p_i(t) = 1$. The first law of thermodynamics in the model follows from the master equation (\ref{Meq}):
\begin{equation}
 \dot{U} = \tr[H_S \dot{\rho}_S] = \dot{Q}_1 + \dot{Q}_2 + \dot{Q}_3, 
\end{equation}
which in steady state conditions read $\dot{Q}_1 + \dot{Q}_2 + \dot{Q}_3 = 0$. In such case, the heat fluxes become
\begin{eqnarray} \label{ss-fluxes1}
 \dot{Q}_1 &= \hbar \omega_1  \Gamma_\uparrow^{(1)} \pi_{g} \left( 1 - e^{-(\beta_1^\prime -\beta_1) \hbar \omega_1} \right), \\
 \dot{Q}_2 &= \hbar \omega_2  \Gamma_\uparrow^{(2)} \pi_{A} \left( 1 - e^{-(\beta_2^\prime -\beta_2) \hbar \omega_2} \right),  \\ \label{ss-fluxes3}
 \dot{Q}_3 &= \hbar \omega_3  \Gamma_\uparrow^{(3)} \pi_{g} \left( 1 - e^{-(\beta_3^\prime -\beta_3) \hbar \omega_3} \right),
 \end{eqnarray}
where we employed the detailed balance relations $ \Gamma_\uparrow^{(r)} = e^{\beta_r \hbar \omega_r} \Gamma_\downarrow^{(r)}$ and the definitions for the effective temperatures $\beta_r^\prime$, 
for $r=1,2,3$. Therefore, since the prefactors in all the three above expressions are always positive, the direction of the heat fluxes are determined by the sign of the respective thermodynamic 
force $X_r \equiv \beta_r^\prime - \beta_r$. Indeed near equilibrium when $X_r \ll 1$, we may expand to first order the exponentials in Eqs. (\ref{ss-fluxes1}-\ref{ss-fluxes3}) and recover the well known result 
of linear irreversible thermodynamics
\begin{equation}
\dot{Q}_r = \alpha_r X_r,
\end{equation}
where $\alpha_r$ is a positive constant, that is, fluxes are proportional to thermodynamic forces. In any case, Eqs. (\ref{ss-fluxes1}-\ref{ss-fluxes3}) show that heat flows from environment to a system transition, 
$\dot{Q}_r \geq 0$, if the latter is at an effective temperature lower than the former, $\beta_r^\prime \geq \beta_r$.

The steady state of the dynamics [Eq. (\ref{ss})] for the simpler case in which $\gamma_1 = \gamma_2 = \gamma_3 \equiv \gamma$ reads:
\begin{align}
\pi_g &= \frac{ e^{\beta_3 \hbar \omega_3}\left( 2 e^{\beta_1 \hbar \omega_1 + \beta_2 \hbar \omega_2} -1 \right) - e^{\beta_1 \hbar \omega_1 + \beta_2 \hbar \omega_2} }{Z_\pi},   \\
\pi_{A} &= \frac{ e^{\beta_2 \hbar \omega_2}\left( e^{\beta_1 \hbar \omega_1} - 2 \right) + e^{\beta_3 \hbar \omega_3}\left( 2 e^{\beta_2 \hbar \omega_2} - 1 \right) }{Z_\pi},   \\
\pi_{B} &= \frac{ e^{\beta_3 \hbar \omega_3} + e^{\beta_1 \hbar \omega_1 + \beta_2 \hbar \omega_2} - 2 }{Z_\pi},
\end{align}
where we defined $Z_\pi \equiv e^{\beta_2 \hbar \omega_2} \left(-2 + e^{\beta_1 \hbar \omega_1} \right)- 2   + ~ e^{\beta_3 \hbar \omega_3} \left(2 e^{\beta_2 \hbar \omega_2}\left((1 + e^{\beta_1 \hbar \omega_1}\right) - 1 \right)$.

At steady state conditions, the fridge or heat pump modes of operation can be obtained by properly tuning the energy level spacings. Inserting the steady state values in the expressions for the heat 
fluxes we obtain:
\begin{equation}\label{ssflows}
\dot{Q}_1^{\rm ss} = \gamma \hbar \omega_1 \Delta/Z_\pi \geq 0, ~~~~~~~  \dot{Q}_2^{\rm ss} = \gamma \hbar \omega_2 \Delta/Z_\pi \geq 0, 
\end{equation}
and $\dot{Q}_3^{\rm ss} = - (\dot{Q}_1^{\rm ss} + \dot{Q}_2^{\rm ss}) \leq 0$, where $Z_\pi \geq 0$ and the quantity $\Delta \equiv \left( e^{\beta_3 \hbar \omega_3} - e^{\beta_1 \hbar \omega_1 + \beta_2 \hbar \omega_2}\right) \geq0$. 
Therefore, for a fridge we need $\Delta \geq 0$. This is guaranteed when the following design condition is met: 
\begin{equation}\label{cooling-window}
 \omega_2 \geq \left( \frac{\beta_1 - \beta_3}{\beta_3 - \beta_2} \right) \omega_1.
\end{equation}
Notice also that when the above inequality is inverted, we obtain $\Delta \leq 0$, and the three heat flows invert its signs, hence generating the heat pump mode of operation.

\section{Transient negativity of the adiabatic entropy production rate} \label{appE}

In this appendix we provide further details on the dynamical evolution of thermodynamic quantities used in the description of the driving cavity model in Sec. \ref{S-cavity}
In particular we give explicit expressions for key quantities $\dot{X}_{\varphi}$, $\dot{W}$ and $\dot{Q}$, and discuss the adiabatic entropy production rate $\dot{S}_\mathrm{a}$, 
showing its transient negativity.

The explicit time evolution of the quantities $\dot{X}_{\varphi}$, $\dot{W}$ and $\dot{Q}$ can be obtained from the master equation (\ref{mastereq}). In order to do that, we first 
obtain the following equations for the evolution of the quantities $A \equiv a - \alpha$, and $A^\dagger A = a^\dagger a - |\alpha| (x_{\varphi} - |\alpha|)$. They read
\begin{align}
& \frac{d}{dt} \langle A \rangle_t = - \frac{\gamma_0}{2} \langle A \rangle_t,  \\
& \frac{d}{dt} \langle A^\dagger A \rangle_t = - \gamma_0 \left(\langle A^\dagger A \rangle_t - \langle A^\dagger A \rangle_\infty \right), 
\end{align}
where $\langle A^\dagger A \rangle_\infty = \tr[A^\dagger A \pi] = \tr[a^\dagger a \frac{e^{-\beta H_0}}{Z_0}] = n^{\mathrm{th}}$. 
Consequently we obtain as a result 
\begin{align}\label{solsmot1}
& \langle A \rangle_t = \langle A \rangle_{0} e^{-\gamma_0 t /2}, \\ \label{solsmot2}
& \langle A^\dagger A \rangle_t = \langle A^\dagger A \rangle_{0} e^{-\gamma_0 t} + n^{\mathrm{th}} (1 - e^{-\gamma_0 t}).
\end{align} 

The transient evolution of the field quadrature $\braket{x_{\varphi}}_t \equiv \tr[x_{\varphi} \rho_t]$ is then easily obtained 
from the above equations
\begin{equation} \label{Xevolution}
\dot{X}_{\varphi} = - \frac{\gamma_0}{2}(\langle x_{\varphi}\rangle_t  - \braket{x_{\varphi}}_{\infty}).
\end{equation}
This means that $\langle x_{\varphi} \rangle_t$ exponentially converges to its steady state value $\braket{x_\alpha}_{\infty} = 2 |\alpha|$. Therefore $\dot{X}_{\varphi}$ will be either positive 
or negative during the evolution depending on the displacement of the initial state. If $\braket{x_{\varphi}}_{0} \leq \braket{x_{\varphi}}_{\infty}$ then $\dot{X}_{\varphi} \geq 0~ \forall t$, 
and the system state increases its coherence in the energy basis, while if $\braket{x_{\varphi}}_{0} \geq \braket{x_{\varphi}}_{\infty}$, we have $\dot{X}_{\varphi} \leq 0~ \forall t$ and the coherence 
decreases. From Eq. (\ref{Xevolution}) we have
\begin{equation} \label{Xsol}
\langle x_{\varphi} \rangle_t = \langle x _{\varphi}\rangle_{0} e^{-\gamma_0 t/2} + \langle x _{\varphi} \rangle_{\infty} (1 - e^{-\gamma_0 t/2}).
\end{equation}
Furthermore we can now calculate the transient input power as
\begin{align} \label{Wevolution}
 \dot{W} & = \hbar \omega \tr[(\epsilon a^\dagger + \epsilon^\ast a) \rho_t] \nonumber \\
 & = \epsilon \langle a^\dagger \rangle_t + \epsilon^\ast \langle a \rangle_t \nonumber \\
 & = |\epsilon|\langle x_{\varphi} \rangle_{0} + \dot{W}_\mathrm{ss} (1 - e^{-\gamma_0 t/2}),
\end{align}
where we used $\langle a \rangle_t = \langle A \rangle_t + \alpha$ together with Eq. (\ref{solsmot1}) and we recall that $\dot{W}_\mathrm{ss} = \hbar \omega |\alpha|^2$. Analogously, having Eq. \eqref{solsmot2}, the heat flow follows from
\begin{align} \label{Qevolution}
 \dot{Q} &= \tr[H_0 \mathcal{L}(\rho_t)] = - \gamma_0(\langle a^\dagger a \rangle_t - n^{\mathrm{th}}) \\ 
 &= - \gamma_0 \hbar \omega \big( |\alpha|^2 (1- e^{-\gamma_0 t /2})^2 \nonumber \\ 
 & +  |\alpha| \langle x_{\varphi} \rangle_{0} (1- e^{- \gamma_0 t /2}) e^{-\gamma_0 t/2} + (\langle a^\dagger a \rangle_{0} - n^{\mathrm{th}}) \big), \nonumber
\end{align}
where in the last equality we also used Eq. (\ref{Xsol}). Notice that for the initial state $\rho_0 = \exp(-\beta H_0)/Z_0$ we have $\langle a \rangle_0 = 0$ and $\langle a^\dagger a \rangle_0 = n^{\mathrm{th}}$, and then, using Eqs. (\ref{Wevolution}) 
and (\ref{Qevolution}), we obtain for this case 
\begin{equation}
 \dot{Q} = - \dot{W}(1 - e^{- \gamma_0 t /2}), ~~~~ \dot{U} = \dot{W} + \dot{Q} = \dot{W} e^{- \gamma_0 t /2}.
\end{equation}

Finally, the adiabatic entropy production rate has been defined in Eq. (\ref{nonadeprate}):
\begin{equation}\label{adrateapp}
\dot{S}_\mathrm{a} = \dot{S}_\mathrm{i} - \dot{S}_\mathrm{na} = \beta(\dot{W} - \hbar \omega |\alpha| \dot{X}_{\varphi}).
\end{equation}
We can obtain an explicit expression for its evolution by noticing that the following equality holds
\begin{equation} 
\dot{W} + \hbar \omega |\alpha| \dot{X}_{\varphi} = \dot{W}_\mathrm{ss}.
\end{equation}
Introducing this relation into Eq.~(\ref{adrateapp}) we obtain:
\begin{equation} \label{adratecal}
 \dot{S}_{\mathrm a} = \beta \dot{W}_{\mathrm{ss}} + \beta \hbar \omega |\alpha| \gamma_0 \left(\langle x_{\varphi} \rangle - \braket{x_{\varphi}}_\infty \right).
\end{equation}
Notice now that Eq. (\ref{adratecal}) is negative for any initial transient for which $\hbar \omega |\alpha| \dot X_{\varphi} < \hbar \omega |\alpha| \braket{x_{\varphi}}_\infty 
+ \dot{W}_{\mathrm{ss}}/\gamma_0$. In particular, if the dynamics starts in any state diagonal in the $H_0$ basis, this happens for $t < t_\mathrm{n} \equiv 2\ln(2)/ \gamma_0$ as shown in  
Fig. \ref{F-cavityentropy} of Sec. \ref{S-cavity}.

\bibliography{qft}

\begin{thebibliography}{89}%
\makeatletter
\providecommand \@ifxundefined [1]{%
 \@ifx{#1\undefined}
}%
\providecommand \@ifnum [1]{%
 \ifnum #1\expandafter \@firstoftwo
 \else \expandafter \@secondoftwo
 \fi
}%
\providecommand \@ifx [1]{%
 \ifx #1\expandafter \@firstoftwo
 \else \expandafter \@secondoftwo
 \fi
}%
\providecommand \natexlab [1]{#1}%
\providecommand \enquote  [1]{``#1''}%
\providecommand \bibnamefont  [1]{#1}%
\providecommand \bibfnamefont [1]{#1}%
\providecommand \citenamefont [1]{#1}%
\providecommand \href@noop [0]{\@secondoftwo}%
\providecommand \href [0]{\begingroup \@sanitize@url \@href}%
\providecommand \@href[1]{\@@startlink{#1}\@@href}%
\providecommand \@@href[1]{\endgroup#1\@@endlink}%
\providecommand \@sanitize@url [0]{\catcode `\\12\catcode `\$12\catcode
  `\&12\catcode `\#12\catcode `\^12\catcode `\_12\catcode `\%12\relax}%
\providecommand \@@startlink[1]{}%
\providecommand \@@endlink[0]{}%
\providecommand \url  [0]{\begingroup\@sanitize@url \@url }%
\providecommand \@url [1]{\endgroup\@href {#1}{\urlprefix }}%
\providecommand \urlprefix  [0]{URL }%
\providecommand \Eprint [0]{\href }%
\providecommand \doibase [0]{http://dx.doi.org/}%
\providecommand \selectlanguage [0]{\@gobble}%
\providecommand \bibinfo  [0]{\@secondoftwo}%
\providecommand \bibfield  [0]{\@secondoftwo}%
\providecommand \translation [1]{[#1]}%
\providecommand \BibitemOpen [0]{}%
\providecommand \bibitemStop [0]{}%
\providecommand \bibitemNoStop [0]{.\EOS\space}%
\providecommand \EOS [0]{\spacefactor3000\relax}%
\providecommand \BibitemShut  [1]{\csname bibitem#1\endcsname}%
\let\auto@bib@innerbib\@empty
\bibitem [{\citenamefont {Seifert}(2005)}]{SeifertPRL}%
  \BibitemOpen
  \bibfield  {author} {\bibinfo {author} {\bibfnamefont {U.}~\bibnamefont
  {Seifert}},\ }\bibfield  {title} {\enquote {\bibinfo {title} {{Entropy
  Production along a Stochastic Trajectory and an Integral Fluctuation
  Theorem}},}\ }\href {\doibase 10.1103/PhysRevLett.95.040602} {\bibfield
  {journal} {\bibinfo  {journal} {Phys. Rev. Lett.}\ }\textbf {\bibinfo
  {volume} {95}},\ \bibinfo {pages} {040602} (\bibinfo {year}
  {2005})}\BibitemShut {NoStop}%
\bibitem [{\citenamefont {Seifert}(2012)}]{SeifertREV}%
  \BibitemOpen
  \bibfield  {author} {\bibinfo {author} {\bibfnamefont {U.}~\bibnamefont
  {Seifert}},\ }\bibfield  {title} {\enquote {\bibinfo {title} {{Stochastic
  thermodynamics, fluctuation theorems and molecular machines}},}\ }\href
  {http://iopscience.iop.org/article/10.1088/0034-4885/75/12/126001/meta}
  {\bibfield  {journal} {\bibinfo  {journal} {Rep. Prog. Phys.}\ }\textbf
  {\bibinfo {volume} {75}},\ \bibinfo {pages} {126001} (\bibinfo {year}
  {2012})}\BibitemShut {NoStop}%
\bibitem [{\citenamefont {Jarzynski}(2011)}]{JarzynskiREV}%
  \BibitemOpen
  \bibfield  {author} {\bibinfo {author} {\bibfnamefont {C.}~\bibnamefont
  {Jarzynski}},\ }\bibfield  {title} {\enquote {\bibinfo {title} {{Equalities
  and Inequalities: Irreversibility and the Second Law of Thermodynamics at the
  Nanoscale}},}\ }\href
  {http://www.annualreviews.org/doi/abs/10.1146/annurev-conmatphys-062910-140506?journalCode=conmatphys}
  {\bibfield  {journal} {\bibinfo  {journal} {Annu. Rev. Condens. Matter
  Phys.}\ }\textbf {\bibinfo {volume} {2}},\ \bibinfo {pages} {329--351}
  (\bibinfo {year} {2011})}\BibitemShut {NoStop}%
\bibitem [{\citenamefont {Breuer}\ and\ \citenamefont
  {Petruccione}(2002)}]{BreuerBook}%
  \BibitemOpen
  \bibfield  {author} {\bibinfo {author} {\bibfnamefont {H.-P.}\ \bibnamefont
  {Breuer}}\ and\ \bibinfo {author} {\bibfnamefont {F.}~\bibnamefont
  {Petruccione}},\ }\href@noop {} {\emph {\bibinfo {title} {{The theory of open
  quantum systems}}}}\ (\bibinfo  {publisher} {Oxford University Press},\
  \bibinfo {address} {Oxford},\ \bibinfo {year} {2002})\BibitemShut {NoStop}%
\bibitem [{\citenamefont {Rivas}\ and\ \citenamefont {Huelga}(2012)}]{Rivas}%
  \BibitemOpen
  \bibfield  {author} {\bibinfo {author} {\bibfnamefont {A.}~\bibnamefont
  {Rivas}}\ and\ \bibinfo {author} {\bibfnamefont {S.~F.}\ \bibnamefont
  {Huelga}},\ }\href@noop {} {\emph {\bibinfo {title} {{Open Quantum Systems :
  An Introduction}}}}\ (\bibinfo  {publisher} {Springer},\ \bibinfo {address}
  {Berlin Heidelberg},\ \bibinfo {year} {2012})\BibitemShut {NoStop}%
\bibitem [{\citenamefont {Campisi}\ \emph {et~al.}(2011)\citenamefont
  {Campisi}, \citenamefont {H{\"a}nggi},\ and\ \citenamefont
  {Talkner}}]{CampisiREV}%
  \BibitemOpen
  \bibfield  {author} {\bibinfo {author} {\bibfnamefont {M.}~\bibnamefont
  {Campisi}}, \bibinfo {author} {\bibfnamefont {P.}~\bibnamefont {H{\"a}nggi}},
  \ and\ \bibinfo {author} {\bibfnamefont {P.}~\bibnamefont {Talkner}},\
  }\bibfield  {title} {\enquote {\bibinfo {title} {{Colloquium: Quantum
  fluctuation relations: Foundations and applications}},}\ }\href {\doibase
  10.1103/RevModPhys.83.771} {\bibfield  {journal} {\bibinfo  {journal} {Rev.
  Mod. Phys.}\ }\textbf {\bibinfo {volume} {83}},\ \bibinfo {pages} {771--791}
  (\bibinfo {year} {2011})}\BibitemShut {NoStop}%
\bibitem [{\citenamefont {Esposito}\ \emph {et~al.}(2009)\citenamefont
  {Esposito}, \citenamefont {Harbola},\ and\ \citenamefont
  {Mukamel}}]{EspositoREV}%
  \BibitemOpen
  \bibfield  {author} {\bibinfo {author} {\bibfnamefont {M.}~\bibnamefont
  {Esposito}}, \bibinfo {author} {\bibfnamefont {U.}~\bibnamefont {Harbola}}, \
  and\ \bibinfo {author} {\bibfnamefont {S.}~\bibnamefont {Mukamel}},\
  }\bibfield  {title} {\enquote {\bibinfo {title} {{Nonequilibrium
  fluctuations, fluctuation theorems, and counting statistics in quantum
  systems}},}\ }\href
  {https://arxiv.org/ct?url=http%3A%2F%2Fdx.doi.org%2F10%252E1103%2FRevModPhys%252E81%252E1665&v=9b321d90}
  {\bibfield  {journal} {\bibinfo  {journal} {Rev. Mod. Phys.}\ }\textbf
  {\bibinfo {volume} {81}},\ \bibinfo {pages} {1665--1702} (\bibinfo {year}
  {2009})}\BibitemShut {NoStop}%
\bibitem [{\citenamefont {Deffner}\ and\ \citenamefont {Lutz}(2011)}]{LutzEP}%
  \BibitemOpen
  \bibfield  {author} {\bibinfo {author} {\bibfnamefont {S.}~\bibnamefont
  {Deffner}}\ and\ \bibinfo {author} {\bibfnamefont {E.}~\bibnamefont {Lutz}},\
  }\bibfield  {title} {\enquote {\bibinfo {title} {{Nonequilibrium Entropy
  Production for Open Quantum Systems}},}\ }\href {\doibase
  10.1103/PhysRevLett.107.140404} {\bibfield  {journal} {\bibinfo  {journal}
  {Phys. Rev. Lett.}\ }\textbf {\bibinfo {volume} {107}},\ \bibinfo {pages}
  {140404} (\bibinfo {year} {2011})}\BibitemShut {NoStop}%
\bibitem [{\citenamefont {Gaspard}(2013)}]{Gaspard2013}%
  \BibitemOpen
  \bibfield  {author} {\bibinfo {author} {\bibfnamefont {P.}~\bibnamefont
  {Gaspard}},\ }\bibfield  {title} {\enquote {\bibinfo {title} {{Multivariate
  fluctuation relations for currents}},}\ }\href {\doibase
  10.1088/1367-2630/15/11/115014} {\bibfield  {journal} {\bibinfo  {journal}
  {New J. Phys.}\ }\textbf {\bibinfo {volume} {15}},\ \bibinfo {pages} {115014}
  (\bibinfo {year} {2013})}\BibitemShut {NoStop}%
\bibitem [{\citenamefont {Watanabe}\ \emph {et~al.}(2014)\citenamefont
  {Watanabe}, \citenamefont {Venkatesh}, \citenamefont {Talkner}, \citenamefont
  {Campisi},\ and\ \citenamefont {H{\"a}nggi}}]{Watanabe:2014fh}%
  \BibitemOpen
  \bibfield  {author} {\bibinfo {author} {\bibfnamefont {G.}~\bibnamefont
  {Watanabe}}, \bibinfo {author} {\bibfnamefont {B.~P.}\ \bibnamefont
  {Venkatesh}}, \bibinfo {author} {\bibfnamefont {P.}~\bibnamefont {Talkner}},
  \bibinfo {author} {\bibfnamefont {M.}~\bibnamefont {Campisi}}, \ and\
  \bibinfo {author} {\bibfnamefont {P.}~\bibnamefont {H{\"a}nggi}},\ }\bibfield
   {title} {\enquote {\bibinfo {title} {{Quantum fluctuation theorems and
  generalized measurements during the force protocol}},}\ }\href
  {https://doi.org/10.1103/PhysRevE.89.032114} {\bibfield  {journal} {\bibinfo
  {journal} {Phys. Rev. E}\ }\textbf {\bibinfo {volume} {89}},\ \bibinfo
  {pages} {032114} (\bibinfo {year} {2014})}\BibitemShut {NoStop}%
\bibitem [{\citenamefont {Cuetara}\ \emph {et~al.}(2014)\citenamefont
  {Cuetara}, \citenamefont {Esposito},\ and\ \citenamefont
  {Imparato}}]{Imparato}%
  \BibitemOpen
  \bibfield  {author} {\bibinfo {author} {\bibfnamefont {G.~B.}\ \bibnamefont
  {Cuetara}}, \bibinfo {author} {\bibfnamefont {M.}~\bibnamefont {Esposito}}, \
  and\ \bibinfo {author} {\bibfnamefont {A.}~\bibnamefont {Imparato}},\
  }\bibfield  {title} {\enquote {\bibinfo {title} {{Exact fluctuation theorem
  without ensemble quantities}},}\ }\href {\doibase 10.1103/PhysRevE.89.052119}
  {\bibfield  {journal} {\bibinfo  {journal} {Phys. Rev. E}\ }\textbf {\bibinfo
  {volume} {89}},\ \bibinfo {pages} {052119} (\bibinfo {year}
  {2014})}\BibitemShut {NoStop}%
\bibitem [{\citenamefont {Dorner}\ \emph {et~al.}(2013)\citenamefont {Dorner},
  \citenamefont {Clark}, \citenamefont {Heaney}, \citenamefont {Fazio},
  \citenamefont {Goold},\ and\ \citenamefont {Vedral}}]{Dorner:2013}%
  \BibitemOpen
  \bibfield  {author} {\bibinfo {author} {\bibfnamefont {R.}~\bibnamefont
  {Dorner}}, \bibinfo {author} {\bibfnamefont {S.~R.}\ \bibnamefont {Clark}},
  \bibinfo {author} {\bibfnamefont {L.}~\bibnamefont {Heaney}}, \bibinfo
  {author} {\bibfnamefont {R.}~\bibnamefont {Fazio}}, \bibinfo {author}
  {\bibfnamefont {J.}~\bibnamefont {Goold}}, \ and\ \bibinfo {author}
  {\bibfnamefont {V.}~\bibnamefont {Vedral}},\ }\bibfield  {title} {\enquote
  {\bibinfo {title} {{Extracting Quantum Work Statistics and Fluctuation
  Theorems by Single-Qubit Interferometry}},}\ }\href {\doibase
  10.1103/PhysRevLett.110.230601} {\bibfield  {journal} {\bibinfo  {journal}
  {Phys. Rev. Lett.}\ }\textbf {\bibinfo {volume} {110}},\ \bibinfo {pages}
  {230601} (\bibinfo {year} {2013})}\BibitemShut {NoStop}%
\bibitem [{\citenamefont {Mazzola}\ \emph {et~al.}(2013)\citenamefont
  {Mazzola}, \citenamefont {{De Chiara}},\ and\ \citenamefont
  {Paternostro}}]{Mazzola:2013}%
  \BibitemOpen
  \bibfield  {author} {\bibinfo {author} {\bibfnamefont {L.}~\bibnamefont
  {Mazzola}}, \bibinfo {author} {\bibfnamefont {G.}~\bibnamefont {{De
  Chiara}}}, \ and\ \bibinfo {author} {\bibfnamefont {M.}~\bibnamefont
  {Paternostro}},\ }\bibfield  {title} {\enquote {\bibinfo {title} {{Measuring
  the Characteristic Function of the Work Distribution}},}\ }\href {\doibase
  10.1103/PhysRevLett.110.230602} {\bibfield  {journal} {\bibinfo  {journal}
  {Phys. Rev. Lett.}\ }\textbf {\bibinfo {volume} {110}},\ \bibinfo {pages}
  {230602} (\bibinfo {year} {2013})}\BibitemShut {NoStop}%
\bibitem [{\citenamefont {Campisi}\ \emph {et~al.}(2013)\citenamefont
  {Campisi}, \citenamefont {Blattmann}, \citenamefont {Kohler}, \citenamefont
  {Zueco},\ and\ \citenamefont {H{\"a}nggi}}]{Campisi:2013}%
  \BibitemOpen
  \bibfield  {author} {\bibinfo {author} {\bibfnamefont {M.}~\bibnamefont
  {Campisi}}, \bibinfo {author} {\bibfnamefont {R.}~\bibnamefont {Blattmann}},
  \bibinfo {author} {\bibfnamefont {S.}~\bibnamefont {Kohler}}, \bibinfo
  {author} {\bibfnamefont {D.}~\bibnamefont {Zueco}}, \ and\ \bibinfo {author}
  {\bibfnamefont {P.}~\bibnamefont {H{\"a}nggi}},\ }\bibfield  {title}
  {\enquote {\bibinfo {title} {{Employing circuit QED to measure
  non-equilibrium work fluctuations}},}\ }\href
  {http://iopscience.iop.org/article/10.1088/1367-2630/15/10/105028/meta}
  {\bibfield  {journal} {\bibinfo  {journal} {New J. Phys.}\ }\textbf {\bibinfo
  {volume} {15}},\ \bibinfo {pages} {105028} (\bibinfo {year}
  {2013})}\BibitemShut {NoStop}%
\bibitem [{\citenamefont {Goold}\ \emph {et~al.}(2014)\citenamefont {Goold},
  \citenamefont {Poschinger},\ and\ \citenamefont {Modi}}]{Goold:2013}%
  \BibitemOpen
  \bibfield  {author} {\bibinfo {author} {\bibfnamefont {J.}~\bibnamefont
  {Goold}}, \bibinfo {author} {\bibfnamefont {U.}~\bibnamefont {Poschinger}}, \
  and\ \bibinfo {author} {\bibfnamefont {K.}~\bibnamefont {Modi}},\ }\bibfield
  {title} {\enquote {\bibinfo {title} {{Measuring the heat exchange of a
  quantum process}},}\ }\href {\doibase 10.1103/PhysRevE.90.020101} {\bibfield
  {journal} {\bibinfo  {journal} {Phys. Rev. E}\ }\textbf {\bibinfo {volume}
  {90}},\ \bibinfo {pages} {020101(R)} (\bibinfo {year} {2014})}\BibitemShut
  {NoStop}%
\bibitem [{\citenamefont {Roncaglia}\ \emph {et~al.}(2014)\citenamefont
  {Roncaglia}, \citenamefont {Cerisola},\ and\ \citenamefont
  {Paz}}]{Roncaglia:2014}%
  \BibitemOpen
  \bibfield  {author} {\bibinfo {author} {\bibfnamefont {A.~J.}\ \bibnamefont
  {Roncaglia}}, \bibinfo {author} {\bibfnamefont {F.}~\bibnamefont {Cerisola}},
  \ and\ \bibinfo {author} {\bibfnamefont {J.~P.}\ \bibnamefont {Paz}},\
  }\bibfield  {title} {\enquote {\bibinfo {title} {{Work Measurement as a
  Generalized Quantum Measurement}},}\ }\href
  {https://link.aps.org/doi/10.1103/PhysRevLett.113.250601} {\bibfield
  {journal} {\bibinfo  {journal} {Phys. Rev. Lett.}\ }\textbf {\bibinfo
  {volume} {113}},\ \bibinfo {pages} {250601} (\bibinfo {year}
  {2014})}\BibitemShut {NoStop}%
\bibitem [{\citenamefont {{De Chiara}}\ \emph {et~al.}(2015)\citenamefont {{De
  Chiara}}, \citenamefont {Roncaglia},\ and\ \citenamefont
  {Paz}}]{Chiara:2015}%
  \BibitemOpen
  \bibfield  {author} {\bibinfo {author} {\bibfnamefont {G.}~\bibnamefont {{De
  Chiara}}}, \bibinfo {author} {\bibfnamefont {A.~J.}\ \bibnamefont
  {Roncaglia}}, \ and\ \bibinfo {author} {\bibfnamefont {J.~P.}\ \bibnamefont
  {Paz}},\ }\bibfield  {title} {\enquote {\bibinfo {title} {{Measuring work and
  heat in ultracold quantum gases}},}\ }\href {\doibase
  10.1088/1367-2630/17/3/035004} {\bibfield  {journal} {\bibinfo  {journal}
  {New J. Phys.}\ }\textbf {\bibinfo {volume} {17}},\ \bibinfo {pages} {035004}
  (\bibinfo {year} {2015})}\BibitemShut {NoStop}%
\bibitem [{\citenamefont {Batalh{\~a}o}\ \emph {et~al.}(2014)\citenamefont
  {Batalh{\~a}o}, \citenamefont {Souza}, \citenamefont {Mazzola}, \citenamefont
  {Auccaise}, \citenamefont {Sarthour}, \citenamefont {Oliveira}, \citenamefont
  {Goold}, \citenamefont {{De Chiara}}, \citenamefont {Paternostro},\ and\
  \citenamefont {Serra}}]{Batalhao:2014ta}%
  \BibitemOpen
  \bibfield  {author} {\bibinfo {author} {\bibfnamefont {T.~B.}\ \bibnamefont
  {Batalh{\~a}o}}, \bibinfo {author} {\bibfnamefont {A.~M.}\ \bibnamefont
  {Souza}}, \bibinfo {author} {\bibfnamefont {L.}~\bibnamefont {Mazzola}},
  \bibinfo {author} {\bibfnamefont {R.}~\bibnamefont {Auccaise}}, \bibinfo
  {author} {\bibfnamefont {R.~S.}\ \bibnamefont {Sarthour}}, \bibinfo {author}
  {\bibfnamefont {I.~S.}\ \bibnamefont {Oliveira}}, \bibinfo {author}
  {\bibfnamefont {J.}~\bibnamefont {Goold}}, \bibinfo {author} {\bibfnamefont
  {G.}~\bibnamefont {{De Chiara}}}, \bibinfo {author} {\bibfnamefont
  {M.}~\bibnamefont {Paternostro}}, \ and\ \bibinfo {author} {\bibfnamefont
  {R.~M.}\ \bibnamefont {Serra}},\ }\bibfield  {title} {\enquote {\bibinfo
  {title} {{Experimental Reconstruction of Work Distribution and Study of
  Fluctuation Relations in a Closed Quantum System}},}\ }\href {\doibase
  10.1103/PhysRevLett.113.140601} {\bibfield  {journal} {\bibinfo  {journal}
  {Phys. Rev. Lett.}\ }\textbf {\bibinfo {volume} {113}},\ \bibinfo {pages}
  {140601} (\bibinfo {year} {2014})}\BibitemShut {NoStop}%
\bibitem [{\citenamefont {An}\ \emph {et~al.}(2015)\citenamefont {An},
  \citenamefont {Zhang}, \citenamefont {Um}, \citenamefont {Lv}, \citenamefont
  {Lu}, \citenamefont {Zhang}, \citenamefont {Yin}, \citenamefont {Quan},\ and\
  \citenamefont {Kim}}]{An:2015}%
  \BibitemOpen
  \bibfield  {author} {\bibinfo {author} {\bibfnamefont {S.}~\bibnamefont
  {An}}, \bibinfo {author} {\bibfnamefont {J.-N.}\ \bibnamefont {Zhang}},
  \bibinfo {author} {\bibfnamefont {M.}~\bibnamefont {Um}}, \bibinfo {author}
  {\bibfnamefont {D.}~\bibnamefont {Lv}}, \bibinfo {author} {\bibfnamefont
  {Y.}~\bibnamefont {Lu}}, \bibinfo {author} {\bibfnamefont {J.}~\bibnamefont
  {Zhang}}, \bibinfo {author} {\bibfnamefont {Z.-Q.}\ \bibnamefont {Yin}},
  \bibinfo {author} {\bibfnamefont {H.~T.}\ \bibnamefont {Quan}}, \ and\
  \bibinfo {author} {\bibfnamefont {K.}~\bibnamefont {Kim}},\ }\bibfield
  {title} {\enquote {\bibinfo {title} {{Experimental test of the quantum
  Jarzynski equality with a trapped-ion system}},}\ }\href
  {https://www.nature.com/nphys/journal/v11/n2/full/nphys3197.html} {\bibfield
  {journal} {\bibinfo  {journal} {Nat. Phys.}\ }\textbf {\bibinfo {volume}
  {11}},\ \bibinfo {pages} {193--199} (\bibinfo {year} {2015})}\BibitemShut
  {NoStop}%
\bibitem [{\citenamefont {Hatano}\ and\ \citenamefont
  {Sasa}(2001)}]{HatanoSasa}%
  \BibitemOpen
  \bibfield  {author} {\bibinfo {author} {\bibfnamefont {T.}~\bibnamefont
  {Hatano}}\ and\ \bibinfo {author} {\bibfnamefont {S.-I.}\ \bibnamefont
  {Sasa}},\ }\bibfield  {title} {\enquote {\bibinfo {title} {{Steady-state
  thermodynamics of Langevin systems}},}\ }\href
  {https://doi.org/10.1103/PhysRevLett.86.3463} {\bibfield  {journal} {\bibinfo
   {journal} {Phys. Rev. Lett.}\ }\textbf {\bibinfo {volume} {86}},\ \bibinfo
  {pages} {3463} (\bibinfo {year} {2001})}\BibitemShut {NoStop}%
\bibitem [{\citenamefont {Esposito}\ and\ \citenamefont {{Van den
  Broeck}}(2010{\natexlab{a}})}]{EspositoFaces}%
  \BibitemOpen
  \bibfield  {author} {\bibinfo {author} {\bibfnamefont {M.}~\bibnamefont
  {Esposito}}\ and\ \bibinfo {author} {\bibfnamefont {C.}~\bibnamefont {{Van
  den Broeck}}},\ }\bibfield  {title} {\enquote {\bibinfo {title} {{Three
  detailed fluctuation theorems}},}\ }\href
  {https://link.aps.org/doi/10.1103/PhysRevLett.104.090601} {\bibfield
  {journal} {\bibinfo  {journal} {Phys. Rev. Lett.}\ }\textbf {\bibinfo
  {volume} {104}},\ \bibinfo {pages} {090601} (\bibinfo {year}
  {2010}{\natexlab{a}})}\BibitemShut {NoStop}%
\bibitem [{\citenamefont {Esposito}\ and\ \citenamefont {{Van den
  Broeck}}(2010{\natexlab{b}})}]{EspositoFacesI}%
  \BibitemOpen
  \bibfield  {author} {\bibinfo {author} {\bibfnamefont {M.}~\bibnamefont
  {Esposito}}\ and\ \bibinfo {author} {\bibfnamefont {C.}~\bibnamefont {{Van
  den Broeck}}},\ }\bibfield  {title} {\enquote {\bibinfo {title} {{Three faces
  of the second law. I. Master equation formulation}},}\ }\href {\doibase
  10.1103/PhysRevE.82.011143} {\bibfield  {journal} {\bibinfo  {journal} {Phys.
  Rev. E}\ }\textbf {\bibinfo {volume} {82}},\ \bibinfo {pages} {011143}
  (\bibinfo {year} {2010}{\natexlab{b}})}\BibitemShut {NoStop}%
\bibitem [{\citenamefont {{Van den Broeck}}\ and\ \citenamefont
  {Esposito}(2010)}]{EspositoFacesII}%
  \BibitemOpen
  \bibfield  {author} {\bibinfo {author} {\bibfnamefont {C.}~\bibnamefont {{Van
  den Broeck}}}\ and\ \bibinfo {author} {\bibfnamefont {M.}~\bibnamefont
  {Esposito}},\ }\bibfield  {title} {\enquote {\bibinfo {title} {{Three faces
  of the second law. II. Fokker-Planck formulation}},}\ }\href
  {http://dx.doi.org/10.1103/PhysRevE.82.011144} {\bibfield  {journal}
  {\bibinfo  {journal} {Phys. Rev. E}\ }\textbf {\bibinfo {volume} {82}},\
  \bibinfo {pages} {011144} (\bibinfo {year} {2010})}\BibitemShut {NoStop}%
\bibitem [{\citenamefont {Bisker}\ \emph {et~al.}(2017)\citenamefont {Bisker},
  \citenamefont {Polettini}, \citenamefont {Gingrich},\ and\ \citenamefont
  {Horowitz}}]{Bisker}%
  \BibitemOpen
  \bibfield  {author} {\bibinfo {author} {\bibfnamefont {G.}~\bibnamefont
  {Bisker}}, \bibinfo {author} {\bibfnamefont {M.}~\bibnamefont {Polettini}},
  \bibinfo {author} {\bibfnamefont {T.~R.}\ \bibnamefont {Gingrich}}, \ and\
  \bibinfo {author} {\bibfnamefont {J.~M.}\ \bibnamefont {Horowitz}},\
  }\bibfield  {title} {\enquote {\bibinfo {title} {{Hierarchical bounds on
  entropy production inferred from partial information}},}\ }\href
  {http://stacks.iop.org/1742-5468/2017/i=9/a=093210} {\bibfield  {journal}
  {\bibinfo  {journal} {J. Stat. Mech.: Theor. and Exp.}\ }\textbf {\bibinfo
  {volume} {2017}},\ \bibinfo {pages} {093210} (\bibinfo {year}
  {2017})}\BibitemShut {NoStop}%
\bibitem [{\citenamefont {Vedral}(2012)}]{Vedral}%
  \BibitemOpen
  \bibfield  {author} {\bibinfo {author} {\bibfnamefont {V.}~\bibnamefont
  {Vedral}},\ }\bibfield  {title} {\enquote {\bibinfo {title} {{An
  information--theoretic equality implying the Jarzynski relation}},}\ }\href
  {\doibase 10.1088/1751-8113/45/27/272001} {\bibfield  {journal} {\bibinfo
  {journal} {J. Phys. A: Math. Theor.}\ }\textbf {\bibinfo {volume} {45}},\
  \bibinfo {pages} {272001} (\bibinfo {year} {2012})}\BibitemShut {NoStop}%
\bibitem [{\citenamefont {Kafri}\ and\ \citenamefont
  {Deffner}(2012)}]{Kafri2012}%
  \BibitemOpen
  \bibfield  {author} {\bibinfo {author} {\bibfnamefont {D.}~\bibnamefont
  {Kafri}}\ and\ \bibinfo {author} {\bibfnamefont {S.}~\bibnamefont
  {Deffner}},\ }\bibfield  {title} {\enquote {\bibinfo {title} {{Holevo's bound
  from a gernal quantum fluctuation theorem}},}\ }\href {\doibase
  10.1103/PhysRevA.86.044302} {\bibfield  {journal} {\bibinfo  {journal} {Phys.
  Rev. A}\ }\textbf {\bibinfo {volume} {86}},\ \bibinfo {pages} {044302}
  (\bibinfo {year} {2012})}\BibitemShut {NoStop}%
\bibitem [{\citenamefont {Chetritie}\ and\ \citenamefont
  {Mallick}(2012)}]{Chetrite2012}%
  \BibitemOpen
  \bibfield  {author} {\bibinfo {author} {\bibfnamefont {R.}~\bibnamefont
  {Chetritie}}\ and\ \bibinfo {author} {\bibfnamefont {K.}~\bibnamefont
  {Mallick}},\ }\bibfield  {title} {\enquote {\bibinfo {title} {{Quantum
  Fluctuation Relations for the Lindblad Master Equation}},}\ }\href {\doibase
  10.1007/s10955-012-0557-z} {\bibfield  {journal} {\bibinfo  {journal} {J.
  Stat. Phys.}\ }\textbf {\bibinfo {volume} {148}},\ \bibinfo {pages}
  {480--501} (\bibinfo {year} {2012})}\BibitemShut {NoStop}%
\bibitem [{\citenamefont {Albash}\ \emph {et~al.}(2013)\citenamefont {Albash},
  \citenamefont {Lidar}, \citenamefont {Marvian},\ and\ \citenamefont
  {Zanardi}}]{Albash:2013fq}%
  \BibitemOpen
  \bibfield  {author} {\bibinfo {author} {\bibfnamefont {T.}~\bibnamefont
  {Albash}}, \bibinfo {author} {\bibfnamefont {D.~A.}\ \bibnamefont {Lidar}},
  \bibinfo {author} {\bibfnamefont {M.}~\bibnamefont {Marvian}}, \ and\
  \bibinfo {author} {\bibfnamefont {P.}~\bibnamefont {Zanardi}},\ }\bibfield
  {title} {\enquote {\bibinfo {title} {{Fluctuation theorems for quantum
  processes}},}\ }\href {\doibase 10.1103/PhysRevE.88.032146} {\bibfield
  {journal} {\bibinfo  {journal} {Phys. Rev. E}\ }\textbf {\bibinfo {volume}
  {88}},\ \bibinfo {pages} {032146} (\bibinfo {year} {2013})}\BibitemShut
  {NoStop}%
\bibitem [{\citenamefont {Rastegin}\ and\ \citenamefont
  {{\.Z}yczkowski}(2014)}]{Rastegin:2014hc}%
  \BibitemOpen
  \bibfield  {author} {\bibinfo {author} {\bibfnamefont {A.~E.}\ \bibnamefont
  {Rastegin}}\ and\ \bibinfo {author} {\bibfnamefont {K.}~\bibnamefont
  {{\.Z}yczkowski}},\ }\bibfield  {title} {\enquote {\bibinfo {title}
  {{Jarzynski equality for quantum stochastic maps}},}\ }\href
  {https://doi.org/10.1103/PhysRevE.89.012127} {\bibfield  {journal} {\bibinfo
  {journal} {Phys. Rev. E}\ }\textbf {\bibinfo {volume} {89}},\ \bibinfo
  {pages} {012127} (\bibinfo {year} {2014})}\BibitemShut {NoStop}%
\bibitem [{\citenamefont {Deffner}(2013)}]{DeffnerPS}%
  \BibitemOpen
  \bibfield  {author} {\bibinfo {author} {\bibfnamefont {S.}~\bibnamefont
  {Deffner}},\ }\bibfield  {title} {\enquote {\bibinfo {title} {{Quantum
  entropy production in phase space}},}\ }\href {\doibase
  10.1209/0295-5075/103/30001/meta} {\bibfield  {journal} {\bibinfo  {journal}
  {Europhys. Lett.}\ }\textbf {\bibinfo {volume} {103}},\ \bibinfo {pages}
  {30001} (\bibinfo {year} {2013})}\BibitemShut {NoStop}%
\bibitem [{\citenamefont {Horowitz}\ and\ \citenamefont
  {Parrondo}(2013)}]{JordanParrondo}%
  \BibitemOpen
  \bibfield  {author} {\bibinfo {author} {\bibfnamefont {J.~M.}\ \bibnamefont
  {Horowitz}}\ and\ \bibinfo {author} {\bibfnamefont {J.~M.~R.}\ \bibnamefont
  {Parrondo}},\ }\bibfield  {title} {\enquote {\bibinfo {title} {{Entropy
  production along nonequilibrium quantum jump trajectories}},}\ }\href
  {\doibase 10.1088/1367-2630/15/8/085028} {\bibfield  {journal} {\bibinfo
  {journal} {New. J. Phys}\ }\textbf {\bibinfo {volume} {15}},\ \bibinfo
  {pages} {085028} (\bibinfo {year} {2013})}\BibitemShut {NoStop}%
\bibitem [{\citenamefont {Funo}\ \emph {et~al.}(2013)\citenamefont {Funo},
  \citenamefont {Watanabe},\ and\ \citenamefont {Ueda}}]{Funo2013}%
  \BibitemOpen
  \bibfield  {author} {\bibinfo {author} {\bibfnamefont {K.}~\bibnamefont
  {Funo}}, \bibinfo {author} {\bibfnamefont {Y.}~\bibnamefont {Watanabe}}, \
  and\ \bibinfo {author} {\bibfnamefont {M.}~\bibnamefont {Ueda}},\ }\bibfield
  {title} {\enquote {\bibinfo {title} {{Integral quantum fluctuation theorems
  under measurement and feedback control}},}\ }\href {\doibase
  10.1103/PhysRevE.88.052121} {\bibfield  {journal} {\bibinfo  {journal} {Phys.
  Rev. E}\ }\textbf {\bibinfo {volume} {88}},\ \bibinfo {pages} {052121}
  (\bibinfo {year} {2013})}\BibitemShut {NoStop}%
\bibitem [{\citenamefont {Manzano}\ \emph {et~al.}(2015)\citenamefont
  {Manzano}, \citenamefont {Horowitz},\ and\ \citenamefont {Parrondo}}]{MHP}%
  \BibitemOpen
  \bibfield  {author} {\bibinfo {author} {\bibfnamefont {G.}~\bibnamefont
  {Manzano}}, \bibinfo {author} {\bibfnamefont {J.~M.}\ \bibnamefont
  {Horowitz}}, \ and\ \bibinfo {author} {\bibfnamefont {J.~M.~R.}\ \bibnamefont
  {Parrondo}},\ }\bibfield  {title} {\enquote {\bibinfo {title}
  {{Nonequilibrium potential and fluctuation theorems for quantum maps}},}\
  }\href {http://dx.doi.org/10.1103/PhysRevE.92.032129} {\bibfield  {journal}
  {\bibinfo  {journal} {Phys. Rev. E}\ }\textbf {\bibinfo {volume} {92}},\
  \bibinfo {pages} {032129} (\bibinfo {year} {2015})}\BibitemShut {NoStop}%
\bibitem [{\citenamefont {Alhambra}\ \emph {et~al.}(2016)\citenamefont
  {Alhambra}, \citenamefont {Masanes}, \citenamefont {Oppenheim},\ and\
  \citenamefont {Perry}}]{Alhambra}%
  \BibitemOpen
  \bibfield  {author} {\bibinfo {author} {\bibfnamefont {{\'A}.~M.}\
  \bibnamefont {Alhambra}}, \bibinfo {author} {\bibfnamefont {L.}~\bibnamefont
  {Masanes}}, \bibinfo {author} {\bibfnamefont {J.}~\bibnamefont {Oppenheim}},
  \ and\ \bibinfo {author} {\bibfnamefont {C.}~\bibnamefont {Perry}},\
  }\bibfield  {title} {\enquote {\bibinfo {title} {{Fluctuating Work: From
  Quantum Thermodynamical Identities to a Second Law Equality}},}\ }\href
  {\doibase 10.1103/PhysRevX.6.041017} {\bibfield  {journal} {\bibinfo
  {journal} {Phys. Rev. X}\ }\textbf {\bibinfo {volume} {6}},\ \bibinfo {pages}
  {041017} (\bibinfo {year} {2016})}\BibitemShut {NoStop}%
\bibitem [{\citenamefont {Park}\ \emph {et~al.}(2017)\citenamefont {Park},
  \citenamefont {Kim},\ and\ \citenamefont {Vedral}}]{Park}%
  \BibitemOpen
  \bibfield  {author} {\bibinfo {author} {\bibfnamefont {J.~J.}\ \bibnamefont
  {Park}}, \bibinfo {author} {\bibfnamefont {S.~W.}\ \bibnamefont {Kim}}, \
  and\ \bibinfo {author} {\bibfnamefont {V.}~\bibnamefont {Vedral}},\
  }\bibfield  {title} {\enquote {\bibinfo {title} {{Fluctuation Theorem for
  Arbitrary Quantum Bipartite Systems}},}\ }\href
  {https://arxiv.org/abs/1705.01750} {\bibfield  {journal} {\bibinfo  {journal}
  {arXiv:1705.01750}\ } (\bibinfo {year} {2017})}\BibitemShut {NoStop}%
\bibitem [{\citenamefont {Pekola}\ \emph {et~al.}(2013)\citenamefont {Pekola},
  \citenamefont {Solinas}, \citenamefont {Shnirman},\ and\ \citenamefont
  {Averin}}]{PekolaCalorimeter}%
  \BibitemOpen
  \bibfield  {author} {\bibinfo {author} {\bibfnamefont {J.~P.}\ \bibnamefont
  {Pekola}}, \bibinfo {author} {\bibfnamefont {P.}~\bibnamefont {Solinas}},
  \bibinfo {author} {\bibfnamefont {A.}~\bibnamefont {Shnirman}}, \ and\
  \bibinfo {author} {\bibfnamefont {D.~V.}\ \bibnamefont {Averin}},\ }\bibfield
   {title} {\enquote {\bibinfo {title} {{Calorimetric measurement of work in a
  quantum system}},}\ }\href
  {http://iopscience.iop.org/article/10.1088/1367-2630/15/11/115006/meta}
  {\bibfield  {journal} {\bibinfo  {journal} {New J. Phys.}\ }\textbf {\bibinfo
  {volume} {15}},\ \bibinfo {pages} {115006} (\bibinfo {year}
  {2013})}\BibitemShut {NoStop}%
\bibitem [{\citenamefont {Gasparinetti}\ \emph {et~al.}(2015)\citenamefont
  {Gasparinetti}, \citenamefont {Viisanen}, \citenamefont {Saira},
  \citenamefont {Faivre}, \citenamefont {Arzeo}, \citenamefont {Meschke},\ and\
  \citenamefont {Pekola}}]{Gaspinaretti}%
  \BibitemOpen
  \bibfield  {author} {\bibinfo {author} {\bibfnamefont {S.}~\bibnamefont
  {Gasparinetti}}, \bibinfo {author} {\bibfnamefont {K.~L.}\ \bibnamefont
  {Viisanen}}, \bibinfo {author} {\bibfnamefont {O.-P.}\ \bibnamefont {Saira}},
  \bibinfo {author} {\bibfnamefont {T.}~\bibnamefont {Faivre}}, \bibinfo
  {author} {\bibfnamefont {M.}~\bibnamefont {Arzeo}}, \bibinfo {author}
  {\bibfnamefont {M.}~\bibnamefont {Meschke}}, \ and\ \bibinfo {author}
  {\bibfnamefont {J.~P.}\ \bibnamefont {Pekola}},\ }\bibfield  {title}
  {\enquote {\bibinfo {title} {{Fast Electron Thermometry for Ultrasensitive
  Calorimetric Detection}},}\ }\href {\doibase 10.1103/PhysRevApplied.3.014007}
  {\bibfield  {journal} {\bibinfo  {journal} {Phys. Rev. Applied}\ }\textbf
  {\bibinfo {volume} {3}},\ \bibinfo {pages} {014007} (\bibinfo {year}
  {2015})}\BibitemShut {NoStop}%
\bibitem [{\citenamefont {Suomela}\ \emph {et~al.}(2016)\citenamefont
  {Suomela}, \citenamefont {Kutvonen},\ and\ \citenamefont
  {Ala-Nissila}}]{Suomela}%
  \BibitemOpen
  \bibfield  {author} {\bibinfo {author} {\bibfnamefont {S.}~\bibnamefont
  {Suomela}}, \bibinfo {author} {\bibfnamefont {A.}~\bibnamefont {Kutvonen}}, \
  and\ \bibinfo {author} {\bibfnamefont {T.}~\bibnamefont {Ala-Nissila}},\
  }\bibfield  {title} {\enquote {\bibinfo {title} {{Quantum jump model for a
  system with a finite-size environment}},}\ }\href {\doibase
  10.1103/PhysRevE.93.062106} {\bibfield  {journal} {\bibinfo  {journal} {Phys.
  Rev. E}\ }\textbf {\bibinfo {volume} {93}},\ \bibinfo {pages} {062106}
  (\bibinfo {year} {2016})}\BibitemShut {NoStop}%
\bibitem [{\citenamefont {Scully}\ \emph {et~al.}(2003)\citenamefont {Scully},
  \citenamefont {Zubairy}, \citenamefont {Agarwal},\ and\ \citenamefont
  {Walther}}]{Scully}%
  \BibitemOpen
  \bibfield  {author} {\bibinfo {author} {\bibfnamefont {M.~O.}\ \bibnamefont
  {Scully}}, \bibinfo {author} {\bibfnamefont {M.~S.}\ \bibnamefont {Zubairy}},
  \bibinfo {author} {\bibfnamefont {G.~S.}\ \bibnamefont {Agarwal}}, \ and\
  \bibinfo {author} {\bibfnamefont {H.}~\bibnamefont {Walther}},\ }\bibfield
  {title} {\enquote {\bibinfo {title} {{Extracting work from a single heat bath
  via vanishing quantum coherence}},}\ }\href
  {https://doi.org/10.1126/science.1078955} {\bibfield  {journal} {\bibinfo
  {journal} {Science}\ }\textbf {\bibinfo {volume} {299}},\ \bibinfo {pages}
  {862--864} (\bibinfo {year} {2003})}\BibitemShut {NoStop}%
\bibitem [{\citenamefont {Hardal}\ and\ \citenamefont
  {M{\"u}stecapl{\"A}±o\u{g}lu}(2015)}]{Superradiant}%
  \BibitemOpen
  \bibfield  {author} {\bibinfo {author} {\bibfnamefont {A.~{\"U}.~C.}\
  \bibnamefont {Hardal}}\ and\ \bibinfo {author} {\bibfnamefont {{\"O}~E.}\
  \bibnamefont {M{\"u}stecapl{\"A}±o\u{g}lu}},\ }\bibfield  {title} {\enquote
  {\bibinfo {title} {{Superradiant Quantum Heat Engine}},}\ }\href {\doibase
  10.1038/srep12953} {\bibfield  {journal} {\bibinfo  {journal} {Sci. Rep.}\
  }\textbf {\bibinfo {volume} {5}},\ \bibinfo {pages} {12953} (\bibinfo {year}
  {2015})}\BibitemShut {NoStop}%
\bibitem [{\citenamefont {Lutz}\ and\ \citenamefont
  {Dillenschneider}(2009)}]{LutzCorr}%
  \BibitemOpen
  \bibfield  {author} {\bibinfo {author} {\bibfnamefont {E.}~\bibnamefont
  {Lutz}}\ and\ \bibinfo {author} {\bibfnamefont {R.}~\bibnamefont
  {Dillenschneider}},\ }\bibfield  {title} {\enquote {\bibinfo {title}
  {{Energetics of quantum correlations}},}\ }\href {\doibase
  10.1209/0295-5075/88/50003} {\bibfield  {journal} {\bibinfo  {journal}
  {Europhys. Lett.}\ }\textbf {\bibinfo {volume} {88}},\ \bibinfo {pages}
  {50003} (\bibinfo {year} {2009})}\BibitemShut {NoStop}%
\bibitem [{\citenamefont {Huang}\ \emph {et~al.}(2012)\citenamefont {Huang},
  \citenamefont {Wang},\ and\ \citenamefont {Yi}}]{JaposSqueez}%
  \BibitemOpen
  \bibfield  {author} {\bibinfo {author} {\bibfnamefont {X.~L.}\ \bibnamefont
  {Huang}}, \bibinfo {author} {\bibfnamefont {T.}~\bibnamefont {Wang}}, \ and\
  \bibinfo {author} {\bibfnamefont {X.~X.}\ \bibnamefont {Yi}},\ }\bibfield
  {title} {\enquote {\bibinfo {title} {{Effects of reservoir squeezing on
  quantum systems and work extraction}},}\ }\href {\doibase
  10.1103/PhysRevE.86.051105} {\bibfield  {journal} {\bibinfo  {journal} {Phys.
  Rev. E}\ }\textbf {\bibinfo {volume} {86}},\ \bibinfo {pages} {051105}
  (\bibinfo {year} {2012})}\BibitemShut {NoStop}%
\bibitem [{\citenamefont {Ro{\ss}nagel}\ \emph {et~al.}(2014)\citenamefont
  {Ro{\ss}nagel}, \citenamefont {Abah}, \citenamefont {Schmidt-Kaler},
  \citenamefont {Singer},\ and\ \citenamefont {Lutz}}]{LutzSqueez}%
  \BibitemOpen
  \bibfield  {author} {\bibinfo {author} {\bibfnamefont {J.}~\bibnamefont
  {Ro{\ss}nagel}}, \bibinfo {author} {\bibfnamefont {O.}~\bibnamefont {Abah}},
  \bibinfo {author} {\bibfnamefont {F.}~\bibnamefont {Schmidt-Kaler}}, \bibinfo
  {author} {\bibfnamefont {K.}~\bibnamefont {Singer}}, \ and\ \bibinfo {author}
  {\bibfnamefont {Eric}\ \bibnamefont {Lutz}},\ }\bibfield  {title} {\enquote
  {\bibinfo {title} {{Nanoscale Heat Engine Beyond the Carnot Limit}},}\ }\href
  {https://link.aps.org/doi/10.1103/PhysRevLett.112.030602} {\bibfield
  {journal} {\bibinfo  {journal} {Phys. Rev. Lett.}\ }\textbf {\bibinfo
  {volume} {112}},\ \bibinfo {pages} {030602} (\bibinfo {year}
  {2014})}\BibitemShut {NoStop}%
\bibitem [{\citenamefont {Correa}\ \emph {et~al.}(2014)\citenamefont {Correa},
  \citenamefont {Palao}, \citenamefont {Alonso},\ and\ \citenamefont
  {Adesso}}]{CorreaSqz}%
  \BibitemOpen
  \bibfield  {author} {\bibinfo {author} {\bibfnamefont {L.~A.}\ \bibnamefont
  {Correa}}, \bibinfo {author} {\bibfnamefont {J.~P.}\ \bibnamefont {Palao}},
  \bibinfo {author} {\bibfnamefont {D.}~\bibnamefont {Alonso}}, \ and\ \bibinfo
  {author} {\bibfnamefont {G.}~\bibnamefont {Adesso}},\ }\bibfield  {title}
  {\enquote {\bibinfo {title} {{Quantum-enhanced absorption refrigerators}},}\
  }\href {https://www.nature.com/articles/srep03949} {\bibfield  {journal}
  {\bibinfo  {journal} {Sci. Rep.}\ }\textbf {\bibinfo {volume} {4}},\ \bibinfo
  {pages} {3949} (\bibinfo {year} {2014})}\BibitemShut {NoStop}%
\bibitem [{\citenamefont {Manzano}\ \emph {et~al.}(2016)\citenamefont
  {Manzano}, \citenamefont {Galve}, \citenamefont {Zambrini},\ and\
  \citenamefont {Parrondo}}]{SqzRes}%
  \BibitemOpen
  \bibfield  {author} {\bibinfo {author} {\bibfnamefont {G.}~\bibnamefont
  {Manzano}}, \bibinfo {author} {\bibfnamefont {F.}~\bibnamefont {Galve}},
  \bibinfo {author} {\bibfnamefont {R.}~\bibnamefont {Zambrini}}, \ and\
  \bibinfo {author} {\bibfnamefont {J.~M.~R.}\ \bibnamefont {Parrondo}},\
  }\bibfield  {title} {\enquote {\bibinfo {title} {{Entropy production and
  thermodynamic power of the squeezed thermal reservoir}},}\ }\href
  {https://doi.org/10.1103/PhysRevE.93.052120} {\bibfield  {journal} {\bibinfo
  {journal} {Phys. Rev. E}\ }\textbf {\bibinfo {volume} {93}},\ \bibinfo
  {pages} {052120} (\bibinfo {year} {2016})}\BibitemShut {NoStop}%
\bibitem [{\citenamefont {Klaers}\ \emph {et~al.}(2017)\citenamefont {Klaers},
  \citenamefont {Faelt}, \citenamefont {Imamoglu},\ and\ \citenamefont
  {Togan}}]{Klaers}%
  \BibitemOpen
  \bibfield  {author} {\bibinfo {author} {\bibfnamefont {J.}~\bibnamefont
  {Klaers}}, \bibinfo {author} {\bibfnamefont {S.}~\bibnamefont {Faelt}},
  \bibinfo {author} {\bibfnamefont {A.}~\bibnamefont {Imamoglu}}, \ and\
  \bibinfo {author} {\bibfnamefont {E.}~\bibnamefont {Togan}},\ }\bibfield
  {title} {\enquote {\bibinfo {title} {{Squeezed Thermal Reservoirs as a
  Resource for a Nanomechanical Engine beyond the Carnot Limit}},}\ }\href
  {\doibase 10.1103/PhysRevX.7.031044} {\bibfield  {journal} {\bibinfo
  {journal} {Phys. Rev. X}\ }\textbf {\bibinfo {volume} {7}},\ \bibinfo {pages}
  {031044} (\bibinfo {year} {2017})}\BibitemShut {NoStop}%
\bibitem [{\citenamefont {Niedenzu}\ \emph {et~al.}(2016)\citenamefont
  {Niedenzu}, \citenamefont {Gelbwaser-Klimovsky}, \citenamefont {Kofman},\
  and\ \citenamefont {Kurizki}}]{Niedenzu}%
  \BibitemOpen
  \bibfield  {author} {\bibinfo {author} {\bibfnamefont {W.}~\bibnamefont
  {Niedenzu}}, \bibinfo {author} {\bibfnamefont {D.}~\bibnamefont
  {Gelbwaser-Klimovsky}}, \bibinfo {author} {\bibfnamefont {A.~G.}\
  \bibnamefont {Kofman}}, \ and\ \bibinfo {author} {\bibfnamefont
  {G.}~\bibnamefont {Kurizki}},\ }\bibfield  {title} {\enquote {\bibinfo
  {title} {{On the operation of machines powered by quantum non-thermal
  baths}},}\ }\href {\doibase 10.1088/1367-2630/18/8/083012} {\bibfield
  {journal} {\bibinfo  {journal} {New J. Phys.}\ }\textbf {\bibinfo {volume}
  {18}},\ \bibinfo {pages} {083012} (\bibinfo {year} {2016})}\BibitemShut
  {NoStop}%
\bibitem [{\citenamefont {Reeb}\ and\ \citenamefont {Wolf}(2014)}]{ReebWolf}%
  \BibitemOpen
  \bibfield  {author} {\bibinfo {author} {\bibfnamefont {D.}~\bibnamefont
  {Reeb}}\ and\ \bibinfo {author} {\bibfnamefont {M.~M.}\ \bibnamefont
  {Wolf}},\ }\bibfield  {title} {\enquote {\bibinfo {title} {{An improved
  Landauer principle with finite-size corrections}},}\ }\href {\doibase
  10.1088/1367-2630/16/10/103011} {\bibfield  {journal} {\bibinfo  {journal}
  {New J. Phys.}\ }\textbf {\bibinfo {volume} {16}},\ \bibinfo {pages} {103011}
  (\bibinfo {year} {2014})}\BibitemShut {NoStop}%
\bibitem [{\citenamefont {Goold}\ \emph {et~al.}(2015)\citenamefont {Goold},
  \citenamefont {Paternostro},\ and\ \citenamefont {Modi}}]{GooldLand}%
  \BibitemOpen
  \bibfield  {author} {\bibinfo {author} {\bibfnamefont {J.}~\bibnamefont
  {Goold}}, \bibinfo {author} {\bibfnamefont {M.}~\bibnamefont {Paternostro}},
  \ and\ \bibinfo {author} {\bibfnamefont {K.}~\bibnamefont {Modi}},\
  }\bibfield  {title} {\enquote {\bibinfo {title} {{Nonequilibrium Quantum
  Landauer Principle}},}\ }\href {\doibase 10.1103/PhysRevLett.114.060602}
  {\bibfield  {journal} {\bibinfo  {journal} {Phys. Rev. Lett.}\ }\textbf
  {\bibinfo {volume} {114}},\ \bibinfo {pages} {060602} (\bibinfo {year}
  {2015})}\BibitemShut {NoStop}%
\bibitem [{\citenamefont {Esposito}\ \emph {et~al.}(2010)\citenamefont
  {Esposito}, \citenamefont {Lindenberg},\ and\ \citenamefont {{Van den
  Broeck}}}]{EspositoNJP}%
  \BibitemOpen
  \bibfield  {author} {\bibinfo {author} {\bibfnamefont {M.}~\bibnamefont
  {Esposito}}, \bibinfo {author} {\bibfnamefont {K.}~\bibnamefont
  {Lindenberg}}, \ and\ \bibinfo {author} {\bibfnamefont {C.}~\bibnamefont
  {{Van den Broeck}}},\ }\bibfield  {title} {\enquote {\bibinfo {title}
  {{Entropy production as correlation between system and reservoir}},}\ }\href
  {\doibase 10.1088/1367-2630/12/1/013013} {\bibfield  {journal} {\bibinfo
  {journal} {New J. Phys.}\ }\textbf {\bibinfo {volume} {12}},\ \bibinfo
  {pages} {013013} (\bibinfo {year} {2010})}\BibitemShut {NoStop}%
\bibitem [{\citenamefont {Sagawa}(2013)}]{SagawaEntropies}%
  \BibitemOpen
  \bibfield  {author} {\bibinfo {author} {\bibfnamefont {T.}~\bibnamefont
  {Sagawa}},\ }\bibfield  {title} {\enquote {\bibinfo {title} {{Second law-like
  inequalitites with quantum relative entropy: An introduction}},}\ }in\
  \href@noop {} {\emph {\bibinfo {booktitle} {{Lectures on quantum computing,
  thermodynamics and statistical physics}}}},\ \bibinfo {series} {{Kinki
  University Series on Quantum Computing}}, Vol.~\bibinfo {volume} {8},\
  \bibinfo {editor} {edited by\ \bibinfo {editor} {\bibfnamefont
  {M.}~\bibnamefont {Nakahara}}}\ (\bibinfo  {publisher} {World Scientific New
  Jersey},\ \bibinfo {year} {2013})\BibitemShut {NoStop}%
\bibitem [{\citenamefont {Horowitz}\ and\ \citenamefont
  {Sagawa}(2014)}]{JordanSagawa}%
  \BibitemOpen
  \bibfield  {author} {\bibinfo {author} {\bibfnamefont {J.~M.}\ \bibnamefont
  {Horowitz}}\ and\ \bibinfo {author} {\bibfnamefont {T.}~\bibnamefont
  {Sagawa}},\ }\bibfield  {title} {\enquote {\bibinfo {title} {{Equivalent
  definitions of the quantum nonadiabatic entropy production}},}\ }\href
  {https://link.springer.com/article/10.1007%2Fs10955-014-0991-1} {\bibfield
  {journal} {\bibinfo  {journal} {J. Stat. Phys.}\ }\textbf {\bibinfo {volume}
  {156}},\ \bibinfo {pages} {55--65} (\bibinfo {year} {2014})}\BibitemShut
  {NoStop}%
\bibitem [{\citenamefont {Leggio}\ \emph {et~al.}(2013)\citenamefont {Leggio},
  \citenamefont {Napoli}, \citenamefont {Messina},\ and\ \citenamefont
  {Breuer}}]{Leggio}%
  \BibitemOpen
  \bibfield  {author} {\bibinfo {author} {\bibfnamefont {B.}~\bibnamefont
  {Leggio}}, \bibinfo {author} {\bibfnamefont {A.}~\bibnamefont {Napoli}},
  \bibinfo {author} {\bibfnamefont {A.}~\bibnamefont {Messina}}, \ and\
  \bibinfo {author} {\bibfnamefont {H.~P.}\ \bibnamefont {Breuer}},\ }\bibfield
   {title} {\enquote {\bibinfo {title} {{Entropy production and information
  fluctuations along quantum trajectories}},}\ }\href {\doibase
  10.1103/PhysRevA.88.042111} {\bibfield  {journal} {\bibinfo  {journal} {Phys.
  Rev. A}\ }\textbf {\bibinfo {volume} {88}},\ \bibinfo {pages} {042111}
  (\bibinfo {year} {2013})}\BibitemShut {NoStop}%
\bibitem [{\citenamefont {Elouard}\ \emph
  {et~al.}(2017{\natexlab{a}})\citenamefont {Elouard}, \citenamefont
  {Herrera-Mart{\'i}}, \citenamefont {Clusel},\ and\ \citenamefont
  {Auff{\`e}ves}}]{Auffeves}%
  \BibitemOpen
  \bibfield  {author} {\bibinfo {author} {\bibfnamefont {C.}~\bibnamefont
  {Elouard}}, \bibinfo {author} {\bibfnamefont {D.~A.}\ \bibnamefont
  {Herrera-Mart{\'i}}}, \bibinfo {author} {\bibfnamefont {M.}~\bibnamefont
  {Clusel}}, \ and\ \bibinfo {author} {\bibfnamefont {A.}~\bibnamefont
  {Auff{\`e}ves}},\ }\bibfield  {title} {\enquote {\bibinfo {title} {{The role
  of quantum measurement in stochastic thermodynamics}},}\ }\href
  {https://www.nature.com/articles/s41534-017-0008-4} {\bibfield  {journal}
  {\bibinfo  {journal} {njp Quantum Information}\ }\textbf {\bibinfo {volume}
  {3}},\ \bibinfo {pages} {9} (\bibinfo {year}
  {2017}{\natexlab{a}})}\BibitemShut {NoStop}%
\bibitem [{\citenamefont {Elouard}\ \emph
  {et~al.}(2017{\natexlab{b}})\citenamefont {Elouard}, \citenamefont
  {Bernardes}, \citenamefont {Carvalho}, \citenamefont {Santos},\ and\
  \citenamefont {Auff{\`e}ves}}]{Auffeves2}%
  \BibitemOpen
  \bibfield  {author} {\bibinfo {author} {\bibfnamefont {C.}~\bibnamefont
  {Elouard}}, \bibinfo {author} {\bibfnamefont {N.~K.}\ \bibnamefont
  {Bernardes}}, \bibinfo {author} {\bibfnamefont {A.~R.~R.}\ \bibnamefont
  {Carvalho}}, \bibinfo {author} {\bibfnamefont {M.~F.}\ \bibnamefont
  {Santos}}, \ and\ \bibinfo {author} {\bibfnamefont {A.}~\bibnamefont
  {Auff{\`e}ves}},\ }\bibfield  {title} {\enquote {\bibinfo {title} {{Probing
  quantum fluctuation theorems in engineered reservoirs}},}\ }\href
  {http://stacks.iop.org/1367-2630/19/i=10/a=103011} {\bibfield  {journal}
  {\bibinfo  {journal} {New J. Phys.}\ }\textbf {\bibinfo {volume} {19}},\
  \bibinfo {pages} {103011} (\bibinfo {year} {2017}{\natexlab{b}})}\BibitemShut
  {NoStop}%
\bibitem [{\citenamefont {Santos}\ \emph {et~al.}(2017)\citenamefont {Santos},
  \citenamefont {Landi},\ and\ \citenamefont {Paternostro}}]{Santos}%
  \BibitemOpen
  \bibfield  {author} {\bibinfo {author} {\bibfnamefont {J.~P.}\ \bibnamefont
  {Santos}}, \bibinfo {author} {\bibfnamefont {G.~T.}\ \bibnamefont {Landi}}, \
  and\ \bibinfo {author} {\bibfnamefont {M.}~\bibnamefont {Paternostro}},\
  }\bibfield  {title} {\enquote {\bibinfo {title} {{Wigner Entropy Production
  Rate}},}\ }\href {\doibase 10.1103/PhysRevLett.118.220601} {\bibfield
  {journal} {\bibinfo  {journal} {Phys. Rev. Lett.}\ }\textbf {\bibinfo
  {volume} {118}},\ \bibinfo {pages} {220601} (\bibinfo {year}
  {2017})}\BibitemShut {NoStop}%
\bibitem [{\citenamefont {Bera}\ \emph {et~al.}(2017)\citenamefont {Bera},
  \citenamefont {Riera}, \citenamefont {Lewestein},\ and\ \citenamefont
  {Winter}}]{Lewestein}%
  \BibitemOpen
  \bibfield  {author} {\bibinfo {author} {\bibfnamefont {M.~N.}\ \bibnamefont
  {Bera}}, \bibinfo {author} {\bibfnamefont {A.}~\bibnamefont {Riera}},
  \bibinfo {author} {\bibfnamefont {M.}~\bibnamefont {Lewestein}}, \ and\
  \bibinfo {author} {\bibfnamefont {A.}~\bibnamefont {Winter}},\ }\bibfield
  {title} {\enquote {\bibinfo {title} {{Thermodynamics as a consequence of
  information conservation}},}\ }\href {https://128.84.21.199/abs/1707.01750}
  {\bibfield  {journal} {\bibinfo  {journal} {arXiv:1707.01750}\ } (\bibinfo
  {year} {2017})}\BibitemShut {NoStop}%
\bibitem [{\citenamefont {Groisman}\ \emph {et~al.}(2005)\citenamefont
  {Groisman}, \citenamefont {Popescu},\ and\ \citenamefont
  {Winter}}]{Groisman}%
  \BibitemOpen
  \bibfield  {author} {\bibinfo {author} {\bibfnamefont {B.}~\bibnamefont
  {Groisman}}, \bibinfo {author} {\bibfnamefont {S.}~\bibnamefont {Popescu}}, \
  and\ \bibinfo {author} {\bibfnamefont {A.}~\bibnamefont {Winter}},\
  }\bibfield  {title} {\enquote {\bibinfo {title} {{Quantum, classical, and
  total amount of correlations in a quantum state}},}\ }\href {\doibase
  10.1103/PhysRevA.72.032317} {\bibfield  {journal} {\bibinfo  {journal} {Phys.
  Rev. A}\ }\textbf {\bibinfo {volume} {72}},\ \bibinfo {pages} {032317}
  (\bibinfo {year} {2005})}\BibitemShut {NoStop}%
\bibitem [{\citenamefont {Kraus}\ \emph {et~al.}(1983)\citenamefont {Kraus},
  \citenamefont {B{\"o}hm}, \citenamefont {Dollard},\ and\ \citenamefont
  {Wootters}}]{Kraus}%
  \BibitemOpen
  \bibfield  {author} {\bibinfo {author} {\bibfnamefont {K.}~\bibnamefont
  {Kraus}}, \bibinfo {author} {\bibfnamefont {A.}~\bibnamefont {B{\"o}hm}},
  \bibinfo {author} {\bibfnamefont {J.~D.}\ \bibnamefont {Dollard}}, \ and\
  \bibinfo {author} {\bibfnamefont {W.~H.}\ \bibnamefont {Wootters}},\
  }\href@noop {} {\emph {\bibinfo {title} {{States, effects, and operations :
  fundamental notions of quantum theory}}}},\ {Lecture notes in physics}\
  (\bibinfo  {publisher} {Springer-Verlag},\ \bibinfo {address} {Berlin},\
  \bibinfo {year} {1983})\BibitemShut {NoStop}%
\bibitem [{\citenamefont {Wiseman}\ and\ \citenamefont
  {Milburn}(2010)}]{Wiseman}%
  \BibitemOpen
  \bibfield  {author} {\bibinfo {author} {\bibfnamefont {H.~M.}\ \bibnamefont
  {Wiseman}}\ and\ \bibinfo {author} {\bibfnamefont {G.~J.}\ \bibnamefont
  {Milburn}},\ }\href@noop {} {\emph {\bibinfo {title} {{Quantum Measurement
  and Control}}}}\ (\bibinfo  {publisher} {Cambridge University Press},\
  \bibinfo {address} {Cambridge},\ \bibinfo {year} {2010})\BibitemShut
  {NoStop}%
\bibitem [{\citenamefont {Parrondo}\ \emph {et~al.}(2015)\citenamefont
  {Parrondo}, \citenamefont {Horowitz},\ and\ \citenamefont
  {Sagawa}}]{ThermoInfo}%
  \BibitemOpen
  \bibfield  {author} {\bibinfo {author} {\bibfnamefont {J.~M.~R.}\
  \bibnamefont {Parrondo}}, \bibinfo {author} {\bibfnamefont {J.~M.}\
  \bibnamefont {Horowitz}}, \ and\ \bibinfo {author} {\bibfnamefont
  {T.}~\bibnamefont {Sagawa}},\ }\bibfield  {title} {\enquote {\bibinfo {title}
  {{Thermodynamics of information}},}\ }\href
  {http://www.nature.com/nphys/journal/v11/n2/full/nphys3230.html} {\bibfield
  {journal} {\bibinfo  {journal} {Nat. Phys.}\ }\textbf {\bibinfo {volume}
  {11}},\ \bibinfo {pages} {131--139} (\bibinfo {year} {2015})}\BibitemShut
  {NoStop}%
\bibitem [{\citenamefont {Nielsen}\ and\ \citenamefont
  {Chuang}(2000)}]{NielsenChuang}%
  \BibitemOpen
  \bibfield  {author} {\bibinfo {author} {\bibfnamefont {M.~A.}\ \bibnamefont
  {Nielsen}}\ and\ \bibinfo {author} {\bibfnamefont {I.~L.}\ \bibnamefont
  {Chuang}},\ }\href@noop {} {\emph {\bibinfo {title} {{Quantum Computation and
  Quantum Information}}}}\ (\bibinfo  {publisher} {Cambridge University
  Press},\ \bibinfo {address} {Cambridge},\ \bibinfo {year} {2000})\BibitemShut
  {NoStop}%
\bibitem [{\citenamefont {Luo}(2008)}]{Luo08}%
  \BibitemOpen
  \bibfield  {author} {\bibinfo {author} {\bibfnamefont {S.}~\bibnamefont
  {Luo}},\ }\bibfield  {title} {\enquote {\bibinfo {title} {{Using
  measurement-induced disturbance to characterize correlations as classical or
  quantum}},}\ }\href {\doibase 10.1103/PhysRevA.77.022301} {\bibfield
  {journal} {\bibinfo  {journal} {Phys. Rev. A}\ }\textbf {\bibinfo {volume}
  {77}},\ \bibinfo {pages} {022301} (\bibinfo {year} {2008})}\BibitemShut
  {NoStop}%
\bibitem [{\citenamefont {Modi}\ \emph {et~al.}(2012)\citenamefont {Modi},
  \citenamefont {Brodutch}, \citenamefont {Cable}, \citenamefont {Paterek},\
  and\ \citenamefont {Vedral}}]{ModiRev}%
  \BibitemOpen
  \bibfield  {author} {\bibinfo {author} {\bibfnamefont {K.}~\bibnamefont
  {Modi}}, \bibinfo {author} {\bibfnamefont {A.}~\bibnamefont {Brodutch}},
  \bibinfo {author} {\bibfnamefont {H.}~\bibnamefont {Cable}}, \bibinfo
  {author} {\bibfnamefont {T.}~\bibnamefont {Paterek}}, \ and\ \bibinfo
  {author} {\bibfnamefont {V.}~\bibnamefont {Vedral}},\ }\bibfield  {title}
  {\enquote {\bibinfo {title} {{The classical-quantum boundary for
  correlations: Discord and related measures}},}\ }\href {\doibase
  10.1103/RevModPhys.84.1655} {\bibfield  {journal} {\bibinfo  {journal} {Rev.
  Mod. Phys.}\ }\textbf {\bibinfo {volume} {84}},\ \bibinfo {pages} {1655}
  (\bibinfo {year} {2012})}\BibitemShut {NoStop}%
\bibitem [{\citenamefont {Partovi}(2008)}]{Par08}%
  \BibitemOpen
  \bibfield  {author} {\bibinfo {author} {\bibfnamefont {M.~Hossein}\
  \bibnamefont {Partovi}},\ }\bibfield  {title} {\enquote {\bibinfo {title}
  {{Entanglement versus Stosszahlansatz: Disappearance of the thermodynamic
  arrow in a high-correlation environment}},}\ }\href {\doibase
  10.1103/PhysRevE.77.021110} {\bibfield  {journal} {\bibinfo  {journal} {Phys.
  Rev. E}\ }\textbf {\bibinfo {volume} {77}},\ \bibinfo {pages} {021110}
  (\bibinfo {year} {2008})}\BibitemShut {NoStop}%
\bibitem [{\citenamefont {Jennings}\ and\ \citenamefont
  {Rudolph}(2010)}]{Jen10}%
  \BibitemOpen
  \bibfield  {author} {\bibinfo {author} {\bibfnamefont {D.}~\bibnamefont
  {Jennings}}\ and\ \bibinfo {author} {\bibfnamefont {T.}~\bibnamefont
  {Rudolph}},\ }\bibfield  {title} {\enquote {\bibinfo {title} {{Entanglement
  and the thermodynamic arrow of time}},}\ }\href {\doibase
  10.1103/PhysRevE.81.061130} {\bibfield  {journal} {\bibinfo  {journal} {Phys.
  Rev. E}\ }\textbf {\bibinfo {volume} {81}},\ \bibinfo {pages} {061130}
  (\bibinfo {year} {2010})}\BibitemShut {NoStop}%
\bibitem [{\citenamefont {Haake}(2010)}]{haake}%
  \BibitemOpen
  \bibfield  {author} {\bibinfo {author} {\bibfnamefont {F.}~\bibnamefont
  {Haake}},\ }\href@noop {} {\emph {\bibinfo {title} {{Quantum signatures of
  chaos}}}},\ \bibinfo {edition} {3rd}\ ed.,\ {Springer series in
  synergetics,}\ (\bibinfo  {publisher} {Springer},\ \bibinfo {address}
  {Berlin},\ \bibinfo {year} {2010})\BibitemShut {NoStop}%
\bibitem [{\citenamefont {Andrieux}\ and\ \citenamefont
  {Gaspard}(2008)}]{Andrieux-Gaspard}%
  \BibitemOpen
  \bibfield  {author} {\bibinfo {author} {\bibfnamefont {D.}~\bibnamefont
  {Andrieux}}\ and\ \bibinfo {author} {\bibfnamefont {P.}~\bibnamefont
  {Gaspard}},\ }\bibfield  {title} {\enquote {\bibinfo {title} {{Quantum Work
  Relations and Response Theory}},}\ }\href {\doibase
  10.1103/PhysRevLett.100.230404} {\bibfield  {journal} {\bibinfo  {journal}
  {Phys. Rev. Lett.}\ }\textbf {\bibinfo {volume} {100}},\ \bibinfo {pages}
  {230404} (\bibinfo {year} {2008})}\BibitemShut {NoStop}%
\bibitem [{\citenamefont {Sagawa}\ and\ \citenamefont
  {Ueda}(2013)}]{StochasticMutual}%
  \BibitemOpen
  \bibfield  {author} {\bibinfo {author} {\bibfnamefont {T.}~\bibnamefont
  {Sagawa}}\ and\ \bibinfo {author} {\bibfnamefont {M.}~\bibnamefont {Ueda}},\
  }\bibfield  {title} {\enquote {\bibinfo {title} {{Role of mutual information
  in entropy production under information exchanges}},}\ }\href {\doibase
  10.1088/1367-2630/15/12/125012} {\bibfield  {journal} {\bibinfo  {journal}
  {New J. Phys.}\ }\textbf {\bibinfo {volume} {15}},\ \bibinfo {pages} {125012}
  (\bibinfo {year} {2013})}\BibitemShut {NoStop}%
\bibitem [{\citenamefont {Crooks}(2008)}]{Crooks}%
  \BibitemOpen
  \bibfield  {author} {\bibinfo {author} {\bibfnamefont {G.~E.}\ \bibnamefont
  {Crooks}},\ }\bibfield  {title} {\enquote {\bibinfo {title} {{Quantum
  operation time reversal}},}\ }\href {\doibase 10.1103/PhysRevA.77.034101}
  {\bibfield  {journal} {\bibinfo  {journal} {Phys. Rev. A}\ }\textbf {\bibinfo
  {volume} {77}},\ \bibinfo {pages} {034101} (\bibinfo {year}
  {2008})}\BibitemShut {NoStop}%
\bibitem [{\citenamefont {Fagnola}\ and\ \citenamefont
  {Umanit{\`a}}(2007)}]{Fagnola:2007hj}%
  \BibitemOpen
  \bibfield  {author} {\bibinfo {author} {\bibfnamefont {F.}~\bibnamefont
  {Fagnola}}\ and\ \bibinfo {author} {\bibfnamefont {V.}~\bibnamefont
  {Umanit{\`a}}},\ }\bibfield  {title} {\enquote {\bibinfo {title} {{Generators
  of detailed balance quantum Markov semigroups}},}\ }\href {\doibase
  10.1142/S0219025707002762} {\bibfield  {journal} {\bibinfo  {journal} {Infin.
  Dimens. Anal. Quantum. Probab. Relat. Top.}\ }\textbf {\bibinfo {volume}
  {10}},\ \bibinfo {pages} {335--363} (\bibinfo {year} {2007})}\BibitemShut
  {NoStop}%
\bibitem [{\citenamefont {Lindblad}(1976)}]{Lindblad}%
  \BibitemOpen
  \bibfield  {author} {\bibinfo {author} {\bibfnamefont {G.}~\bibnamefont
  {Lindblad}},\ }\bibfield  {title} {\enquote {\bibinfo {title} {{On the
  generators of quantum dynamical semigroups}},}\ }\href
  {https://projecteuclid.org/euclid.cmp/1103899849} {\bibfield  {journal}
  {\bibinfo  {journal} {Commun. Math. Phys.}\ }\textbf {\bibinfo {volume}
  {48}},\ \bibinfo {pages} {119--130} (\bibinfo {year} {1976})}\BibitemShut
  {NoStop}%
\bibitem [{\citenamefont {Szczygielski}\ \emph {et~al.}(2013)\citenamefont
  {Szczygielski}, \citenamefont {Gelbwaser-Klimovsky},\ and\ \citenamefont
  {Alicki}}]{Szczygielski:2013wc}%
  \BibitemOpen
  \bibfield  {author} {\bibinfo {author} {\bibfnamefont {K.}~\bibnamefont
  {Szczygielski}}, \bibinfo {author} {\bibfnamefont {D.}~\bibnamefont
  {Gelbwaser-Klimovsky}}, \ and\ \bibinfo {author} {\bibfnamefont
  {R.}~\bibnamefont {Alicki}},\ }\bibfield  {title} {\enquote {\bibinfo {title}
  {{Markovian master equation and thermodynamics of a two-level system in a
  strong laser field}},}\ }\href {https://doi.org/10.1103/PhysRevE.87.012120}
  {\bibfield  {journal} {\bibinfo  {journal} {Phys. Rev. E}\ }\textbf {\bibinfo
  {volume} {87}},\ \bibinfo {pages} {012120} (\bibinfo {year}
  {2013})}\BibitemShut {NoStop}%
\bibitem [{\citenamefont {Horowitz}(2012)}]{JordanPRA}%
  \BibitemOpen
  \bibfield  {author} {\bibinfo {author} {\bibfnamefont {J.~M.}\ \bibnamefont
  {Horowitz}},\ }\bibfield  {title} {\enquote {\bibinfo {title}
  {{Quantum-trajectory approach to the stochastic thermodynamics of a forced
  harmonic oscillator}},}\ }\href {https://doi.org/10.1103/PhysRevE.85.031110}
  {\bibfield  {journal} {\bibinfo  {journal} {Phys. Rev. E}\ }\textbf {\bibinfo
  {volume} {85}},\ \bibinfo {pages} {031110} (\bibinfo {year}
  {2012})}\BibitemShut {NoStop}%
\bibitem [{\citenamefont {Spohn}(1978)}]{Spohn}%
  \BibitemOpen
  \bibfield  {author} {\bibinfo {author} {\bibfnamefont {H.}~\bibnamefont
  {Spohn}},\ }\bibfield  {title} {\enquote {\bibinfo {title} {{Entropy
  production for quantum dynamical semigroups}},}\ }\href {\doibase
  10.1063/1.523789} {\bibfield  {journal} {\bibinfo  {journal} {J. Math.
  Phys.}\ }\textbf {\bibinfo {volume} {19}},\ \bibinfo {pages} {1227--1230}
  (\bibinfo {year} {1978})}\BibitemShut {NoStop}%
\bibitem [{\citenamefont {Wu}\ \emph {et~al.}(2009)\citenamefont {Wu},
  \citenamefont {Poulsen},\ and\ \citenamefont {M{\o}lmer}}]{Simdisc1}%
  \BibitemOpen
  \bibfield  {author} {\bibinfo {author} {\bibfnamefont {S.}~\bibnamefont
  {Wu}}, \bibinfo {author} {\bibfnamefont {U.~V.}\ \bibnamefont {Poulsen}}, \
  and\ \bibinfo {author} {\bibfnamefont {K.}~\bibnamefont {M{\o}lmer}},\
  }\bibfield  {title} {\enquote {\bibinfo {title} {{Correlations in local
  measurements on a quantum state, and complementarity as an explanation of
  nonclassicality}},}\ }\href {\doibase 10.1103/PhysRevA.80.032319} {\bibfield
  {journal} {\bibinfo  {journal} {Phys. Rev. A}\ }\textbf {\bibinfo {volume}
  {80}},\ \bibinfo {pages} {032319} (\bibinfo {year} {2009})}\BibitemShut
  {NoStop}%
\bibitem [{\citenamefont {Girolami}\ \emph {et~al.}(2011)\citenamefont
  {Girolami}, \citenamefont {Paternostro},\ and\ \citenamefont
  {Adesso}}]{Simdisc2}%
  \BibitemOpen
  \bibfield  {author} {\bibinfo {author} {\bibfnamefont {D.}~\bibnamefont
  {Girolami}}, \bibinfo {author} {\bibfnamefont {M.}~\bibnamefont
  {Paternostro}}, \ and\ \bibinfo {author} {\bibfnamefont {G.}~\bibnamefont
  {Adesso}},\ }\bibfield  {title} {\enquote {\bibinfo {title} {{Faithful
  nonclassicality indicators and extremal quantum correlations in two-qubit
  states}},}\ }\href {http://stacks.iop.org/1751-8121/44/i=35/a=352002}
  {\bibfield  {journal} {\bibinfo  {journal} {J. Phys. A: Math. Theor.}\
  }\textbf {\bibinfo {volume} {44}},\ \bibinfo {pages} {352002} (\bibinfo
  {year} {2011})}\BibitemShut {NoStop}%
\bibitem [{\citenamefont {Ollivier}\ and\ \citenamefont
  {Zurek}(2001)}]{ZurekDiscord}%
  \BibitemOpen
  \bibfield  {author} {\bibinfo {author} {\bibfnamefont {H.}~\bibnamefont
  {Ollivier}}\ and\ \bibinfo {author} {\bibfnamefont {W.~H.}\ \bibnamefont
  {Zurek}},\ }\bibfield  {title} {\enquote {\bibinfo {title} {{Quantum Discord:
  A Measure of the Quantumness of Correlations}},}\ }\href {\doibase
  10.1103/PhysRevLett.88.017901} {\bibfield  {journal} {\bibinfo  {journal}
  {Phys. Rev. Lett.}\ }\textbf {\bibinfo {volume} {88}},\ \bibinfo {pages}
  {017901} (\bibinfo {year} {2001})}\BibitemShut {NoStop}%
\bibitem [{\citenamefont {Scovil}\ and\ \citenamefont
  {Schulz-DuBois}(1959)}]{Scovil}%
  \BibitemOpen
  \bibfield  {author} {\bibinfo {author} {\bibfnamefont {H.~E.~D.}\
  \bibnamefont {Scovil}}\ and\ \bibinfo {author} {\bibfnamefont {E.~O.}\
  \bibnamefont {Schulz-DuBois}},\ }\bibfield  {title} {\enquote {\bibinfo
  {title} {{Three-Level Masers as Heat Engines}},}\ }\href {\doibase
  10.1103/PhysRevLett.2.262} {\bibfield  {journal} {\bibinfo  {journal} {Phys.
  Rev. Lett.}\ }\textbf {\bibinfo {volume} {2}},\ \bibinfo {pages} {262}
  (\bibinfo {year} {1959})}\BibitemShut {NoStop}%
\bibitem [{\citenamefont {Geusic}\ \emph {et~al.}(1967)\citenamefont {Geusic},
  \citenamefont {Schulz-DuBois},\ and\ \citenamefont {Scovil}}]{Geusic}%
  \BibitemOpen
  \bibfield  {author} {\bibinfo {author} {\bibfnamefont {J.~E.}\ \bibnamefont
  {Geusic}}, \bibinfo {author} {\bibfnamefont {E.~O.}\ \bibnamefont
  {Schulz-DuBois}}, \ and\ \bibinfo {author} {\bibfnamefont {H.~E.~D.}\
  \bibnamefont {Scovil}},\ }\bibfield  {title} {\enquote {\bibinfo {title}
  {{Quantum Equivalent of the Carnot Cycle}},}\ }\href {\doibase
  10.1103/PhysRev.156.343} {\bibfield  {journal} {\bibinfo  {journal} {Physical
  Review}\ }\textbf {\bibinfo {volume} {156}},\ \bibinfo {pages} {343}
  (\bibinfo {year} {1967})}\BibitemShut {NoStop}%
\bibitem [{\citenamefont {Palao}\ \emph {et~al.}(2001)\citenamefont {Palao},
  \citenamefont {Kosloff},\ and\ \citenamefont {Gordon}}]{Palao}%
  \BibitemOpen
  \bibfield  {author} {\bibinfo {author} {\bibfnamefont {J.~P.}\ \bibnamefont
  {Palao}}, \bibinfo {author} {\bibfnamefont {R.}~\bibnamefont {Kosloff}}, \
  and\ \bibinfo {author} {\bibfnamefont {J.~M.}\ \bibnamefont {Gordon}},\
  }\bibfield  {title} {\enquote {\bibinfo {title} {{Quantum thermodynamic
  cooling cycle}},}\ }\href {\doibase 10.1103/PhysRevE.64.056130} {\bibfield
  {journal} {\bibinfo  {journal} {Phys. Rev. E}\ }\textbf {\bibinfo {volume}
  {64}},\ \bibinfo {pages} {056130} (\bibinfo {year} {2001})}\BibitemShut
  {NoStop}%
\bibitem [{\citenamefont {Kosloff}\ and\ \citenamefont {Levy}(2014)}]{Kosloff}%
  \BibitemOpen
  \bibfield  {author} {\bibinfo {author} {\bibfnamefont {R.}~\bibnamefont
  {Kosloff}}\ and\ \bibinfo {author} {\bibfnamefont {A.}~\bibnamefont {Levy}},\
  }\bibfield  {title} {\enquote {\bibinfo {title} {{Quantum Heat Engines and
  Refrigerators: Continuous Devices}},}\ }\href
  {https://doi.org/10.1146/annurev-physchem-040513-103724} {\bibfield
  {journal} {\bibinfo  {journal} {Annu. Rev. Phys. Chem.}\ }\textbf {\bibinfo
  {volume} {65}},\ \bibinfo {pages} {365--393} (\bibinfo {year}
  {2014})}\BibitemShut {NoStop}%
\bibitem [{\citenamefont {Mitchison}\ \emph {et~al.}(2015)\citenamefont
  {Mitchison}, \citenamefont {Woods}, \citenamefont {Prior},\ and\
  \citenamefont {Huber}}]{Mitchison}%
  \BibitemOpen
  \bibfield  {author} {\bibinfo {author} {\bibfnamefont {M.~T.}\ \bibnamefont
  {Mitchison}}, \bibinfo {author} {\bibfnamefont {M.~P.}\ \bibnamefont
  {Woods}}, \bibinfo {author} {\bibfnamefont {J.}~\bibnamefont {Prior}}, \ and\
  \bibinfo {author} {\bibfnamefont {M.}~\bibnamefont {Huber}},\ }\bibfield
  {title} {\enquote {\bibinfo {title} {{Coherence-assisted single-shot cooling
  by quantum absorption refrigerators}},}\ }\href {\doibase
  10.1088/1367-2630/17/11/115013} {\bibfield  {journal} {\bibinfo  {journal}
  {New. J. Phys}\ }\textbf {\bibinfo {volume} {17}},\ \bibinfo {pages} {115013}
  (\bibinfo {year} {2015})}\BibitemShut {NoStop}%
\bibitem [{\citenamefont {Brask}\ and\ \citenamefont {Brunner}(2015)}]{Brask}%
  \BibitemOpen
  \bibfield  {author} {\bibinfo {author} {\bibfnamefont {J.~B.}\ \bibnamefont
  {Brask}}\ and\ \bibinfo {author} {\bibfnamefont {N.}~\bibnamefont
  {Brunner}},\ }\bibfield  {title} {\enquote {\bibinfo {title} {{Small quantum
  absorption refrigerator in the transient regime: Time scales, enhanced
  cooling, and entanglement}},}\ }\href {\doibase 10.1103/PhysRevE.92.062101}
  {\bibfield  {journal} {\bibinfo  {journal} {Phys. Rev. E}\ }\textbf {\bibinfo
  {volume} {92}},\ \bibinfo {pages} {062101} (\bibinfo {year}
  {2015})}\BibitemShut {NoStop}%
\bibitem [{\citenamefont {Brunner}\ \emph {et~al.}(2012)\citenamefont
  {Brunner}, \citenamefont {Linden}, \citenamefont {Popescu},\ and\
  \citenamefont {Skrzypczyk}}]{VirtualQubits}%
  \BibitemOpen
  \bibfield  {author} {\bibinfo {author} {\bibfnamefont {N.}~\bibnamefont
  {Brunner}}, \bibinfo {author} {\bibfnamefont {N.}~\bibnamefont {Linden}},
  \bibinfo {author} {\bibfnamefont {S.}~\bibnamefont {Popescu}}, \ and\
  \bibinfo {author} {\bibfnamefont {P.}~\bibnamefont {Skrzypczyk}},\ }\bibfield
   {title} {\enquote {\bibinfo {title} {{Virtual qubits, virtual temperatures,
  and the foundations of thermodynamics}},}\ }\href {\doibase
  10.1103/PhysRevE.85.051117} {\bibfield  {journal} {\bibinfo  {journal} {Phys.
  Rev. E}\ }\textbf {\bibinfo {volume} {85}},\ \bibinfo {pages} {051117}
  (\bibinfo {year} {2012})}\BibitemShut {NoStop}%
\bibitem [{\citenamefont {Skrzypczyk}\ \emph {et~al.}(2015)\citenamefont
  {Skrzypczyk}, \citenamefont {Silva},\ and\ \citenamefont
  {Brunner}}]{VirtualTemperatures}%
  \BibitemOpen
  \bibfield  {author} {\bibinfo {author} {\bibfnamefont {P.}~\bibnamefont
  {Skrzypczyk}}, \bibinfo {author} {\bibfnamefont {R.}~\bibnamefont {Silva}}, \
  and\ \bibinfo {author} {\bibfnamefont {N.}~\bibnamefont {Brunner}},\
  }\bibfield  {title} {\enquote {\bibinfo {title} {{Passivity, complete
  passivity, and virtual temperatures}},}\ }\href {\doibase
  10.1103/PhysRevE.91.052133} {\bibfield  {journal} {\bibinfo  {journal} {Phys.
  Rev. E}\ }\textbf {\bibinfo {volume} {91}},\ \bibinfo {pages} {052133}
  (\bibinfo {year} {2015})}\BibitemShut {NoStop}%
\bibitem [{\citenamefont {Silva}\ \emph {et~al.}(2016)\citenamefont {Silva},
  \citenamefont {Manzano}, \citenamefont {Skrzypczyk},\ and\ \citenamefont
  {Brunner}}]{Ral16}%
  \BibitemOpen
  \bibfield  {author} {\bibinfo {author} {\bibfnamefont {R.}~\bibnamefont
  {Silva}}, \bibinfo {author} {\bibfnamefont {G.}~\bibnamefont {Manzano}},
  \bibinfo {author} {\bibfnamefont {P.}~\bibnamefont {Skrzypczyk}}, \ and\
  \bibinfo {author} {\bibfnamefont {N.}~\bibnamefont {Brunner}},\ }\bibfield
  {title} {\enquote {\bibinfo {title} {{Performance of autonomous quantum
  thermal machines: Hilbert space dimension as a thermodynamical resource}},}\
  }\href {\doibase 10.1103/PhysRevE.94.032120} {\bibfield  {journal} {\bibinfo
  {journal} {Phys. Rev. E}\ }\textbf {\bibinfo {volume} {94}},\ \bibinfo
  {pages} {032120} (\bibinfo {year} {2016})}\BibitemShut {NoStop}%
\bibitem [{\citenamefont {Rivas}\ \emph {et~al.}(2010)\citenamefont {Rivas},
  \citenamefont {Plato}, \citenamefont {Huelga},\ and\ \citenamefont
  {Plenio}}]{rivas2010}%
  \BibitemOpen
  \bibfield  {author} {\bibinfo {author} {\bibfnamefont {A.}~\bibnamefont
  {Rivas}}, \bibinfo {author} {\bibfnamefont {A.~D.~K.}\ \bibnamefont {Plato}},
  \bibinfo {author} {\bibfnamefont {S.~F.}\ \bibnamefont {Huelga}}, \ and\
  \bibinfo {author} {\bibfnamefont {M.~B.}\ \bibnamefont {Plenio}},\ }\bibfield
   {title} {\enquote {\bibinfo {title} {{Markovian master equations: a critical
  study}},}\ }\href {http://stacks.iop.org/1367-2630/12/i=11/a=113032}
  {\bibfield  {journal} {\bibinfo  {journal} {New Journal of Physics}\ }\textbf
  {\bibinfo {volume} {12}},\ \bibinfo {pages} {113032} (\bibinfo {year}
  {2010})}\BibitemShut {NoStop}%
\bibitem [{\citenamefont {Horowitz}\ and\ \citenamefont
  {Esposito}(2016)}]{Horowitz2016}%
  \BibitemOpen
  \bibfield  {author} {\bibinfo {author} {\bibfnamefont {J.~M.}\ \bibnamefont
  {Horowitz}}\ and\ \bibinfo {author} {\bibfnamefont {M.}~\bibnamefont
  {Esposito}},\ }\bibfield  {title} {\enquote {\bibinfo {title} {{Work
  producing reservoirs: Stochastic thermodynamics with generalized Gibbs
  ensembles}},}\ }\href@noop {} {\bibfield  {journal} {\bibinfo  {journal}
  {Phys. Rev. E}\ }\textbf {\bibinfo {volume} {94}},\ \bibinfo {pages}
  {020102(R)} (\bibinfo {year} {2016})}\BibitemShut {NoStop}%
\end{thebibliography}%

\end{document}